\documentclass[%
 amsmath,amssymb,
 aps,
pra,
]{revtex4-2}

\usepackage{graphicx}
\usepackage{dcolumn}
\usepackage{bm}

\begin{document}

\title{The exact solution of the Koga-Widom-Indekeu model and related models of wetting in fluid mixtures}%

\author{A.O.\ Parry}
 \affiliation{Department of Mathematics, Imperial College London, London SW7 2BZ, UK}%
 
\author{C.\ Rasc\'{o}n}
\affiliation{GISC, Departamento de Matem\'aticas, Universidad Carlos III de Madrid, 28911 Legan\'es, Madrid, Spain
}%

\begin{abstract}

We show how a broad class of two-component square-gradient models of wetting may be solved exactly for the surface tensions and density profile paths, and clarify how the presence or absence of critical point wetting, in binary and ternary mixtures, is related to universality and symmetry principles at critical end points. We begin by solving a model of fluid interfaces, first introduced by Koga and Widom, in ternary mixtures showing three phase coexistence. Numerical studies had revealed interesting wetting transitions, as well as curious geometrical properties of the profile paths in the density plane, and led these authors to conjecture expressions for the surface tensions. These conjectures were extended by Koga and Indekeu and predicted that partial wetting may persist up to the line of critical end points, i.e.\ critical point wetting was, unexpectedly, absent. Here, we obtain the exact density profiles and surface tensions for the Koga-Widom-Indekeu (KWI) model using complex analysis and drawing on the theory of algebraic curves. The exact solution determines the location and order of wetting transitions in the surface phase diagram, confirming that critical point wetting is absent. The model also displays the remarkable property that microscopic density profiles are mapped, by a conformal transform, onto the shape of a macroscopic drop near the contact line whose tensions satisfy the Neumann triangle. Two related models, which illustrate the role of the component isotropy, are also discussed: First, wetting by a critical layer where the critical singularities reflect an XY-like Casimir interaction mediated by the wetting film. Secondly, wetting in a binary mixture near a wall where critical point wetting is also absent and component density profiles can be obtained exactly revealing unexpected symmetries and connections to the mesoscopic disjoining pressure. These models suggest that a universality principle governs wetting in fluid mixtures, resolving contradicting results from earlier studies: Critical point wetting is present if the order-parameter components of the mixture describe Ising-like criticality, but is absent if there is a local XY symmetry. Implications for wetting transitions in more microscopic models and in experiments are discussed.\\

\end{abstract}

\maketitle

\tableofcontents

\section{\label{sec:level1}Introduction}

\subsection{Young's equation and critical point wetting in simple fluids}

\begin{figure}[h]
\includegraphics[height=4.cm]{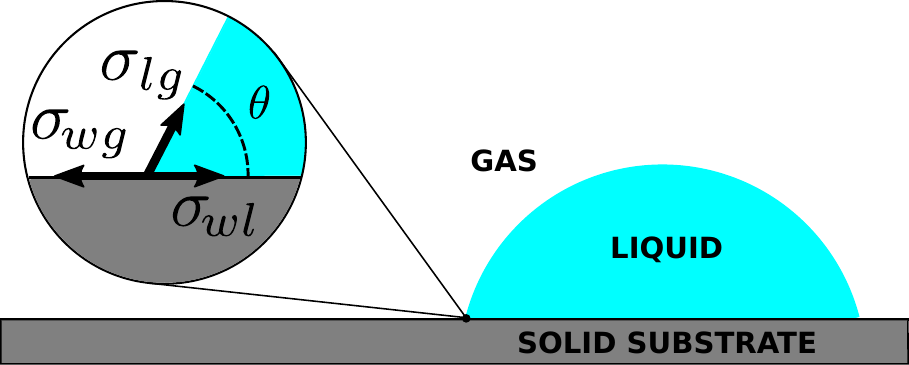}
\caption{\label{Fig01} A macroscopic sessile liquid drop on a smooth substrate. The surface tensions $\sigma_{wg}$, $\sigma_ {wl}$ and $\sigma_{lg}$ satisfy Young's equation, $\sigma_{wg}=\sigma_{wl}+\sigma_{lg}\cos\theta$, which determines the contact angle $\theta$ from the force balance at the contact line (magnified circle).}
\end{figure}

 The phase separation of coexisting states of matter leads naturally to interfacial phenomena governed by the surface tension. A familiar example of this is the contact angle $\theta$ of a sessile drop of liquid on a solid substrate, usually modeled as a smooth wall exerting an external potential. In this case, the equilibrium value of the contact angle is determined by Young's equation
 \begin{equation}
 \sigma_{wg}=\sigma_ {wl}+\sigma_{lg}\cos\theta
 \end{equation}
 which expresses the force balance between the tensions of the wall-gas, wall-liquid and liquid-gas interfaces, which pull the contact line along their own plane - see Fig.\ \ref{Fig01}. The modern theory of wetting began in 1977 when, independently Cahn \cite{Cahn1977}, and Ebner and Saam \cite{Ebner1977} showed, using classical density functional theory \cite{Evans1979}, that a wetting transition may occur in which the contact angle vanishes as the temperature $T$ is increased to a wetting temperature $T_w$. For $T>T_w$, the wall-gas interface is then completely wet by liquid and the surface tensions satisfy Antonov's rule $\sigma_{wg}=\sigma_ {wl}+\sigma_{lg}$. At the wetting transition, the adsorption of liquid at the planar wall-gas interface therefore changes from microscopic to macroscopic and may be first-order or continuous depending on the details of the intermolecular forces and also fluctuation effects - for general reviews see   \cite{Sullivan1986,Dietrich1988,Schick1990,Forgacs1991,Bonn2009}.

 Despite the simplicity of the above macroscopic picture,  several key theoretical issues have remained unclear for much of the last half century. For example, it has only recently been appreciated that, in three-dimensions and for systems with short-ranged forces, critical singularities arise from both the capillary-wave like fluctuations of the liquid-gas interface {\it{and}} bulk fluctuations associated with the thermal Casimir effect, which are needed to resolve longstanding discrepancies between theory and simulation studies \cite{Squarcini2022}. A second example concerns the status of Cahn's original speculation of {\it{critical point wetting}}, i.e.\ that a wetting transition {\it{must}} happen on approaching the bulk critical temperature $T_c$, where liquid-gas coexistence ends. Cahn's reasoning was as follows: The difference between the wall-gas and wall-liquid surface tensions vanishes, on approaching $T_c$, with the same critical exponent as that of the bulk order-parameter, $\rho_l-\rho_g\propto (T_c-T)^\beta$, with $\beta\approx 0.32$, while the liquid-gas tension vanishes as $\sigma_{lg}\propto (T_c-T)^\mu$ with a critical exponent $\mu\approx 1.26$ \cite{Rowlinson1982}. Thus, Cahn argued, if partial wetting prevails at low temperatures, so that $\sigma_{wg}-\sigma_{wl}<\sigma_{lg}$, the difference between the exponents means that Antonov's rule for complete wetting must be obeyed at some temperature prior to $T_c$. This argument has long been known to be wrong since the difference between the surface tensions of the wall-gas and wall-liquid interfaces is not governed by the order-parameter exponent $\beta$. Nevertheless, critical point wetting does occur in simple one-component fluids when the wall-fluid and fluid-fluid intermolecular forces are either both short-ranged \cite{Evans2019} or both long-ranged \cite{Parry2023b}, the understanding of which has only recently been precised. The correct reasoning why this is the case can be traced back to the insights of Nakanishi and Fisher who expounded the deep connection between wetting phase boundaries and surface criticality using a simple square-gradient (Landau) theory \cite{Nakanishi1982}. They showed that the lines of wetting transitions and drying transitions (equivalent to wetting by gas) in the global phase diagram necessarily converge to a surface phase transition, separating regimes of critical adsorption and critical desorption \cite{Nakanishi1982}, at the limiting point of wetting neutrality where the contact angle is $\pi/2$ -- details we shall return to later. This has recently being shown to be the case also when the wall-fluid and fluid-fluid forces are both long-ranged, reflecting universal properties in the critical region \cite{Parry2023b}. For simple fluids, critical point wetting is only absent when there is a mismatch between their ranges \cite{DeGennes1983,Nightingale1985,Evans2019} although even here there are common universal features of the surface phase diagrams linking the wetting phase boundaries with surface criticality \cite{Parry2024}.

 \subsection{The Neumann triangle and wetting in fluid mixtures: critical point wetting or not?}

\begin{figure}[ht]
\includegraphics[width=.8\columnwidth]{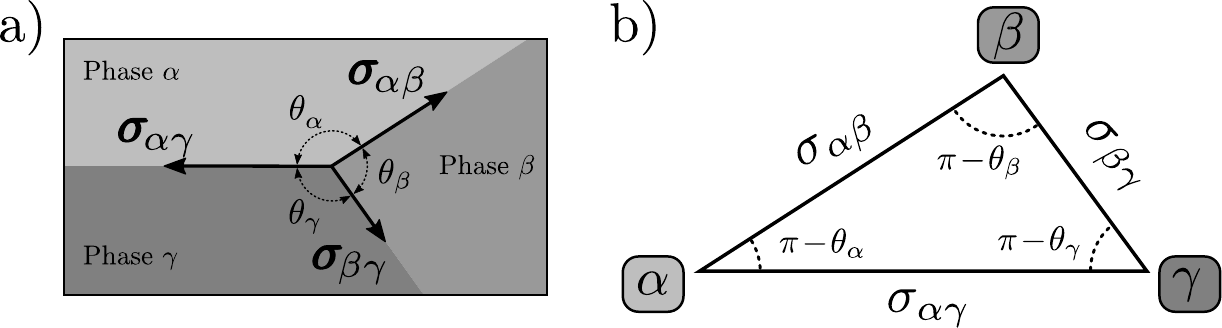}
\caption{\label{Fig02} 
 (a) The macroscopic contact angles, $\theta_\alpha$, $\theta_\beta$, and $\theta_\gamma$ at the contact line of three coexisting fluid phases. The contact line is stabilized by the surface tensions, ${\bm \sigma}_{\alpha\beta}$, ${\bm \sigma}_{\alpha\gamma}$, and ${\bm \sigma}_{\beta\gamma}$, which pull the interfaces along the direction of their planes. (b) The Neumann triangle expressing the relation between the magnitudes of the surface tensions and the complementary of each contact angle.}
\end{figure}

Wetting transitions may also occur in the absence of a solid substrate/wall, involving only fluid-fluid interfaces, as happens, for example, in binary and ternary mixtures, for which the necessity of critical point wetting is much less clear. In this scenario, there may be three coexisting fluid phases, $\alpha$, $\beta$ and $\gamma$, and therefore, three different planar interfaces, labeled $\alpha\beta$, $\beta\gamma$ and $\alpha\gamma$, each with a surface tension. 
The meeting of the three bulk phases along a contact or triple line is then characterised by three macroscopic contact angles $\theta_\alpha$, $\theta_\beta$ and $\theta_\gamma$ -- see Fig.\ \ref{Fig02}a. The values of these angles are determined by the surface tension forces of the planar interfaces, which each pull the contact line along their own plane. The stability of the contact line means that, at equilibrium, the three surface tensions satisfy the condition for force balance 
\begin{equation}
\bm{\sigma}_{\alpha\beta}+\bm{\sigma}_{\beta\gamma}+\bm{\sigma}_{\alpha\gamma}=\bm{0}
\end{equation}
This simple vector relation means that the magnitudes of the surface tensions, $\sigma_{\alpha\beta}$, $\sigma_{\beta\gamma}$ and $\sigma_{\alpha\gamma}$, can be identified with the side lengths of a triangle -- the Neumann triangle \cite{Rowlinson1982} -- whose interior angles are the complementary contact angles $\pi-\theta_\alpha$, $\pi-\theta_\beta$ and $\pi-\theta_\gamma$ -- see Fig.\ref{Fig02}b. The Neumann triangle, therefore, directly encodes the shape of a macroscopic drop in the near vicinity of the contact line, where all three interfaces are planar. 
The elementary cosine rule then identifies that, for example, the contact angle $\theta_\beta$ is given by
\begin{equation}
    \sigma_{\alpha\gamma}^2=\sigma_{\alpha\beta}^2+\sigma_{\beta\gamma}^2+2\:\sigma_{\alpha\beta}\;\sigma_{\beta\gamma}\:\cos\theta_\beta
\end{equation}
 which is the analogue of Young's equation for $\theta_\beta$.
Complete wetting by $\beta$ in this situation means that $\theta_\beta=0$ so the tensions satisfy Antonov's rule,
\begin{equation}\sigma_{\alpha\gamma}\; =\; \sigma_{\alpha\beta}\,+\,\sigma_{\beta\gamma}
\label{Ant}
\end{equation}
implying that $\beta$ intervenes between the $\alpha$ and $\gamma$ phases, i.e.\ the Neumann triangle collapses to a line.

For a ternary fluid mixture, critical point wetting would then refer to the necessity of complete wetting, by one of the phases, as we approach anywhere on the two critical end point (CEP) lines which demarcate the three-phase region and meet at the bulk tricritical point (TCP) \cite{Griffiths1974}. However, the situation is different from that of a wall-fluid interface because all three fluid interfaces must be treated on an equal footing and there is no obvious analogue of an external potential, as for wall-fluid interfaces, whose strength that can be used to induce a wetting transition. In the classic text by Rowlinson and Widom, it was argued that, near the TCP, we can expect that the phase with intermediate density (by convention the $\beta$ phase) completely wets the $\alpha\gamma$ interface. However, the necessity of critical point wetting near the CEP lines was left open \cite{Rowlinson1982}. 

\begin{figure}[h]
\includegraphics[width=.8\columnwidth]{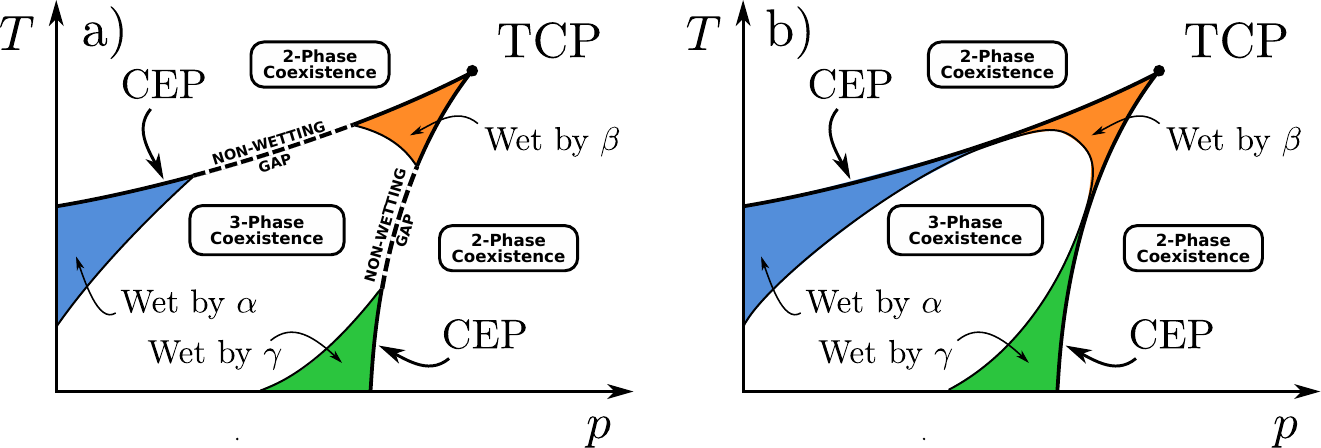}
\caption{\label{Fig03} Schematic surface phase diagrams showing the regions of complete wetting by the three coexisting phases $\alpha$, $\beta$, and $\gamma$ in the temperature $T$ and pressure $p$ plane, with: a) Non-wetting gaps, and  b) Critical point wetting at the critical end point (CEP) lines. The two CEP lines meet at the bulk tricritical point (TCP).}
\end{figure}

The uncertainty regarding critical point wetting in fluid mixtures has recently come to a head, where even for systems with short-ranged forces, different mean-field studies have reached different conclusions. These studies begin from a model grand potential functional theory or similar square-gradient Landau theory which is minimized to obtain the equilibrium planar density profiles, surface tensions and contact angles. Indekeu and Koga have argued against the necessity of critical point wetting, and illustrate this using an elegant two-component square-gradient model for which they conjecture exact results \cite{Indekeu2022}. The model was first introduced, and studied numerically, by Koga and Widom who presented strong evidence that a second-order wetting transition (by $\beta$) at the $\alpha\gamma$ interface occurs prior to the TCP \cite{Koga2008,Koga2016}. This is already much more interesting than the one-component square-gradient description which only shows complete wetting \cite{Rowlinson1982}. Koga and Widom also noted some intriguing geometrical properties of the profile paths, when drawn as a tricuspid in the density plane, which hint at a deeper mathematical structure and guessed expressions for the surface tensions along a line of symmetry in the phase diagram. These guesses were generalized further by Koga and Indekeu into a simple, and apparently exact, formula for the surface tension \cite{Koga2019}, and later applied it to the whole three-phase region and {\it{imposed}} constraints on the values of the bulk component densities due to Griffiths \cite{Griffiths1974}. Indekeu and Koga showed that combining the Griffiths constraints with the conjecture for the surface tension leads to  non-wetting gaps in the surface phase diagram; specifically that there are sections along the CEP lines where there is only partial wetting \cite{Indekeu2022}, i.e.\ where critical point wetting is absent. This is a remarkable prediction coming from a Landau-like, square-gradient, theory modeling systems with only short-ranged forces. In contrast to this, a numerical study of a self-consistent mean-field theory of a ternary mixture with Flory-Huggins interactions, by Leermakers and Ergorov, indicates that critical point wetting is still present and the lines of wetting transition meet at the point of wetting neutrality at the $\alpha\beta$ CEP line \cite{Leermakers2025}.
Wetting neutrality here means that the limiting value of the contact angles are $\theta_\alpha=\theta_\beta=\pi/2$ while $\theta_\gamma=\pi$, mirroring the analogue of wetting neutrality for a simple one- component liquid where $\theta=\pi/2$. This prediction of critical point wetting is consistent with older density functional studies by Telo da Gamma and Evans, of three phase coexistence in a binary mixture, where a wetting transition was found to occur prior to the CEP \cite{TeloDaGama1983a,TeloDaGama1983b}. The implication of these studies is that, in contrast to the non-wetting gaps that appear in the Indekeu-Koga phase diagram, critical point wetting is still present in ternary mixtures. These conflicting possibilities are shown schematically in Fig.\ \ref{Fig03}.

\subsection{Outline}

The purpose of this paper is to show how the Koga-Widom-Indekeu (KWI) model may be solved exactly, deriving the surface tension, contact angles and density profiles, and also to pinpoint exactly what aspect of the model leads to the absence of critical point wetting. We shall show that critical point wetting is absent in this model because the two-components have a local XY symmetry arising from an assumed perfect isotropy in the free-energy -- a symmetry which also affects the universality of Casimir forces and surface critical behaviour.
This explains the discrepancy between the findings of Indekeu and Koga \cite{Indekeu2022} and that of Leermakers and Ergorov \cite{Leermakers2025}, whose model free-energy does not have this symmetry. Our central conclusion is that critical point wetting is to be expected in ordinary fluid mixtures because critical end points belong to the Ising universality class. A brief account of some of our results has appeared in \cite{Parry2024a}. 

In the next section, we recall the Griffiths scaling theory of bulk tricriticality, the simple one-component square-gradient theory and introduce the  KWI model and earlier conjectures for its properties. We then solve the KWI model in two ways. First, in Section III, we show how the density profile trajectories may be determined by integration of the coupled Euler-Lagrange equations yielding the wetting phase boundaries. This demonstrates that critical point wetting is indeed absent in sections of the phase diagram, therefore confirming the earlier conjectures of Koga and Widom and allowing us to derive the Koga-Indekeu conjecture for the surface tension after observing that the trajectories are conformally invariant. We also show how the explicit spatial dependence of the two-components may be determined using the rational parameterization of algebraic curves. In Section IV, we reformulate the problem using complex analysis, which directly exploits the underlying conformal invariance in component space, revealing more clearly what quantities are conserved in the mechanical analogy. This approach shows that the KWI model belongs to a much broader class of exactly solvable two-component Landau theories, which all have a local XY symmetry due to the assumed component isotropy. These models display a remarkable feature in which the tricuspid of the three possible microscopic density profiles in component space can be mapped directly onto the Neumann triangle, and hence onto the macroscopic shape of a drop, by a conformal transform. The consequences of the local XY symmetry in the KWI and related models are then explored in detail. For example, we show that it affects the universality of Casimir forces leading to a wetting transition, governed by distinct critical exponents, when a critical phase is adsorbed between two non-critical phases. In Section V, we show that precisely the same non-wetting gap predicted by the KWI model also occurs in a model of wetting in a binary liquid mixture near a wall, where we are able to derive exact results for the component density profiles, which reveal unexpected symmetries and establish connections to the mesoscopic disjoining pressure. In Section VI, we critically reassess the symmetries present in these models showing that the local XY symmetry leads to different surface critical behaviour compared to that for Ising-like systems. However, these symmetries are not expected for ordinary fluids, something we illustrate explicitly using the Sullivan model of a binary mixture. We show that the  inclusion of anisotropy breaks the local XY symmetry and restores both Ising-like surface critical behaviour and critical point wetting, explaining the discrepancy between the two phase diagrams in Fig.\ \ref{Fig03}, and clarifying what should be expected experimentally. Further calculation details and connections with other works are presented in the Appendices.

\section{The KWI model and its conjectures}

\subsection{Griffiths scaling, Griffiths constraints and the one-component square-gradient theory}

\begin{figure}[t]
\includegraphics[width=.8\columnwidth]{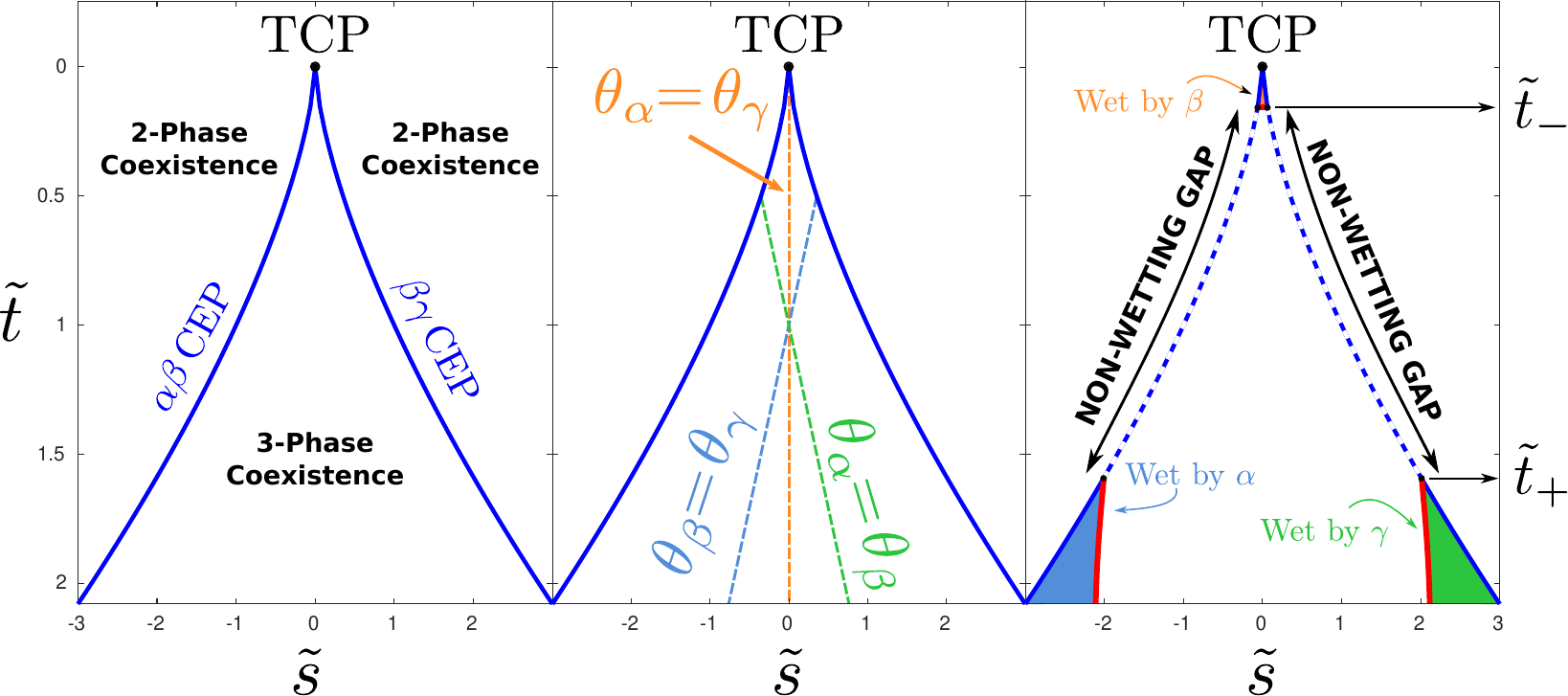}
\caption{\label{Fig04} a) The Griffiths bulk phase diagram, showing the region of three phase coexistence in the $(\tilde s,\tilde t)$ plane, where $\tilde s$ and $\tilde t$ are the pressure-like and temperature-like scaling variables. The CEP lines are located  at $\tilde s\tilde t^{-3/2}=\pm 1$, with the TCP at $\tilde t=\tilde s=0$. b) The lines of symmetry (dashed) in the phase diagram where two tensions are equal. In the one-component description, there is a line of symmetry at $\tilde s=0$, at which $\rho_1^\beta=0$ and $\rho_1^\alpha=-\rho_1^\gamma\equiv a=\sqrt{3\tilde t}$, for which $\sigma_{\alpha\beta}=\sigma_{\beta\gamma}$, implying $\theta_\alpha=\theta_\gamma$. In the two-component KWI description, there are two other lines of symmetry, where two other tensions and contact angles are equal, which are formed when the bulk densities $(\rho_1^\mu,\rho_2^\mu)$, with $\mu=\alpha,\beta,\gamma$ form an isosceles triangle in the density plane. These two additional lines of symmetry terminate at $\tilde t=1/2$ at each CEP line, which are limiting points of wetting neutrality. c) The Indekeu-Koga phase diagrams showing the regions of complete wetting by $\alpha$, $\beta$ and $\gamma$ and non-wetting gaps along the CEP lines between $\tilde t_-=(7-\sqrt{33})/8$ and $\tilde t_+=(7+\sqrt{33})/8$, where critical point wetting is absent.}
\end{figure}

 For a ternary fluid mixture of three different component species, the Gibbs phase rule allows that three bulk fluid phases, labeled $\alpha$, $\beta$ and $\gamma$, may coexist in a region of the temperature and pressure plane bounded by two CEP lines which meet at a bulk TCP, where all phase coexistence ends. Along one CEP line, the $\alpha$ and $\beta$ phases become identical and $\gamma$ remains non-critical, while along the other CEP $\beta$ and $\gamma$ are identical and $\alpha$ is non-critical. A very elegant Landau-like description determining the scaling properties and associated bulk mean-field critical singularities near the TCP and CEPs, was developed by Griffiths \cite{Griffiths1974}. The bulk free-energy density depends on the microscopic densities $n_1$, $n_2$ and $n_3$ of the three different chemical components. This is expressed as $\omega(\rho_1,\rho_2,\rho_3)$, in terms of three scaling densities $\rho_1$, $\rho_2$ and $\rho_3$ which are linear combinations of the microscopic densities relative to their values at the TCP. These scaling densities are coordinates along axes which span the original space $(n_1,n_2,n_3)$ of microscopic densities and which vanish at the TCP, i.e.\ they are the components of an order-parameter appearing in a Landau-like expansion of the free-energy, describing phase coexistence in the three-phase region. The most important of these is the primary density $\rho_1$, the axis of which is fixed by the coexistence curve and determines the critical singularities \cite{Rowlinson1982}. The free-energy density necessarily contains terms up the sixth power of $\rho_1$, as required to describe tricritical singularities. The bulk values of the primary density are solutions of the cubic equation
 \begin{equation}
(\rho_1^\mu)^3-3\,\tilde t\rho_1^\mu+2\tilde s=0
\label{Griff}
\end{equation}
with $\mu=\alpha,\beta,\gamma$. Here, $\tilde t$ and $\tilde s$ are the temperature-like and pressure-like scaling fields respectively, centered at the TCP. Three phase coexistence occurs in the regime $-1\le \tilde s\tilde t^{-3/2}\le 1$, bounded by the two CEP lines which meet at the TCP, where $\tilde s=\tilde t=0$. The primary densities satisfy the useful relation
\begin{equation}
\rho_1^\alpha+\rho_1^\beta+\rho_1^\gamma=0
\end{equation}
and by convention are ordered $\rho_1^\alpha\le\rho_1^\beta\le\rho_1^\gamma$, so that $\beta$ is the phase of intermediate density. The values of the bulk secondary and tertiary densities were also determined by Griffiths from the scaling of the free-energy and the shape of the coexistence curve,  and must behave as the square and cube of the primary density respectively, in order to preserve analytical properties \cite{Griffiths1974,Rowlinson1982} -- something we will return to later. In particular, for discussion of the KWI model, we will need the Griffiths constraint on the bulk secondary density
 \begin{equation}
\rho_2^\mu = -(\rho_1^\mu)^2
 \end{equation}
 which applies to all the phases. The densities of the three bulk phases $\alpha$, $\beta$ and $\gamma$, therefore, lie on the parabola $\rho_2 = -(\rho_1)^2$ in the component plane. Along the $\alpha\beta$ CEP, $\rho_1^\alpha=\rho_1^\beta=-\tilde t^{1/2}$ and $\rho_1^\gamma=2\tilde t^{1/2}$, while along the $\beta\gamma$ CEP, $\rho_1^\beta=\rho_1^\gamma=\tilde t^{1/2}$ and $\rho_1^\alpha=-2\tilde t^{1/2}$. At the TCP, all three densities vanish and $\rho_1^\alpha=\rho_1^\beta=\rho_1^\gamma=0$. The phase diagram has a line of symmetry at $\tilde s=0$, where $\rho_1^\beta=0$ and $\rho_1^\gamma=-\rho_1^\alpha=(3\tilde t)^{1/2}$, along which the surface tensions of the $\alpha\beta$ and $\beta\gamma$ interfaces must be the same - see Fig.\ \ref{Fig04}.

 The simplest square-gradient theory of interfacial properties in the three-phase region is based on just the primary density field, as discussed in detail by Rowlinson and Widom \cite{Rowlinson1982}. Within this one-component description, the grand potential functional, modeling the excess surface contribution to the free-energy per unit area, of a planar interface is written
 \begin{equation}
\Omega[\rho_1]=\int_{-\infty }^{\infty}\!\!\! dz \; \left(\frac{1}{2}\left(\frac{d\rho_1}{dz}\right)^2+\omega(\rho_1)\right)
\label{oneKWI}
\end{equation}
 where $z$ is the coordinate normal to the plane of the interface. The  bulk potential $\omega(\rho_1)$ is the minimum of $\omega(\rho_1,\rho_2,\rho_3)$ with $\rho_1$ fixed and taken to be
 \begin{equation}
 \omega(\rho_1)=(\rho_1-\rho_1^\alpha)^2\,(\rho_1-\rho_1^\beta)^2\,(\rho_1-\rho_1^\gamma)^2
 \end{equation}
 or, equivalently,
 \begin{equation}
 \omega(\rho_1)=(\rho_1^3-3\tilde t\rho_1+2\tilde s)^2
 \label{omega1}
 \end{equation}
   Here, we have conveniently subtracted the bulk contribution to the potential so that $\omega(\rho_1^\mu)=0$, and also set the coefficient of the gradient term to a half without loss of generality. The square-gradient term must, of course, be present to ensure that fluctuations away from a uniform, bulk, density are stable, and also to generate a positive surface tension. We note that the expansion of $\omega(\rho_1)$ is therefore a sixth-order polynomial, as required to describe the coexistence of up to three-phases below the TCP. We remark that the bulk free-energy per unit volume, $f_b$,  simply corresponds to the smallest of the two maxima of $\omega(\rho_1)$. A simple calculation determines this as $f_b=-4(t^{3/2}-|s|)^2$, which identifies the CEP lines and the second-order nature of the bulk transitions that occur at them (at mean-field level).

\begin{figure}[h]
\includegraphics[width=.45\columnwidth]{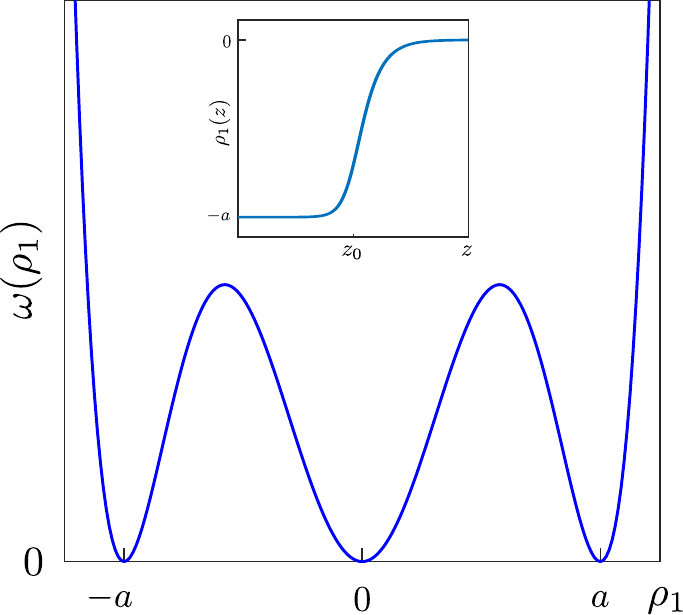}
\caption{\label{Fig05} The potential $\omega(\rho_1)=\rho_1^2 \,(\rho_1^2-a^2)^2$ appearing in the one-component description of wetting in a ternary mixture at three-phase coexistence, at $\tilde s=0$ with $\rho_1^\gamma=a\equiv\sqrt{3\tilde t}$. The middle phase $\beta$ (corresponding to $\rho_1^\beta=0$) completely wets the $\alpha\gamma$ interface, which decomposes into separate $\alpha\beta$ and $\beta\gamma$ interfaces, macroscopically far apart from each other. Inset: The equilibrium density profile $\rho_1(z)$ for the $\alpha\beta$ interface.  }
\end{figure}

 Minimization of the grand potential functional, by requiring
\begin{equation}
    \frac{\delta\Omega}{\delta\rho_1}=0
\end{equation}
leads to the Euler-Lagrange equation for the equilibrium density profile $\rho_1(z)$,
\begin{equation}
\frac{d^2\rho_1}{dz^2}=\frac{d\omega}{d\rho_1}
\end{equation}
which must be solved subject to boundary conditions that have a pair of bulk phase at $z=\pm\infty$. The solution is very easily determined since the Euler-Lagrange equation has a first integral
 \begin{equation}
     \frac{1}{2}\Big(\frac{d\rho_1}{dz}\Big)^2=\omega(\rho_1)
 \end{equation}
As has been noted many times, this is equivalent to the conservation of energy in classical mechanics, with the density $\rho_1$ equivalent to the "coordinate" $x$, and the distance from the interface $z$, equivalent to "time" $t$. The particle has just enough energy, here $E=0$, to traverse between stationary points situated at the bulk densities. This one-component theory implies that the $\alpha\gamma$ interface is complete wet by $\beta$ since the potential $\omega(\rho_1)$ has two barriers situated between the three minima at the bulk densities - see Fig.\ \ref{Fig05}. Therefore, the $\alpha\gamma$ density profile decomposes into an $\alpha\beta$ interface and a $\beta\gamma$ interface, which are infinitely separated. For example, along the line of symmetry in the phase diagram at $\tilde s=0$, for which  $\rho_1^\beta=0$ and $\rho_1^\gamma=-\rho_1^\alpha\equiv a$ with $a=\sqrt{3\tilde t}$, the density profile of the $\alpha\beta$ interface is given by
\begin{equation}
    \rho_1^{\alpha\beta}(z)=-\frac{a}{\sqrt{1+e^{\kappa_\alpha(z-z_0)}}}
    \label{onecompkwi}
\end{equation}
and $\rho_1^{\beta\gamma}(z)=-\rho_1^{\alpha\beta}(-z)$ for the $\beta\gamma$ interface. Here, $\kappa_\alpha=2\sqrt{2}a^2$ is the inverse bulk correlation length of the $\alpha$ phase, with $\kappa_\beta=\kappa_\alpha/2$, and $z_0$ is the interface location which may be arbitrary. As discussed by Rowlinson and Widom \cite{Rowlinson1982}, this one-component description should be accurate near the TCP, since contributions from the variation of the secondary ($\rho_2$) and tertiary ($\rho_3$) densities becomes negligible. Therefore, we can be confident that in more microscopic descriptions, or if more components are included in the square-gradient description, the phase of middle density $\beta$, completely wets the $\alpha\gamma$ interface near the TCP.

\subsection{The KWI model}

The KWI model generalizes this square-gradient description allowing for a secondary density profile $\rho_2(z)$, but ignoring the contribution from the tertiary scaling density. The free-energy functional then depends on two-components similar to earlier discussions of wetting in the three-state Potts models \cite{Hauge1986}, and in anisotropic ferromagnetic systems \cite{Walden1989}. The grand potential functional, which again models the excess contribution to the free-energy per unit area, is written as
\begin{equation}
\Omega[\rho_1,\rho_2]=\int_{-\infty}^\infty\!\!\! dz \left(\frac{1}{2}\Big(\frac{d\rho_1}{dz}\Big)^2+\frac{1}{2}\Big(\frac{d\rho_2}{dz}\Big)^2+\omega(\rho_1,\rho_2)\right)
\label{KWI}
\end{equation}
so that the gradient terms are treated in exactly the same way for both components with no cross term. With this assumption, there will again be a mechanical analogy where the gradient terms behave as a kinetic energy. Ignoring the contribution from the variation of the third density component means the bulk potential $\omega(\rho_1,\rho_2)$ is now the minimum of the three density function $\omega(\rho_1,\rho_2,\rho_3)$ for fixed values of $\rho_1$ and $\rho_2$. Within the KWI model, this potential is assumed to be a product of three isotropic paraboloids given by
\begin{equation}
\omega(\rho_1,\rho_2)=\left((\rho_1-\rho_1^\alpha)^2+(\rho_2-\rho_2^\alpha)^2\right)\,\big((\rho_1-\rho_1^\beta)^2+(\rho_2-\rho_2^\beta)^2\big)\,\left((\rho_1-\rho_1^\gamma)^2+(\rho_2-\rho_2^\gamma)^2\right)
\label{omega}
\end{equation}
 generating a sixth-order polynomial in the two density fields. Again, the only parameters in the model are the three bulk density pairs $(\rho_1^\mu,\rho_2^\mu)$, which take their Griffiths values in terms of $\tilde t$ and $\tilde s$. \\

 The equilibrium density profiles for different interfacial configurations are found from minimization of the grand potential functional by requiring
\begin{equation}
\frac{\delta\Omega}{\delta\rho_1}=0,\hspace{1cm}\frac{\delta\Omega}{\delta\rho_2}=0
\end{equation}
yielding two coupled Euler-Lagrange equations
\begin{equation}
\frac{d^2\rho_1}{dz^2}=\frac{\partial\omega(\rho_1,\rho_2)}{\partial\rho_1},\hspace{1cm}\frac{d^2\rho_2}{dz^2}=\frac{\partial\omega(\rho_1,\rho_2)}{\partial\rho_2}
\end{equation}
These are solved subject to bulk boundary conditions at $z=\pm\infty$
and determine the equilibrium density profiles $\rho_1(z)$ and $\rho_2(z)$ for the three possible $\alpha\beta$, $\beta\gamma$ and $\alpha\gamma$ interfaces. Each of these profiles can be represented by a path  $\rho_2(\rho_1)$ connecting each pair of the two bulk densities in the $(\rho_1,\rho_2)$ plane. This path is the analogue of the trajectory or orbit $y(x)$ of the dynamical motion of a particle with time dependent coordinates $x(t), y(t)$ moving in the 2D plane - an analogy which we shall return to shortly. The properties of this profile path will be crucial in the solution of the model. 

\begin{figure}[h]
\includegraphics[width=.8\columnwidth]{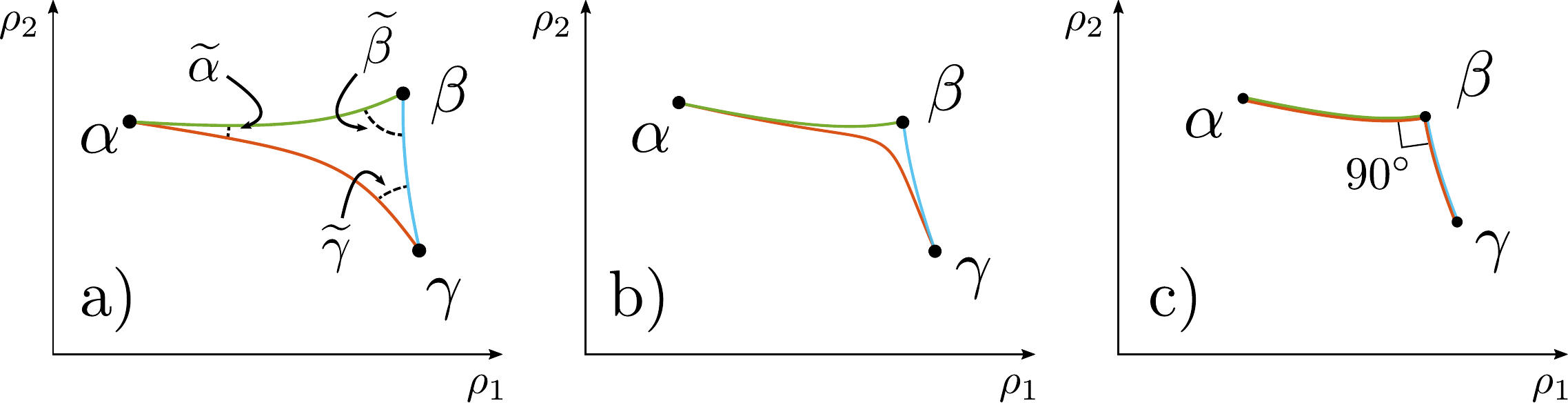}
\caption{\label{Fig06} a) and b) A tricuspid in the $(\rho_1,\rho_2)$ density plane, together with  its interior angles $\widetilde\alpha$, $\widetilde\beta$ and $\widetilde\gamma$, formed from the trajectories of the $\alpha\gamma$, $\alpha\beta$ and $\beta\gamma$ interfaces for the case of partial wetting. The tricuspid angles satisfy $\widetilde\alpha+\widetilde\beta+\widetilde\gamma=\pi/2$. c) At a wetting transition (by $\beta$ in this example), the tricuspid disappears, leaving only the trajectories of the $\alpha\beta$ and $\beta\gamma$ interfaces, which meet at a right angle, so $\widetilde\alpha=\widetilde\gamma=0$ and $\widetilde\beta=\pi/2$.  }
\end{figure}

The KWI model is much more interesting than the single component theory because it exhibits both partial and complete wetting. For partial wetting, the density profile paths outline the shape of a curvilinear tricuspid with internal angles $\widetilde\alpha$, $\widetilde\beta$ and $\widetilde\gamma$, see Fig.\ \ref{Fig06}, which will play a crucial part in the exact solution. The tricuspid is interior to the simple triangle connecting the bulk densities in the two dimensional $(\rho_1,\rho_2)$ plane. From their numerical results, Koga and Widom conjectured that, regardless of where the bulk densities lie in the 2D component plane, the tricuspid angles always satisfy \cite{Koga2008,Koga2016}
\begin{equation}
\widetilde\alpha+\widetilde\beta+\widetilde\gamma=\frac{\pi}{2}
\end{equation}
so that on approaching the wetting transition, which their numerical results indicate is continuous, the tricuspid vanishes, leaving a single trajectory that passes through all three bulk vertices. For wetting by $\beta$ at the $\alpha\gamma$ interface, for example, this would mean that $\widetilde\alpha=\widetilde\gamma=0$ and $\widetilde\beta=\pi/2$, i.e.\ the $\alpha\gamma$ profile decomposes into separate $\alpha\beta$ and $\beta\gamma$ paths that meet at a right angle. 

Integration of the Euler-Lagrange equations leads to 
\begin{equation}
\frac{1}{2}\Big(\frac{d\rho_1}{dz}\Big)^2+\frac{1}{2}\Big(\frac{d\rho_2}{dz}\Big)^2=\omega(\rho_1,\rho_2)
\end{equation}
equivalent to the conservation of energy of a particle of unit mass, moving in a two dimensional plane with total energy $E=0$, that takes infinite time to traverse between two peaks of the potential.  Using the conservation of energy, it follows that the surface tension of the $\alpha\gamma$ interface (say), which is simply the equilibrium value of $\Omega$, reduces to
\begin{equation}
    \sigma_{\alpha\gamma}=\int_{-\infty}^{\infty}\!\!\! dz \;\,\left(\left(\frac{d\rho_1}{dz}\right)^2+\left(\frac{d\rho_2}{dz}\right)^2\right)
\end{equation}
and hence to the integral
\begin{equation}
\sigma_{\alpha\gamma}=\sqrt{2}\int_{\rho_1^\alpha}^{\rho_1^{\gamma}}d\rho_1 \;\,\sqrt{1+(d\rho_2/d\rho_1)^2}\,\sqrt{\omega(\rho_1,\rho_2(\rho_1))}
\end{equation}
where we must use the function $\rho_2(\rho_1)$ representing the $\alpha\gamma$ path. From the three surface tensions, $\sigma_{\alpha\beta}$, $\sigma_{\alpha\gamma}$ and $\sigma_{\beta\gamma}$, the three contact angles can then be obtained using the Neumann triangle i.e. the equilibrium profiles and tensions of the three, separate, planar profiles determine the macroscopic contact angles, without having to the compute the properties of the density profile near the three-phase contact line. That, much more difficult task would be required, for example, to determine the line tension. \\

\begin{figure}[h]
\includegraphics[height=5.cm]{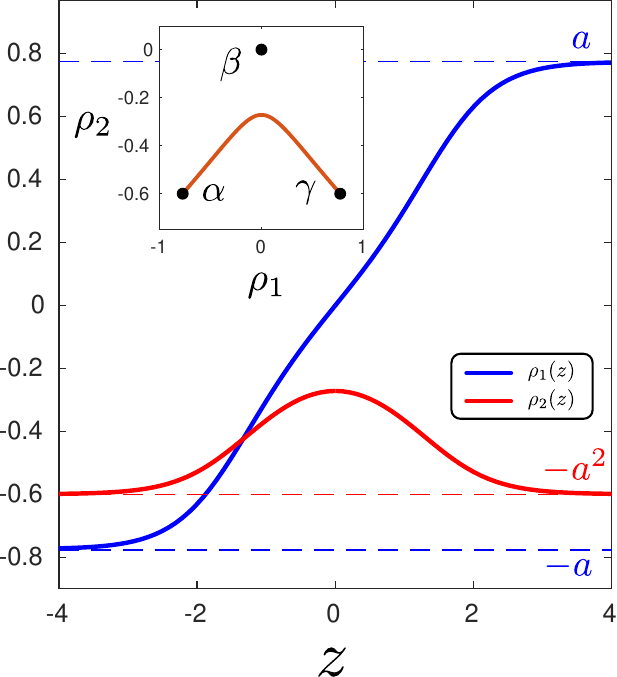}
\caption{\label{Fig07} Primary and secondary density profiles, $\rho_1(z)$ and $\rho_2(z)$, obtained from numerical solution of the Euler-Lagrange equations, for the $\alpha\gamma$ interface along the line of symmetry $\tilde s=0$, at $\tilde t=0.2$ (corresponding to $a\approx 0.775$), showing a thick wetting layer of $\beta$ emerging around $z=0$. Inset: The associated profile path $\rho_2(\rho_1)$ for the $\alpha\gamma$ interface.  As the path gets closer to $\beta$, the intruding $\beta$ phase becomes thicker.}
\end{figure}

The KWI model has three lines of symmetry, where two of the three tensions are the same, which occur whenever the bulk densities form the vertices of an isosceles triangle - see Fig.\ \ref{Fig04}b. For example, along $\tilde s=0$, for which $\rho_1^\beta=0$ and $\rho_1^\gamma=-\rho_1^\alpha=a$ with $a\equiv\sqrt{3\tilde t}$, the congruent sides of the triangle meet at $\beta$ so the tensions of the $\alpha\beta$ and $\beta\gamma$ interfaces are identical implying $\theta_\alpha=\theta_\gamma$. An example, is given in Fig.\ \ref{Fig07}, showing the component density profiles $\rho_1(z)$ and $\rho_2(z)$ and the associated profile path $\rho_2(\rho_1)$, obtained from numerical solution of the Euler-Lagrange equations, showing a thick partial wetting layer of $\beta$ at the $\alpha\gamma$ interface.

This thermodynamic path was the main focus of the original numerical studies of Koga and Widom \cite{Koga2008,Koga2016} who predicted that the $\alpha\gamma$ interface is wet by $\beta$ at a specific value of $a=a_w$, (or equivalently  $\tilde t=\tilde t_w$) and that the wetting transition is likely second-order -- this is essentially an illustration of tricritical point wetting. From their numerical results, they were able to conjecture highly non-trivial analytic expressions for the relevant surface tensions. For the $\alpha\beta$ and $\beta\gamma$ interfaces, they proposed that
\begin{equation}
\sigma_{\alpha\beta}=\sigma_{\beta\gamma}=\frac{1}{6\sqrt{2}}a^4(1+a^2)^{3/2}(9+a^2)^{1/2}
\label{KW1}
\end{equation}
valid for all $a$. In the partial wetting regime $a>a_w$, they guessed the somewhat simpler expression 
\begin{equation}
\sigma_{\alpha\gamma}=\frac{4\sqrt{2}}{3}a^5
\label{KW2}
\end{equation}
for the $\alpha\gamma$ tension. Application of Antonov's rule (\ref{Ant}) then leads to the prediction that the wetting transition occurs at
\begin{equation}
a_w=\sqrt{2\sqrt{3}-3}
\end{equation}
or equivalently $\tilde t_w=2/\sqrt{3}-1\approx 0.1547$, in perfect agreement with their numerical findings $a_w\approx 0.681$. The same conjectures determine that the singular part of the surface tension, $\sigma_{sing}=\sigma_{\alpha\beta}+\sigma_{\beta\gamma}-\sigma_{\alpha\gamma}$,   
vanishes as $\sigma_{sing}\propto (a-a_w)^2$. This means that the wetting transition is second-order, i.e.\ the surface specific heat exponent is $\alpha_s=0$, or equivalently that the contact angle vanishes linearly as $\theta_\beta\propto (a-a_w)$. We note that there are two other lines of symmetry in the KWI model, where the $\alpha$ or $\gamma$ vertex is the tip of an isosceles triangle, which meet the CEP lines at $\tilde t=1/2$. For example, along the $\alpha\beta$ CEP line, $\tilde s\,\tilde t^{-3/2}=-1$, the limiting values of the contact angle at this point are $\theta_\gamma=\pi$ and $\theta_\beta=\theta_\alpha=\pi/2$, corresponding to perfect wetting neutrality. The three lines of symmetry intersect at $\tilde s=0$ and $\tilde t=1$, where the triangle is equilateral, implying the three tensions are equal and $\theta_\alpha=\theta_\beta=\theta_\gamma=2\pi/3$ - see Fig.\ \ref{Fig04}b.

\begin{figure}[h]
\includegraphics[height=5.cm]{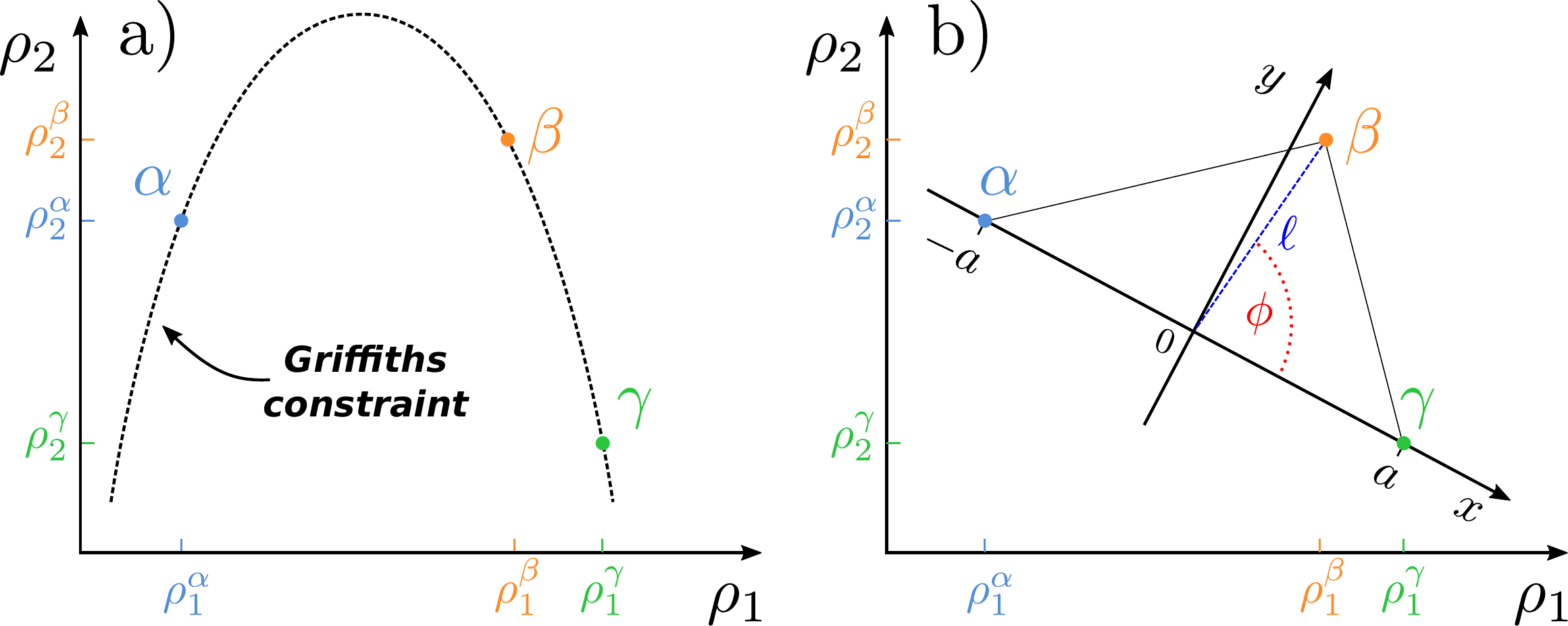}
\caption{\label{Fig08} a) In the two dimensional density plane, the bulk vertices $\alpha$, $\beta$ and $\gamma$ sit at $(\rho_1^\alpha,\rho_2^\alpha)$, $(\rho_1^\beta,\rho_2^\beta)$ and $(\rho_1^\gamma,\rho_2^\gamma)$. Within the KWI model, these lie on the parabola $\rho_2=-(\rho_1)^2$, following the Griffiths constraint. b) The coordinates $(x,y)$ relative to those of the scaling densities $(\rho_1,\rho_2)$, used to describe the trajectory of the $\alpha\gamma$ interface. The length scales $a$ and $\ell$ and the median angle $\phi$, characterizing the shape of the simple triangle connecting the bulk vertices $\alpha$, $\beta$ and $\gamma$ are also shown. The Koga-Indekeu conjecture determines the tension of the $\alpha\gamma$ interface in terms of these geometrical quantities with the intriguingly simple formula $\sigma_{\alpha\gamma}=4\sqrt{2}\,a^3\ell/3$, provided it is only partially wet by $\beta$. }
\end{figure}

 The above guesses for the surface tensions were subsequently generalized by Koga and Indekeu away from the lines of symmetry \cite{Koga2019}. Provided that the $\alpha\gamma$ interface is not wet by $\beta$, they conjectured that
\begin{equation}
\sigma_{\alpha\gamma}=\frac{4\sqrt{2}}{3}a^3 \ell
\label{KI}
\end{equation}
where, in the geometry of the 2D density plane,  $2a$ is the Euclidean distance between the $\alpha$ and $\gamma$ vertices  and $\ell$ is the distance of the midpoint between them to $\gamma$ -- see Fig.\ \ref{Fig08}. This formula applies equally to the two other tensions, provided the interface is not wet by the remaining phase. 

It is important to note that this expression does not rely on the Griffiths constraints for the bulk densities and purports to be the exact expression for the tension of the partially wet interface, for arbitrary locations of the bulk densities in 2D component space. This ansatz recovers the earlier Koga-Widom conjectures for an isosceles triangle (with $\ell=a^2$), works for the special case when the bulk densities lie on a line, and is supported by their extensive numerical results. 

Assuming it to be generally valid, Indekeu and Koga then applied it to the whole phase diagram, using the Griffiths constraints to determine the values of $a$ and $\ell$ as a function of $\tilde t$ and $\tilde s$. In this way they obtained three wetting phase boundaries by $\alpha$, $\beta$ and $\gamma$, from application of Antonov's rule \cite{Indekeu2022}. Their phase diagram is illustrated in Fig.\ \ref{Fig04}c and shows lines of second-order wetting transitions, by $\alpha$, $\beta$ and $\gamma$ that intersect the CEP lines, at $\tilde t_-$ and $\tilde t_+$ where
\begin{equation}
\tilde t_{\pm}=\frac{7\pm \sqrt{33}}{8}
\end{equation}
leaving a non-wetting gap between them. For example, on approaching the $\alpha\beta$ CEP line at $\tilde s\tilde t^{-3/2}=-1$, the contact angle of the non-critical phase $\theta_\gamma\to\pi$, while
\begin{equation}
\cos\theta_\beta=\frac{(2\tilde t-1)(1-31\tilde t+4\tilde t^2)}{[(1+\tilde t)(1+4\tilde t)]^{3/2}}
\label{IKbeta}
\end{equation}
for $\tilde t_-\le \tilde t\le \tilde t_+$. The contact angle $\theta_\beta$ varies continuously between $0$, at $\tilde t=\tilde t_-$ and $\pi$ at $\tilde t=\tilde t_+$, with $\theta_\beta=\pi/2$ at the point of neutrality $\tilde t=1/2$. We note that the contact angle behaves linearly near $\tilde t_\pm$, indicating that the wetting transitions, occurring at these points, involving macroscopically diffuse fluid interfaces are still second-order.\\

\section{The exact solution to the KWI model}

Here, we show how the KWI model may be solved exactly by considering conserved quantities which allow the determination of the density profile paths. We then employ some complex analysis which determines the surface tension and reveals an unexpectedly simple mapping between the microscopic profile paths and the Neumann triangle. Finally, we show how the individual component density profiles may be determined using methods drawn from the theory of algebraic curves. The solution is quite involved and is divided into self-contained subsections, for ease of reading.

\subsection{The mechanical analogy and scattering angles}

 Ahead of the complex analysis, and also to emphasis the very useful mechanical analogy, we first simplify the notation replacing the coordinate $z$ with “time” $t$, and the densities profiles $\rho_1(z)$ and $\rho_2(z)$ with 2D Cartesian coordinates $x(t)$ and $y(t)$ respectively. The grand potential potential functional is then the classical action
 \begin{equation}
     \Omega[x,y]=\int_{-\infty}^\infty\!\!\! dt\;\Big(\,\frac{1}{2}(\dot x^2+\dot y^2)+\omega(x,y)\Big)
     \label{genSGT}
 \end{equation}
 corresponding to the integral of the kinetic energy minus the potential $V=-\omega(x,y)$. The Euler-Lagrange equations are therefore,
\begin{equation}
\ddot x=\omega_x(x,y), \hspace{1cm} \ddot y=\omega_y(x,y)
\label{genEL}
\end{equation}
equivalent to the Cartesian components of Newton's law. Here, the dot denotes differentiation w.r.t.\ the "time" $t$, the subscript denotes a partial derivative and
\begin{equation}
\omega(x,y)=\big((x-x_\alpha)^2+(y-y_\alpha)^2\big)\,\big((x-x_\beta)^2+(y-y_\beta)^2\big)\,\big((x-x_\gamma)^2+(y-y_\gamma)^2\big)
\label{KWIomegaxy}
\end{equation}
is the KWI potential where the $(x_\mu,y_\mu)$ are the locations of the three bulk vertices. We leave these as arbitrary, but ordered such that $x_\alpha\le x_\beta \le x_\gamma$, and only use the appropriate Griffiths values of the bulk densities at the end of our calculations. This is important because it will allow us to distinguish between the aspects of the KWI model that stem from the component isotropy and aspects that are related to the imposition of the Griffiths constraints on the primary and secondary bulk densities. Our task here is first to derive the trajectories and then the general Koga-Indekeu conjecture for the surface tension, when the bulk densities are at arbitrary positions in 2D component space. This will allow us to understand how wetting transitions, and non-wetting gaps along the CEP lines, may occur in the model. The imposition of the Griffiths constraints then determines at what values of $\tilde t$ and $\tilde s$ these occur in the phase diagram.

The conservation of energy relation has the familiar form
\begin{equation}
\frac{1}{2}\,(\dot x ^2 +\dot y ^2)=\omega(x,y)
\label{CoE}
\end{equation}
while the expressions for surface tension of the $\alpha\gamma$ interface (say), which is the minimum value of the classical action, read
\begin{equation}
\sigma_{\alpha\gamma}=\int_{-\infty}^{\infty}\!\!\!dt\; (\dot x^2+\dot y^2)
\label{ST1}
\end{equation}
or
\begin{equation}
\sigma_{\alpha\gamma}=\sqrt{2}\int_{x_\alpha}^{x_{\gamma}}\!\!\!dx\; \sqrt{(1+y'^2)\,\omega(x,y(x))}
\label{ST}
\end{equation}
where $y(x)$ is the, as yet unknown, trajectory of the $\alpha\gamma$ path and $y'=dy/dx$. The path starts at $(x_\alpha,y_\alpha)$, ends at $(x_\gamma, y_\gamma)$ and does not pass through the $\beta$ vertex provided that the $\beta$ phase partially wets the $\alpha\gamma$ interface.

\begin{figure}[h]
\includegraphics[height=4.cm]{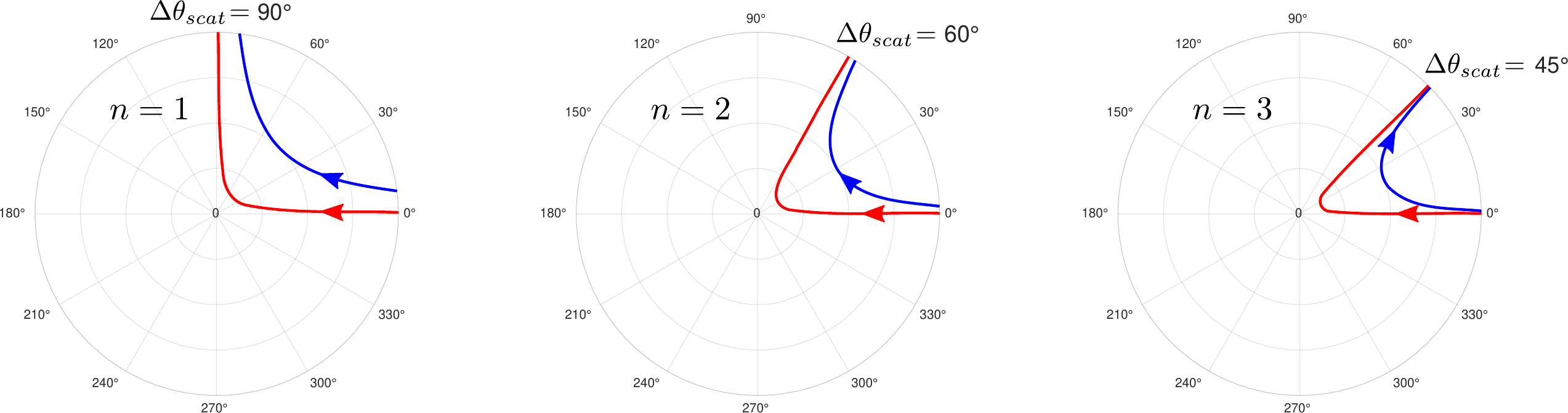}
\caption{\label{Fig09} 
Trajectories with high (blue) and low (red) angular momentum for the motion of a classical particle in a radial potential $\omega(x,y)=(x^2+y^2)^n$, for $n=1$, $2$ and $3$. The scattering angles $\Delta\theta_{scat}(n)$ for each case are shown, corresponding to $\Delta\theta_{scat}(n)=\pi/(n+1)$.
In the limit when the angular momentum $h\to 0$, the trajectories become two straight lines separated by these angles.
}
 \end{figure}

The potential $\omega(x,y)$ clearly does not have a radial symmetry and, hence, the angular momentum is not conserved. However, there are three limiting cases, or rather scales, in which this second familiar conservation law applies to some aspect of the trajectory: a)  when the $\alpha\gamma$ trajectory (say), gets close to the $\beta$ vertex, the potential behaves as the quadratic $\omega(x,y)\propto (x^2+y^2)$ (local to the origin placed at $\beta$), b) when the trajectory is extended through the bulk vertices in which case, at the largest scales, the potential behaves as $\omega(x,y)=(x^2+y^2)^3$ and finally, c) in the limit of approaching a CEP where the vanishingly small distance between the $\alpha$ and $\beta$ vertices is rescaled to unity and the $\gamma$ vertex is removed to infinity. In this case, the potential far from $\alpha$ and $\beta$ behaves as the quartic $\omega(x,y)\propto (x^2+y^2)^2$. In these three limits, the potential is effectively radial, scaling as $\omega(x,y)=C(x^2+y^2)^n$, where the coefficient $C$ is not important, and can be set to $C=1/2$, with $n=1$, $n=3$ and $n=2$ for cases a), b) and c), respectively. Using polar coordinates $(r,\theta)$ in place of $(x,y)$ the conservation of energy (with $E=0$) then reads 
\begin{equation}
\dot r ^2+\frac{h^2}{r^2}-r^{2n}=0
\end{equation}
where $h=r^2\dot\theta$ is the modulus of the angular momentum per unit mass. Using only the conservations of energy and angular momentum, we can obtain the value of the scattering angle, $\Delta\theta_{scat}(n)=\int_{-\infty}^\infty dt\, \dot \theta$, between the incoming and outgoing directions of motion as the integral
\begin{equation}
\Delta\theta_{scat}(n)=2\int_{r_0}^\infty\!\!\! dr\;\frac{h}{r\sqrt{r^{2(n+1)}-h^2} }    \end{equation}
which evaluates as
\begin{equation}\label{tscat}
\Delta\theta_{scat}(n)=\frac{\pi}{n+1}
\end{equation}
yielding $\Delta\theta_{scat}(1)=\pi/2$, $\Delta\theta_{scat}(2)=\pi/3$ and $\Delta\theta_{scat}(3)=\pi/4$ for our three cases. Trajectories that just touch the origin, corresponding to angular momentum $h=0$, are straight lines separated by these angles -- see Fig.\ \ref{Fig09}. For these radial potentials, it is not difficult to determine the exact trajectories, which evaluate as 
\begin{equation}
 \left(\frac{r_0}{r}\right)^{n+1}=\cos\,\large((n+1)\,\theta\large)
 \label{radialtraj}
\end{equation}
where $r_0$ is the distance of closest approach, satisfying $h=r_0^{(n+1)}$. The values of the scattering angles can also be read directly from this. The result for the quadratic potential $n=1$ already explains the numerical observation of Koga and Widom that, at a continuous wetting transition, the $\alpha\gamma$ path separates into $\alpha\beta$ and $\beta\gamma$ trajectories that meet the $\beta$ vertex at a right angle $\widetilde\beta=\pi/2$,
with the analogous result applying for wetting by $\alpha$ and $\gamma$. The significance of the two other scattering angles will emerge later.

\subsection{Conservation of angles along the path}

 As for gravitational motion in celestial mechanics, it is convenient to focus on the path $y(x)$ rather than the full solution to the equations of motion, which we will seek later. Taking the derivative of $y'(x)=\dot y/\dot x$ with respect to $x$ determines that the trajectory satisfies the non-linear ODE
 \begin{equation}
\frac{y''}{1+y'^2}=\frac{\omega_y(x,y)-y'\omega_x(x,y)}{2\;\omega(x,y)}
\label{pathODE1}
\end{equation}
Substituting for the KWI potential (\ref{KWIomegaxy}) then gives
\begin{equation}
\frac{y''}{1+y'^2}=\sum_\mu \frac{(y-y_\mu)-y' (x-x_\mu)}{(x-x_\mu)^2+(y-y_\mu)^2}
\label{pathODE}
\end{equation}
which conveniently separates into contributions from each bulk scattering centre. We can recognize that the left-hand side is the derivative of $\tan^{-1} y' $, while each term on the right-hand side is homogeneous and can be integrated using the standard substitution $\tau_\mu= (y-y_\mu)/(x-x_\mu)$. The equation for the path therefore has a first integral
\begin{equation}
\tan^{-1} y' +\sum_\mu\tan^{-1} \Big(\frac{y-y_\mu}{x-x_\mu}\Big)=K
\label{tans}
\end{equation}
where $K$ is a constant to be determined, and which is different for the $\alpha\beta$, $\beta\gamma$ and $\alpha\gamma$ paths. Geometrically, this result expresses a second conservation law between the direction of the trajectory and the sum of the angles made with the bulk vertices. From this, we can determine the angle with which the path meets the bulk vertex at each end. Cyclically applying this to all three trajectories and adding, determines that the sum of the interior angles of the tricuspid satisfies
\begin{equation}\label{sumang}
\widetilde\alpha+\widetilde\beta+\widetilde\gamma=\frac{\pi}{2}
\end{equation}
as conjectured by Koga and Widom. We shall prove this in another way later by a conformal mapping of the tricuspid onto the Neumann triangle, which determines each individual tricuspid angle.  This result confirms that, at a wetting transition, two of the angles vanish and the remaining one must be $\pi/2$, which, as discussed above, is just the local scattering angle $\Delta\theta_{scat}(1)$ of a quadratic potential.

\subsection{Density profile trajectories and the wetting phase boundary}

The next step is to integrate Eq.\ (\ref{tans}) to obtain the desired density profile paths. We do this first for the line of symmetry in the phase diagram which establishes the method and reveals the wetting transition on approaching the TCP.

\subsubsection*{The line of symmetry $\tilde s=0$}

\begin{figure}[h]
\includegraphics[height=5.cm]{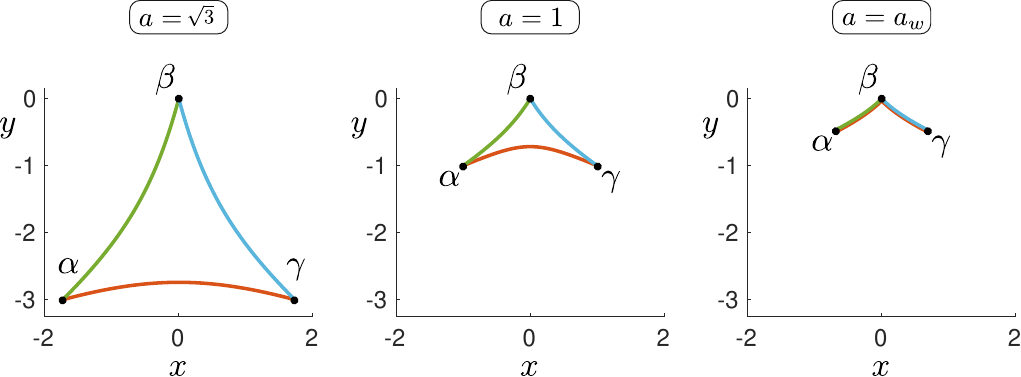}
\caption{\label{Fig10} Exact profile trajectories $y(x)$ of the $\alpha\gamma$ interface (orange), the $\alpha\beta$ interface (green) and the $\beta\gamma$ interface (blue), with $x\equiv \rho_1$ and $y\equiv\rho_2$. They are shown for $\tilde s=0$
(the central line of symmetry where $\theta_\alpha=\theta_\gamma$), for three different values of $a\equiv\sqrt{3\tilde t}\equiv\rho_1^\gamma$. LEFT: $a=\sqrt{3}$, where all three tensions are identical. MIDDLE: $a=1$, in the approach to wetting. RIGHT: At the wetting phase boundary $a_w=\sqrt{2\sqrt{3}-3}$. At the wetting phase boundary, the tricuspid disappears and the $\alpha\gamma$ trajectory decomposes into $\alpha\beta$ and $\beta\gamma$ trajectories, which meet at a right angle. }
 \end{figure}

 Along the lines of symmetry in the phase diagram the bulk vertices sit at the corners of an isosceles triangle. Let us consider the central line of symmetry, at $\tilde s=0$, as first studied by Koga and Widom \cite{Koga2008}, in which case the Griffiths constraint on the bulk densities places them at $(x_\alpha,y_\alpha)=(-a,-a^2)$, $(x_\beta,y_\beta)=(0,0)$ and $(x_\gamma,y_\gamma)=(a,-a^2)$ with $a\equiv \rho_1^\alpha=\sqrt{3\tilde t}$. The first integral, therefore, reads
\begin{equation}
\tan^{-1} y'(x) +\tan^{-1}\Big( \frac{y+a^2}{x+a}\Big)+\tan^{-1} \frac{y}{x}+\tan^{-1} \Big(\frac{y+a^2}{x-a}\Big)=\tan^{-1}k
\end{equation}
where we have introduced $k\equiv\tan^{-1}K$. We now combine the four inverse tangents, which leads to
\begin{equation}
\frac{(xy'+y)(x^2-a^2-(y+a^2)^2)+2x(y+a^2)(x-yy')}{(x-yy')(x^2-a^2-(y+a^2)^2)-2x(y+a^2)(xy'+y)}=k
\label{tank}
\end{equation}
Consider, first, the partially wet $\alpha\gamma$ path, i.e.\ one that does not pass through the $\beta$ vertex at the origin. This is the simplest to consider since it is symmetric, with $y(x)=y(-x)$, implying that $y'(0)=0$, and hence identifies $K_{\alpha\gamma}=\pi/2$ or equivalently $k_{\alpha\gamma}=\infty$. Thus, the denominator on the L.H.S of (\ref{tank}) must vanish, and rearranging we arrive at the first-order equation
\begin{equation}
y'(x)=\frac{x(x^2-a^2-(y+a^2)^2)-2xy(y+a^2)}{2x^2(y+a^2)+y(x^2-a^2-(y+a^2)^2)}
\end{equation}
Following standard methods we now rewrite this as the differential $P(x,y) dx+Q(x,y)dy=0$ and observe that $
P_y(x,y)=Q_x(x,y)=-6xy - 4a^2 x$
implying that the differential is an {\it{exact}} derivative, i.e.\ it is the total derivative of a function $u(x,y)$, with $u_x=P$ and $u_y=Q$. That is,
\begin{equation}
u_x(x,y)=x(x^2-a^2-(y+a^2)^2)-2xy(y+a^2)
\end{equation}
and
\begin{equation}
u_y(x,y)=-2x^2(y+a^2)-y(x^2-a^2-(y+a^2)^2)
\end{equation}
Integration and application of the boundary condition, that the path passes through the $\alpha$ vertex at $(-a,-a^2)$, identifies the equation for the trajectory as 
\begin{equation}
x^4+y^4+\frac{8}{3}a^2y^3+2a^2(1+a^2)(y^2-x^2) -6x^2y^2-8a^2x^2y=\frac{a^4(a^4+6a^2-3)}{3}
\label{ualphagammasym}
\end{equation}
The trajectory is unique implying that any wetting transition is continuous. In the partial wetting regime, $a>a_w$, the trajectory passes directly from $\alpha$ to $\gamma$ and $y(0)<0$. However, on approaching the phase boundary, the distance of closest approach to the $\beta$ vertex, analogous to the perihelion in celestial mechanics, vanishes. The additional condition that $y(0)=0$ at the wetting phase boundary requires that $a_w^4+6a_w^2-3=0$,
determining that $
a_w=\sqrt{2\sqrt{3}-3}$, precisely as conjectured by Koga and Widom. A direct trajectory from $\alpha$ to $\gamma$ does not exist for $a<a_w$, corresponding to the complete wetting regime, and the path decomposes into two separate $\alpha\beta$ and $\beta\gamma$ trajectories. The exact trajectories of the $\alpha\gamma$ path, for values of $a=\sqrt{3}$ , $a=1$ and $a_w=\sqrt{2\sqrt{3}-3}$ are shown in Fig.\ \ref{Fig10}.

The same method determines the $\alpha\beta$ (and symmetric $\beta\gamma$) path. In this case, the calculation is a little more unwieldy because the value of the constant $k_{\alpha\beta}$ is left undetermined until we apply the final boundary conditions that the path passes through $\alpha$ and $\beta$. Rearranging (\ref{tank}) into the form $P(x,y)dx+Q(x,y)dy=0$ shows again that it is an exact differential 
 with $P=u_x$ and $Q=u_y$. Integrating, and apply the boundary conditions that the path passes through $(-a,-a^2)$ and $(0,0)$ determines the equation for the trajectory as
\begin{equation}
xy(a^2+a^4+y^2-x^2+2a^2y)-\frac{2}{3}a^2x^3 +\frac{k_{\alpha\beta}}{4}\Big(x^4+y^4+\frac{8}{3}a^2y^3+2a^2(1+a^2)(y^2-x^2) -6y^2x^2-8a^2yx^2\Big)=0
\label{ualphabeta}
\end{equation}
where
\begin{equation}
k_{\alpha\beta}=\frac{8a}{3-6a^2-a^4}
\end{equation}
At the wetting phase boundary, $k_{\alpha\beta}=\infty$, and the $\alpha\beta$ and $\alpha\gamma$ paths are identical. At this transition, the $\alpha\beta$ and $\beta\gamma$ paths meet at the required right angle, since $y'(0)=1$, for the $\alpha\beta$ path, and $y'(0)=-1$ for the $\beta\gamma$ path.  The exact trajectories of the symmetric $\alpha\beta$ and $\beta\gamma$ paths, for values of $a=\sqrt{3}$ , $a=1$ and $a=\sqrt{2\sqrt{3}-3}$ are shown in Fig.\ \ref{Fig10}.

\subsubsection*{The general trajectory}

\begin{figure}[t]
\includegraphics[height=5cm]{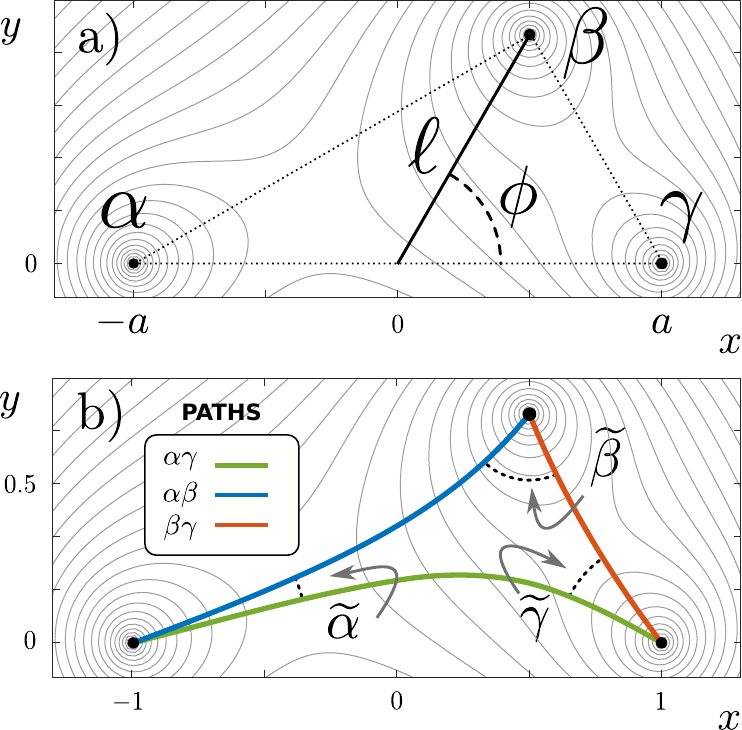}
\caption{\label{Fig11} (a) Rotated coordinates in the density plane used to define the trajectory $y(x)$ of the $\alpha\gamma$ interface (See also Fig.\ \ref{Fig08}). The contours of the potential $\omega(x,y)$ are shown in gray. The bulk phases $\alpha$, $\beta$ and $\gamma$ are located at $(-a,0)$, $(\ell\cos\phi,\ell\sin\phi)$, and $(a,0)$, respectively. (b) A tricuspid in the ($x,y$) density plane, with its interior angles $\widetilde{\alpha}$, $\widetilde{\beta}$ and $\widetilde{\gamma}$, for a case of partial wetting, with $\ell=a=1$ and $\phi=\pi/3$. }
 \end{figure}

Since the KWI model is isotropic in the density components $\rho_1$ and $\rho_2$, we can always rotate and translate the coordinates so that the $\alpha$ and $\gamma$ bulk vertices lie at $(-a,0)$ and $(a,0)$ in the $(x,y)$ plane
leaving the $\beta$ vertex at some general position $\ell(\cos\phi,\sin\phi)$ - see Fig.\ \ref{Fig11}.  In this case, the potential is written
\begin{equation}
 \omega(x,y)=\big((x+a)^2+y^2\big)\,\big((x-\ell\cos\theta)^2+(y-\ell\sin\theta)^2\big)\,\big((x-a)^2+y^2\big)
 \label{omegagen}
\end{equation}
where the coordinates $x$ and $y$ are related to the original primary and secondary densities by
\begin{equation}
\rho_1=Ax+By+\frac{\rho_1^\alpha+\rho_1^\gamma}{2}
\label{rot1}
\end{equation}
and
\begin{equation}
\rho_2=-Bx+Ay +\frac{\rho_2^\alpha+\rho_2^\gamma}{2}
\label{rot2}
\end{equation}
with $A=(\rho_1^\gamma-\rho_1^\alpha)/2a$ and $B=(\rho_2^\alpha-\rho_2^\gamma)/2a$. This choice of coordinates conveniently allow us to consider the trajectory of an arbitrary $\alpha\gamma$ interface, starting at $(-a,0)$ and ending at $(a,0)$, provided the interface is partially wet by $\beta$. All trajectories can be determined this way, with the other two paths obtained using the lengths $a$ and $\ell$, and the median angle $\phi$, appropriate for that side of the triangle determined by the bulk vertices - see Fig.\ \ref{Fig11}. The value of $\phi$ used here is implicitly $\phi_{\alpha\gamma}$, but there are also the median angles $\phi_{\alpha\beta}$ and $\phi_{\beta\gamma}$ defined for the $\alpha\beta$ and $\beta\gamma$ paths. The lines of symmetry in the phase diagram are all obtained using the choice $\phi=\pi/2$ with the appropriate values of $\ell$ and $a$ determined by $\tilde t$ and $\tilde s$. This is particularly straight forward for $\tilde s=0$, where the Griffiths constraint means that we must choose $\ell=a^2$. In this case, to which we will return multiple times, the coordinate $y$ is a trivial translation, by $\ell=a^2$ of that used in the previous subsection.

\begin{figure}[h]
\includegraphics[height=5.cm]{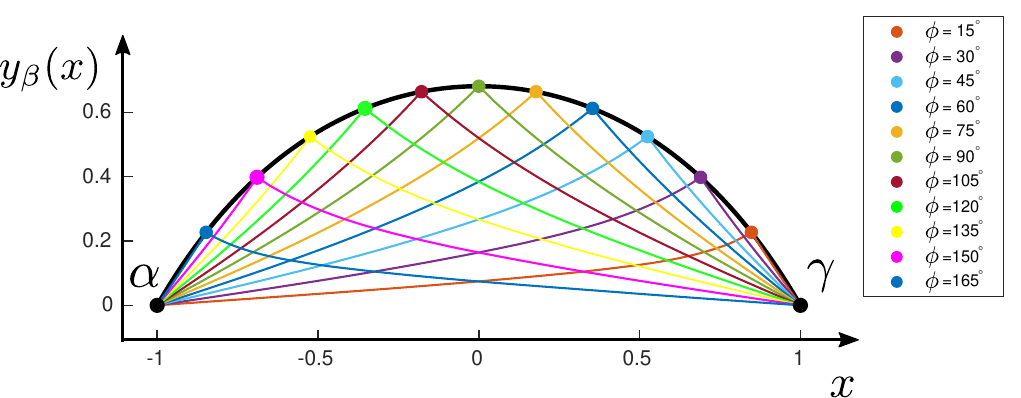}
\caption{\label{Fig12} The $\alpha\gamma$ path at the wetting transition for different values of the median angle $\phi$ (coloured lines). All of them form a right angle at their corresponding $\beta$ vertex (coloured circles).
The black curve $y_\beta(x)$ represents the scaled location of the $\beta$ vertex at the wetting phase boundary, with $\alpha$ and $\gamma$ at $(\mp 1,0)$, respectively. If the $\beta$ vertex lies above this curve, the interface is partially wet, and it is completely wet if lies below. The value at the midpoint $y_\beta(0)=\sqrt{2\sqrt{3}-3}$ is the same as the value of $a_w$ for the line of symmetry. The value of the derivative at the end point, $y_\beta '(- 1)= \sqrt{3}$ determines that the angle between the tangent and the horizontal is $\pi/3$ (the scattering angle $\Delta\theta_{scat}(2)$ for an isotropic quartic potential), and gives rise to the non-wetting gap.}
 \end{figure}

 The method is the same as that described for the symmetric case. The first integral of the differential equation for the path leads to the conservation law for the path direction
\begin{equation}
\tan^{-1}y'(x)+\sum_\mu\tan^{-1}\Big(\frac{y-y_\mu}{x-x_\mu}\Big)=\phi    \end{equation}
so that the constant of integration, $K_{\alpha\gamma}=\phi$, is particularly simple for this choice of coordinates (although this only emerges after we have done the next integral and used both boundary conditions). Combining the inverse tangents leads to the exact derivative
\begin{equation}
    u_x\,dx+u_y\,dy=0
\end{equation}
with
\begin{equation}
    \frac{u_x}{\sqrt2}=\cos\phi\,\big(3x^2y-y^3-a^2y\big)-2\ell xy+\sin\phi\,\big(3xy^2-x^3+a^2x\big)
    \label{ux}
\end{equation}
and
\begin{equation}
    \frac{u_y}{\sqrt2}=\cos\phi(x^3-3xy^2-a^2x)+\ell\,(y^2-x^2+a^2)+\sin\phi(3x^2y-y^3-a^2y)
    \label{uy}
\end{equation}
where we have introduced a factor of $\sqrt{2}$, which will prove convenient later. Integration then determines the general equation for the trajectory as 
\begin{equation}
    u(x,y)=0
\end{equation}
with
\begin{equation}
\frac{u(x,y)}{\sqrt{2}}=\cos\phi\,\big(x^3y-xy^3-a^2xy\big)+\ell y\left(\frac{y^2}{3}-x^2+a^2\right)+\sin\phi\,\left(\frac{3}{2}x^2y^2-\frac{x^4+y^4}{4}+\frac{a^2}{2}(x^2-y^2)\right)-\frac{a^4\sin\phi}{4}
\end{equation}
As stated earlier, the path is unique so that any wetting transition is continuous. The phase boundary is obtained by requiring that the path touches the $\beta$ vertex at $\ell(\cos\phi,\sin\phi)$, yielding the condition
\begin{equation}
    \frac{a}{\ell}=\sqrt{1+\frac{2}{\sqrt3}\sin\phi}
\end{equation}
which locates the position of the $\beta$ vertex required for wetting. Setting $a=1$ and converting to Cartesian coordinates, we obtain the curve $y_\beta(x)$, with
\begin{equation}
    y_\beta^2(x)=x^2-3+2\sqrt{x^4-3x^2+3}
    \label{KWIybeta}
\end{equation}
describing the loci, $(x,y_\beta(x))$, of the $\beta$ vertex 
in the scaled density plane, so that the $\alpha\gamma$ trajectory just meets it. The same result was obtained by Indekeu and Koga  using their ansatz for the surface tension together with Antonov's rule \cite{Indekeu2022}. If the $\beta$ vertex lies outside of the curve, then the $\alpha\gamma$ interface is partially wet, while it is completely wet if it lies within it. The curve, along with representative trajectories for the $\alpha\gamma$ path, which separates into $\alpha\beta$ and $\beta\gamma$ paths that meet at a right angle, are shown in Fig.\ \ref{Fig12}.

The value at the midpoint $y_\beta(0)=\sqrt{2\sqrt{3}-3}$ recovers our earlier result, and the original conjecture of Koga and Widom, for the value of $a_w$ along the line of symmetry. The derivative at the end point, $y'(-1)=\sqrt{3}$, requires that the $\gamma$ vertex makes the specific angle of $\pi/3$ with the straight line between $\alpha$ and $\beta$, whose length vanishes at the CEP. Thus, the intersection of the lines of wetting and the CEP occurs when the limiting value of the median angle $\phi_{\alpha\beta}$, for an infinitely narrow triangle, is equal to the scattering angle $\Delta\theta_{scat}(2)$ of a quartic radial potential. Using the Griffiths values of the bulk densities, which identify the location of the vertices, this determines that the intercept of the lines of wetting and CEP is at $\tilde t_-=(7-\sqrt{33})/8$. A similar argument determines $\tilde t_+=(7+\sqrt{33})/8$ for the intercepts of wetting by the other phases with the CEP lines \cite{Indekeu2022} (See Fig.\ \ref{Fig04}). We shall return to this in much greater detail when we consider the scaling behavior of the trajectories near the CEP lines. This will reveal more clearly how the non-wetting gap arises from the local XY symmetry of the KWI model.

\subsection{The genus of algebraic curves}

\begin{figure}[th]
\includegraphics[height=9.cm]{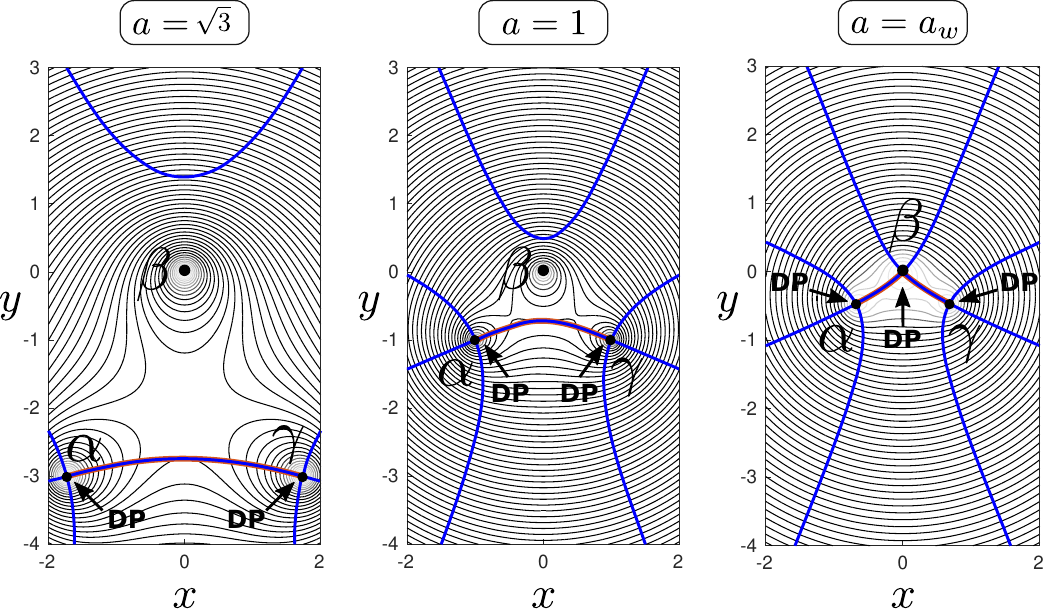}
\caption{\label{Fig13} The quartic algebraic curve $u(x,y)=0$ (blue) of eqn.\ (\ref{ualphagammasym}), with the section corresponding to the $\alpha\gamma$ trajectory highlighted in orange. They are shown along the line of symmetry $\tilde s=0$, for $a=\sqrt{3}$, $a=1$ (both partial wetting), and at the wetting phase boundary $a_w=\sqrt{2\sqrt{3}-3}$, against the contours of the potential $\omega(x,y)$ (thin lines). Double point (DP) singularities occur at the bulk vertices where the curve crosses itself. For partial wetting there are two double points so the curve is elliptical of genus $g=1$. At the wetting transition the curve has three double points so the genus is $g=0$. In this case, the curve is rational and has the parameterization $(x(s), y(s))$ with $y=\tau(s)x(s)$ with $\tau(s)$ given (\ref{taus}) and $x(s)$ by (\ref{xtau}).}
 \end{figure}

In general, the trajectory is an irreducible quartic algebraic curve 
possessing two singularities corresponding to two double points (where $u_x=u_y=0$ but the second derivatives are non vanishing) which are situated at the $\alpha$ and $\gamma$ bulk vertices. There are many famous examples of quartic curves, such as Bernoulli's lemniscate, that have been studied for centuries and often have a simple geometrical interpretation. The curves defined here satisfy the constraint (\ref{tans}), implying that the sum of the angles defined by the tangent and with the lines to the three vertices at $\alpha,\beta,\gamma$, must always be conserved. While we are only interested in the portion of the algebraic curve which directly connects $\alpha$ and $\gamma$, we note the algebraic curve has other branches, some of which are disconnected, and others that intersect at the singularities where the tangent can not be defined -- see Fig.\ \ref{Fig13}. The algebraic curve asymptotes to eight straight lines, each separated by an angle of $\pi/4$, which is the scattering angle $\Delta\theta_{scat}(3)$ of an $r^6$ potential, mentioned earlier.

The degree-genus theorem determines the genus $g$ of the curve as
\begin{equation}\label{genus}
    g=\frac{(d-1)(d-2)}{2}-\delta_{tot}
\end{equation}
where, in our cases, $d=4$ is the degree (since the algebraic curve is a quartic) and $\delta_{tot}$ is the total number of singularities, arising from the self intersections, each of which has a singularity index $n_\mu$. For partial wetting, we have $\delta_{tot}=n_\alpha+n_\beta$ where $n_\alpha=n_\beta=1$, since the self intersections at $\alpha$ and $\gamma$, are double points where $u_x=u_y=0$, for which the singularity index is unity. Thus, in general $\delta_{tot}=2$, implying $g=1$ so the curve is elliptical, i.e.\ it can be parameterized on a rhombus, the fundamental domain of a torus, by Weierstrass elliptic functions. However, at the wetting transition, all the branches connect and the curve has three intersections and  $\delta_{tot}=n_\alpha+n_\beta+n_\gamma$ implying $\delta_{tot}=3$ so that $g=0$, since additionally there is a double point where, $u_x=u_y=0$, at the $\beta$ vertex. This means the curve can be rationally parameterized, i.e.\ it may be expressed as a quotient of polynomials in some parameter, which we write here as $s$ (but not to be confused with the scaling field $\tilde s$). Algorithms for determining these have long been studied and have received great interest in recent years \cite{Sendra2007,Falkensteiner2024}. For example, along the line of symmetry $\tilde s=0$, the algebraic curve $u(x,y)=0$ with $u(x,y)$ given by (\ref{ualphagammasym}) may be written $(x(s),y(s))$ with $y=\tau(s) x(s)$ where
\begin{equation}
    \tau =\frac{4 b^2s}{s^2+4b^2}
    \label{taus}
\end{equation}
and 
\begin{equation}
    x(s)=\frac{4a_w^2\tau(3-\tau^2)(s^2+4b^2)\pm 6c(\tau^2-a_w^2)(s^2-4b^2)}{3(\tau^4-6\tau^2+1)(s^2+4b^2)}
    \label{xtau}
\end{equation}
where the $\pm$ generate the $\alpha\beta$ and $\beta\gamma$ parts of the $\alpha\gamma$ trajectory, while $b^2=2\sqrt3+3$ and $ c^2=1+\frac{1}{\sqrt{3}}$. The case of more general paths will be considered later, which we will derive making use of complex analysis. The rational parameterization is very useful because it leads to the exact solution for the component density profiles, along the whole line of wetting transitions, equivalent to determining the time dependence of the coordinates $x(t)$ and $y(t)$, thus completely specifying the classical motion and, hence, the density profiles (recall eqns.\ (\ref{rot1}) and (\ref{rot2})).

\subsection{Conformal invariance and the surface tension}

The determination of the exact trajectories confirms the surface phase diagram of the KWI model, predicted by Indekeu and Koga using their ansatz (\ref{KI}) for the surface tension. However, it does not reveal where this intriguingly simple expression for the surface tension comes from. Even with the trajectories determined, the evaluation of the integral formula (\ref{ST}) for the tension remains terribly complicated, suggesting that a deeper mathematical structure is present, linked to the geometrical properties of the density profile tricuspid. To see this, we seek the differential equation for the algebraic curve $u(x,y)$ itself, rather than the path $y(x)$.

There is a clue to what the differential equation for $u(x,y)$ may look like. The condition $P_y=Q_x$, that the differential $Pdx+Qdy$ is exact, is simply a check that the mixed derivatives satisfy $u_{xy}=u_{yx}$. It is therefore natural to ask if the remaining second derivatives are also related. From (\ref{ux}) and (\ref{uy}), we immediately see that 
\begin{equation}
    \frac{u_{xx}}{\sqrt{2}}=6xy\cos\phi-2\ell y+(3y^2-3x^2+a^2)\sin\phi
\end{equation}
and 
\begin{equation}
    \frac{u_{yy}}{\sqrt{2}}=-6xy\cos\phi+2\ell y+(3x^2-3y^2-a^2)\sin\phi
\end{equation}
so that the algebraic curve always satisfies the Laplace equation
\begin{equation}
    \nabla^2u(x,y)=0
\end{equation}
The trajectories are therefore conformal invariant throughout the entire phase diagram, not just at wetting phase boundaries or along the CEP lines. The origin of this property is not obvious at this stage -- this is something which we return to when we solve for a more general class of square-gradient models. Putting aside this question for the moment, the fact that the paths satisfy the Laplace equation, i.e. that they are harmonic functions, is a further clue to the underling structure of the KWI model and, in particular, the geometry of the tricuspid. 

\begin{figure}[h]
\includegraphics[height=8.cm]{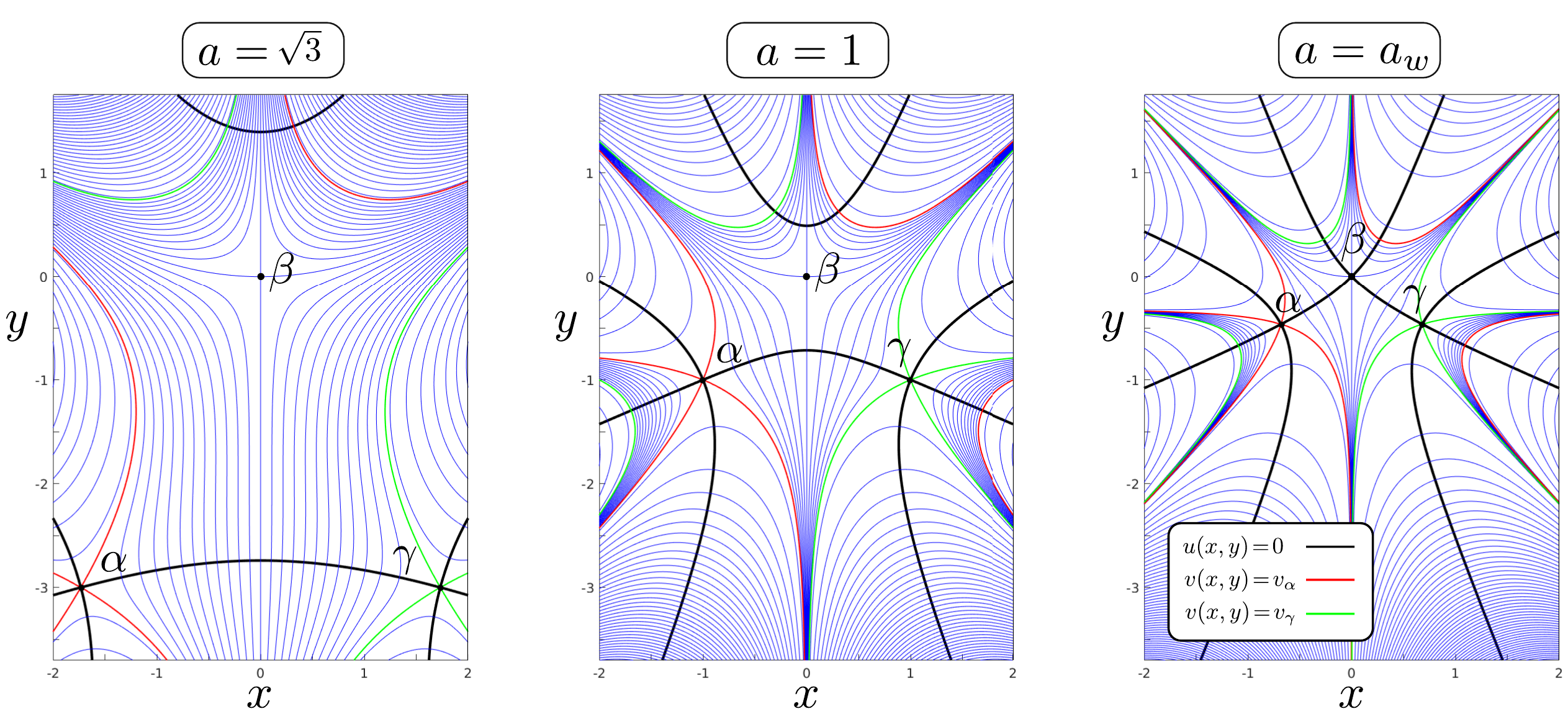}
\caption{\label{Fig14} The quartic algebraic curve $u(x,y)=0$ for the $\alpha\gamma$ interface (black) and the family of quartic algebraic curves (blue) defined by the harmonic conjugate $v(x,y)=constant$. Here, the quartic curve is shown for $\tilde s=0$ (along the line of symmetry $\theta_\alpha=\theta_\gamma$), for $a=\sqrt{3}$, $a=1$ and at the wetting phase boundary $a_w=\sqrt{2\sqrt{3}-3}$. The harmonic conjugate curves are orthogonal to $u$ at every crossing point, except for $v(x,y)=v_\alpha$ (red) and $v(x,y)=v_{\gamma}$ (green) when they pass through the double point singularities at the bulk $\alpha$ and $\gamma$ vertices, and intercept at an angle of $\pi/4$ instead.}
 \end{figure}

Within complex analysis, a harmonic function $u(x,y)$ is the real part of a holomorphic function,
\begin{equation}
    f(z)=u(x,y)+iv(x,y)
\end{equation}
 of a single complex variable $z=x+iy$. To determine this function we just need to identify the harmonic conjugate $v(x,y)$, which is similarly harmonic and also satisfies the Laplace equation $\nabla^2 v(x,y)=0$. To do this we use the Cauchy-Riemann conditions
\begin{equation}
  u_x=v_y,\hspace{1cm}u_y=-v_x 
\end{equation}
and simple integration gives 
\begin{equation}
    \frac{v(x,y)}{\sqrt{2}}=\cos\phi\,\left(\frac{3}{2}x^2y^2-\frac{x^4+y^4}{4}+\frac{a^2}{2}(x^2-y^2)\right)+\ell x\,\left(\frac{x^2}{3}-y^2-a^2\right)+\sin\phi\,\big(xy^3-x^3y+a^2xy\big)
    \label{vharm}
\end{equation}
which is defined up to an unimportant constant which we can ignore. This defines a family of quartic algebraic curves that are orthogonal to $u(x,y)$, i.e.\ intersecting it at each point at a right angle -- see Fig.\ \ref{Fig14}. The values of $v(x,y)$ at the ends of the trajectory $v_\alpha=v(-a,0)$ and $v_\gamma=v(a,0)$ are given by 
\begin{equation}
    v_\alpha=\frac{a^4\cos\phi}{2\sqrt{2}}+\frac{2\sqrt{2} a^3\ell}{3},\hspace{1cm}
    v_\gamma=\frac{a^4\cos\phi}{2\sqrt{2}}-\frac{2\sqrt{2}a^3\ell}{3}
\end{equation}
Notice that the difference between these values is precisely the Koga-Indekeu ansatz for the surface tension, which hints strongly they are related. 

From the harmonic functions $u(x,y)$ and $v(x,y)$ it is now a simple matter to determine  the holomorphic function $f(z)=u+iv$, and term by term comparison gives
\begin{equation}
    f(z)=-i\sqrt{2}e^{-i\phi}\left(\frac{z^4}{4}-\frac{\;a^2}{2}z^2-\ell e^{i\phi}\left(\frac{z^3}{3}-a^2z\right)\right)
\end{equation}
which is defined up to an arbitrary additive (complex) constant. The $\alpha\gamma$ trajectory and the two conjugate curves that pass through its end points, at $\alpha$ and $\gamma$, are then just the real $\Re$ and imaginary $\Im$ parts of $f(z)=u_{\alpha\gamma}+iv_\alpha$ and $f(z)=u_{\alpha\gamma}+iv_\gamma$. The real part of the constant term in $f(z)$ can then always be chosen so that $u_{\alpha\gamma}=0$. We note that the derivative of the function is
\begin{equation}
    f'(z)=-i\sqrt{2} e^{-i\phi}(z-z_\alpha)(z-z_\beta)(z-z_\gamma)
\end{equation}
and has the form of a simple product very similar to the potential $\omega(x,y)$. In fact, we can see that the two functions are related by
\begin{equation}
|f'(z)|^2=2\;\omega(x,y)
\end{equation}
Thus, the integral formula for the surface tension, (\ref{ST}), can be rewritten very simply in the complex plane as
\begin{equation}
    \sigma_{\alpha\gamma}=\int_{z_\alpha}^{z_\gamma}\!\!\! dz\;\, |f'(z)|
    \label{STf}
\end{equation}
Next, we note that the modulus appearing in the integrand does not matter. The reason for this is that $|f'(z)|$ is simply the scale factor of the conformal map $f(x+iy)=u+iv$, and since $u$ is constant along the curve, while $v$ decreases, this change of variables means that  $dv=-|f'|dz$. Consequently,
provided the path between $\alpha$ and $\gamma$ is direct, the tension is 
\begin{equation}
\sigma_{\alpha\gamma}=v_\alpha-v_\gamma
\label{STf2}
\end{equation}
which is equivalent to $\sigma_{\alpha\gamma}=|f(z_\gamma)-f(z_\alpha)|$. Substituting for $v_\alpha$ and $v_\gamma$ gives
\begin{equation}
 \sigma_{\alpha\gamma}=\frac{4\sqrt{2}a^3\ell}{3}
 \label{STv}
\end{equation}
which is the Koga-Indekeu conjecture for the surface tension, Eq.\ (\ref{KI}). 

\begin{figure}[h]
\includegraphics[height=5.cm]{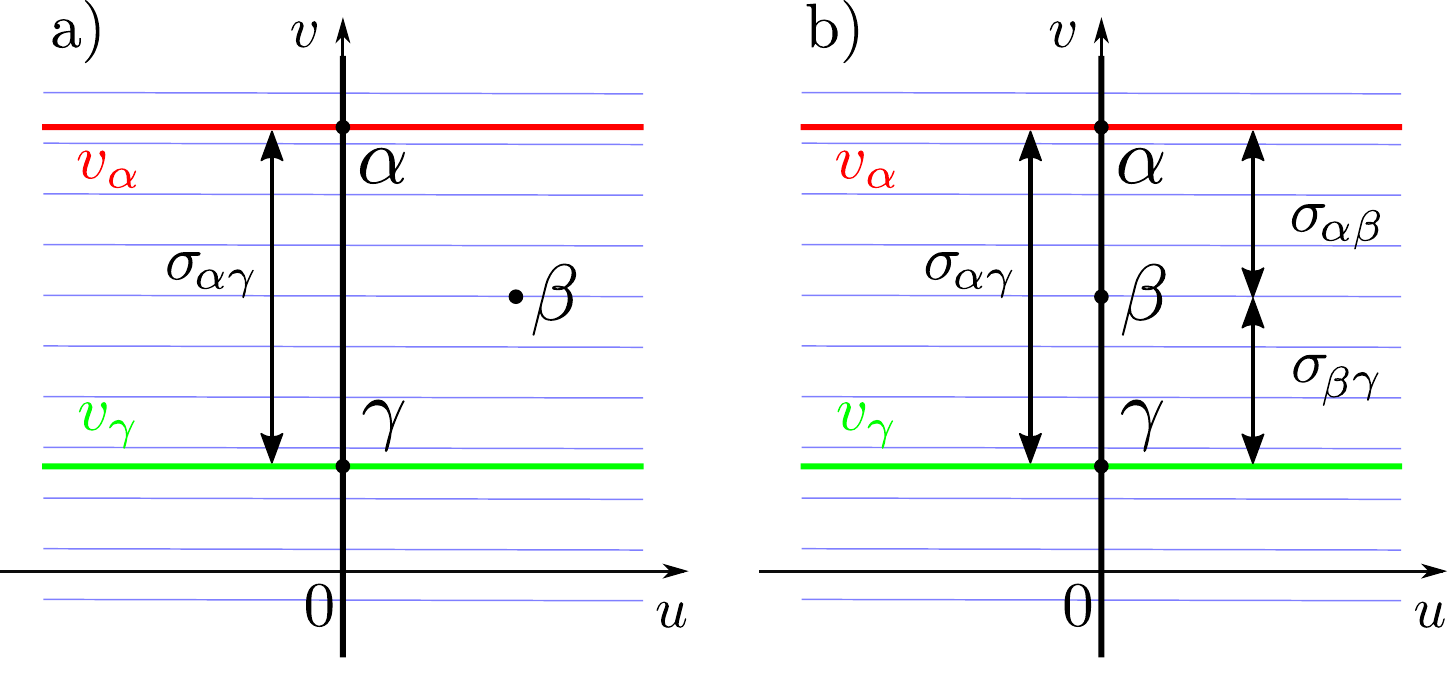}
\caption{\label{Fig15} The conformal map $f(x+iy)=u+iv$ of the $\alpha\gamma$ trajectory, $u(x,y)=0$ and the family of conjugate curves $v(x,y)=constant$, shown in Fig.\ 14, onto the vertical line $u=0$ and horizontal lines (blue), respectively. The distance between the  conjugate curves that pass through the bulk vertices $v(x,y)=v_\alpha$ (red), $v(x,y)=v_\gamma$ (green) identify the surface tension $\sigma_{\alpha\gamma}=v_\alpha-v_\gamma$: a) At partial wetting, b) At the wetting transition, where the bulk vertex $\beta$ also maps onto the line $u=0$ describing the $\alpha\gamma$ trajectory. Here, the addition of the lengths $
   |v_\gamma-v_\alpha|=|v_\gamma-v_\beta|+|v_\beta-v_\alpha|$ is equivalent to Antonov's rule $\,\sigma_{\alpha\gamma}=\sigma_{\alpha\beta}+\sigma_{\beta\gamma}$.}
 \end{figure}

The above analysis has a very simple and elegant geometric interpretation which exploits fully the conformal invariance. Any analytic function may be used to transform the trajectory into new coordinates, but by using the function $f(z)$ itself, we arrive at a remarkable simplicity. In this case, the conformal map, 
\begin{equation}
    f(x+iy)=u+iv
\end{equation}
 projects the $\alpha\gamma$ trajectory $u(x,y)=0$ (as shown in Figs.\ \ref{Fig13} and \ref{Fig14}) onto the vertical straight line $u=0$, since the trajectory in the $(x,y)$ plane is defined by a constant value of $u$. The same mapping also projects the two conjugate curves that pass through the bulk vertices, at the end points, as shown in Fig.\ \ref{Fig14}, onto the horizontal straight lines $v=v_\alpha$ and $v=v_\gamma$. The mapping of the trajectory and family of conjugate curves is shown in Fig.\ \ref{Fig15}.

 The conformal mapping is particularly revealing at a wetting transition, since it not only flattens the $\alpha\gamma$ trajectory, but also maps the $\beta$ vertex onto the same line, i.e.\ it simultaneously maps the $\alpha\gamma$, $\alpha\beta$ and $\beta\gamma$ trajectories onto the same straight line path. This is, of course, consistent with the expression (\ref{STv}), that the surface tension is the length of the line in $(u,v)$ plane. In this case, we can see the trivial geometrical result for the addition of the lengths $
   |v_\gamma-v_\alpha|=|v_\gamma-v_\beta|+|v_\beta-v_\alpha|$
is equivalent to Antonov's rule $
    \sigma_{\alpha\gamma}=\sigma_{\alpha\beta}+\sigma_{\beta\gamma}$.
The conformal mapping therefore ensures there is perfect thermodynamic consistency between the wetting phase boundaries as determined by the condition that the trajectory touches the $\beta$ vertex, or from the application of Antonov's rule.

\subsection{The tricuspid-Neumann triangle mapping}

\begin{figure}[h]
\includegraphics[height=5.cm]{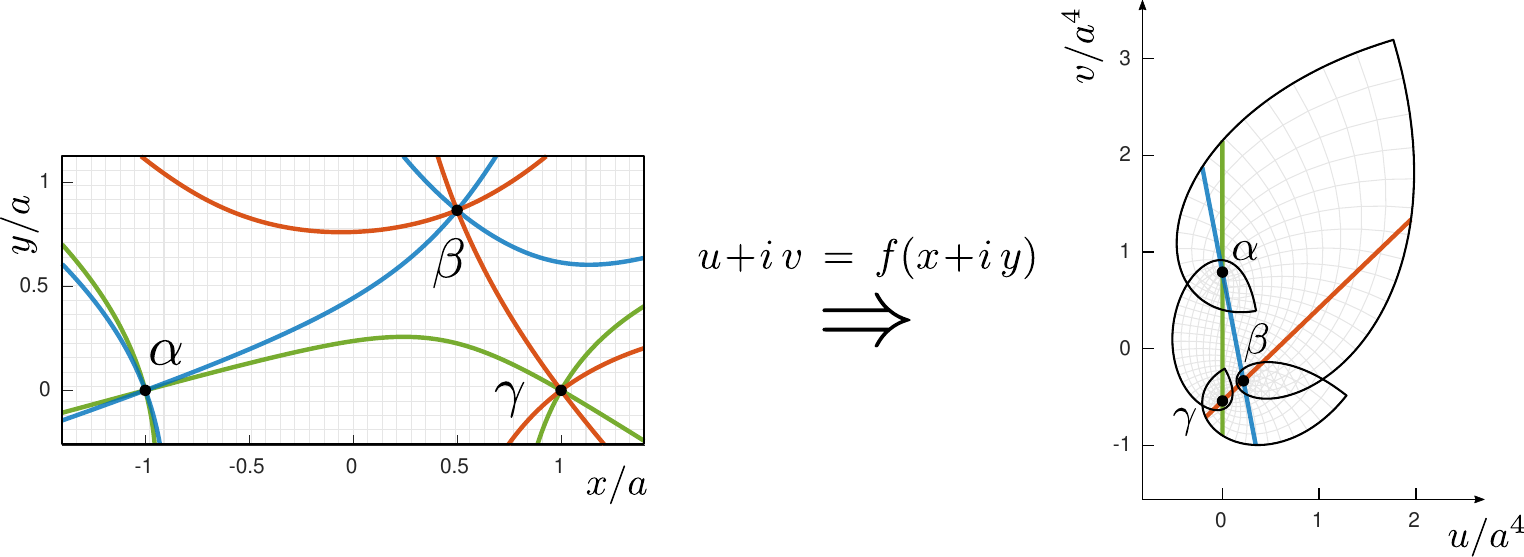}
\caption{\label{Fig16} The conformal map $f(x+iy)=u+iv$ of a rectangle containing the three quartic curves for the $\alpha\gamma$ (green), $\alpha\beta$ (blue) and $\beta\gamma$ (red) trajectories and a square grid, illustrating the conservation of the local angles except for the bulk $\alpha,\beta$ and $\gamma$ vertices, which correspond to double point singularities, where the angles double. The conformal map transforms each quartic curve onto a straight line, and the three trajectories onto a triangle connecting the bulk vertices.}
 \end{figure}

The straightening of the paths under the conformal map $f(z)$ now reveals the full significance of the tricuspid shape, explaining all its geometrical properties. Since $f'(z)=-i\sqrt{2}e^{-i\phi}\prod_\mu(z-z_\mu)$, the function $f(z)$ is, up to a rotation due to the value of $\phi$, the same for all three paths. Therefore, the conformal transformation maps the tricuspid onto a triangle whose lengths are precisely the associated surface tension, i.e.\ the tricuspid is a curvilinear representation of the Neumann triangle for the magnitudes of the tensions required to stabilize the contact line. The vertices of the triangle, at $z_\mu=x_\mu+iy_\mu$, sit at the singularities where the conformal map fails, because $f'(z_\mu)=0$, and angles are no longer preserved. At those points, the map is not invertible. Instead, near these singularities $f(z)\propto (z-z_\mu)^2$, a simple application of de Moivre's theorem implies that the tricuspid angles are doubled under the mapping. For example, it is easy to see that under the mapping $z\to w(z)=u+iv$ with $w(z)=z^2$, the new coordinates satisfy $u=x^2-y^2$ and $v=2xy$. This means that straight lines $y=x\tan\theta $ become straight lines $v=u\tan2\theta$ in the new coordinates. We note that this doubling of the angles means that the different branches of the algebraic curves, which meet at the singularities, are mapped onto the same straight lines in the $(u,v)$ plane - see Fig.\ \ref{Fig16}, where we also show how a square grid is transformed.

\begin{figure}[h]
\includegraphics[height=5.cm]{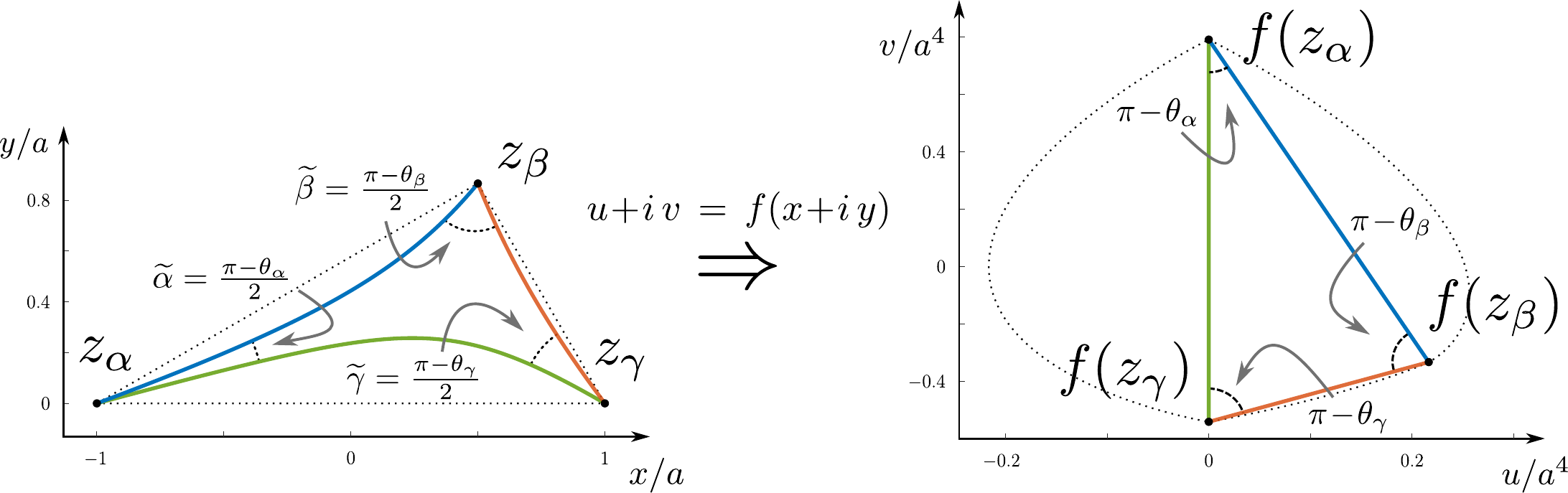}
\caption{\label{Fig17} The conformal map $f(x+iy)=u+iv$ of the three profile paths for $\ell=a$ and $\phi=\pi/3$. The paths transform onto straight lines whose lengths are the corresponding surface tensions, i.e.\ the tricuspid is a curvilinear representation of the Neumann triangle with internal angles $\widetilde\alpha=(\pi-\theta_\alpha)/2$, $\widetilde\beta=(\pi-\theta_\beta)/2$ and $\widetilde\gamma=(\pi-\theta_\gamma)/2$. After the transformation, these angles are doubled. The triangle connecting the bulk densities is also mapped (dotted lines).}
 \end{figure}

The doubling of the angles at the double point singularities means the geometry of the tricuspid shape is very simply expressed through the relations
\begin{equation}
    \widetilde\alpha=\frac{\pi-\theta_\alpha}{2},\hspace{1cm}
    \tilde \beta=\frac{\pi-\theta_\beta}{2},\hspace{1cm}
 \widetilde\gamma=\frac{\pi-\theta_\gamma}{2}
\end{equation}
which is the central result behind the integrability of the KWI model. The mapping of the tricuspid is shown in Fig.\ \ref{Fig17}.

The mapping provides a remarkable direct link between the microscopic density profile paths and the Neumann triangle which, recall, simply describes the macroscopic orientation of the interfaces defined by the contact angles. We illustrate this explicitly for the two most important cases - the line of symmetry and line of critical end points.\\

\subsubsection*{The line of symmetry}

\begin{figure}[h]
\includegraphics[height=5.cm]{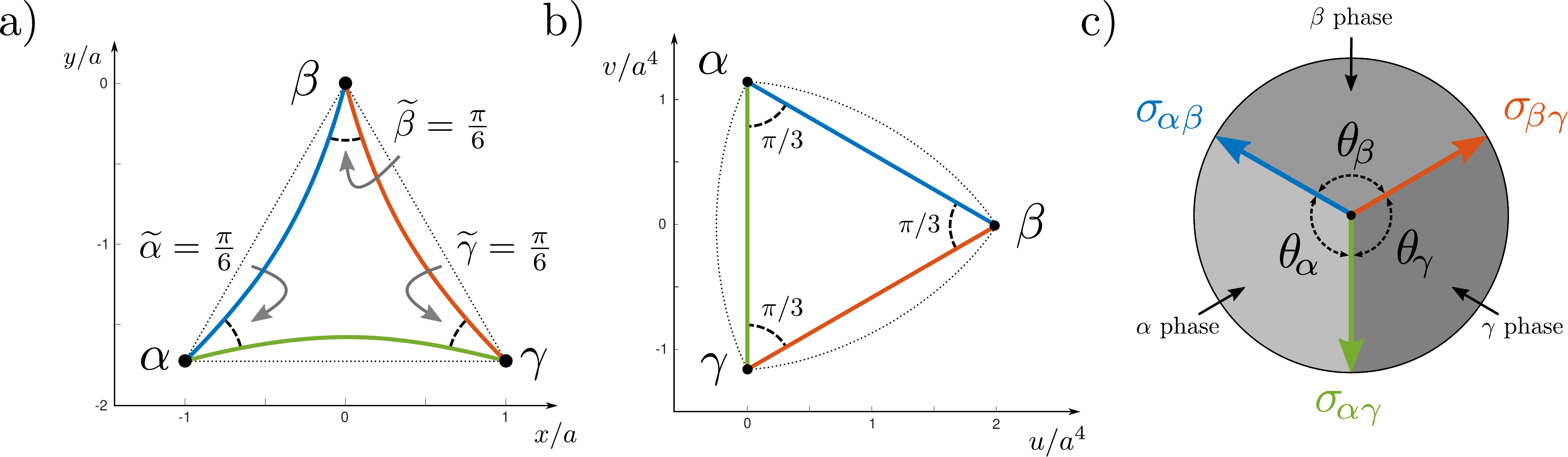}
\caption{\label{Fig18} 
The point of maximum symmetry of the phase diagram, corresponding to $\tilde s=0$ and  $\tilde t=1$ (i.e.\ $a=\sqrt{3}$), where all three surface tensions are the same. a) The tricuspid of microscopic trajectories, b) Their mapping onto straight lines to form the Neumann triangle after the conformal transformation $u+vi=f(x+yi)$. c) The macroscopic drop near the contact line showing the surface tensions, which correspond to the lengths of the sides of the Neumann triangle. At this point of symmetry, $\widetilde\alpha=\widetilde\beta=\widetilde\gamma=\pi/6$ and $\theta_\alpha=\theta_\beta=\theta_\gamma=2\pi/3$. The simple triangle connecting the bulk vertices in the 2D component plane and its conformal map are also shown in a) and b) (dotted lines).}

 \end{figure}

Along the line of symmetry $\tilde s=0$, the bulk densities $\alpha$, $\beta$ and $\gamma$ sit at $(-a,-a^2)$, $(0,0)$ and $(a,-a^2)$, respectively, with $a=\sqrt{3\tilde t}$. The equality of the $\sigma_{\alpha\beta}$ and $\sigma_{\alpha\gamma}$ tensions means that $\theta_\alpha=\theta_\gamma$, with the remaining contact angle $\theta_\beta$, determined by $
\sigma_{\alpha\gamma}=2\sigma_{\alpha\beta}\cos(\theta_\beta/2) $. The surface tensions are exactly given by (\ref{KW1}) and (\ref{KW2}) so that $\theta_\beta$ satisfies
\begin{equation}
  \cos\left(\frac{\theta_\beta}{2}\right)  =\frac{8a}{(1+a^2)^{3/2}(9+a^2)^{1/2}}
  \label{betasymm}
\end{equation}
for $a\ge a_w$. We now compare $\theta_\beta$ with $\widetilde\alpha$, whose value is simply determined by the derivatives of the paths $y_{\alpha\beta}(x)$ and $y_{\alpha\gamma}(x)$ at the $\alpha$ vertex from $
    \widetilde\alpha=\tan^{-1}y'_{\alpha\beta}(-a)-\tan^{-1}y'_{\alpha\gamma}(-a)$. For the $\alpha\gamma$ path, eqn.\ (\ref{tans}) with $K_{\alpha\gamma}=\pi/2$ determines the value of the derivative at $x=-a$ as
\begin{equation}
    \tan^{-1}y'_{\alpha\gamma}=\frac{\pi}{4}-\frac{1}{2}\tan^{-1}a
\end{equation}
and similarly for the $\alpha\beta$ path with $K_{\alpha\beta}=\pi-\tan^{-1}{8a}/(a^4+6a^2-3)$, obtaining
 \begin{equation}
     \tan^{-1}y'_{\alpha\beta}=\frac{\pi}{2}-\frac{1}{2}\tan^{-1}\frac{8a}{a^4+6a^2-3}-\frac{1}{2}\tan^{-1}a
 \end{equation}
 Together, these determine the tricuspid angle $\widetilde\alpha$
 and comparison with (\ref{betasymm}) shows that $\widetilde\alpha=\theta_\beta/4$, which is equivalent to
\begin{equation}
    \widetilde\alpha=\frac{\pi-\theta_\alpha}{2}
    \end{equation}
At the point of maximum symmetry in the phase diagram ($\tilde s=0$, $\tilde t=1$ so $a=\sqrt{3}$), where all three tensions are equal and hence $\theta_\alpha=\theta_\beta=\theta_\gamma=2\pi/3$, the  tricuspid angles are $\widetilde\alpha=\widetilde\beta=\widetilde\gamma=\pi/6$ - see Fig.\ \ref{Fig18}

\subsubsection*{Along the CEP}

\begin{figure}[h]
\includegraphics[height=6.cm]{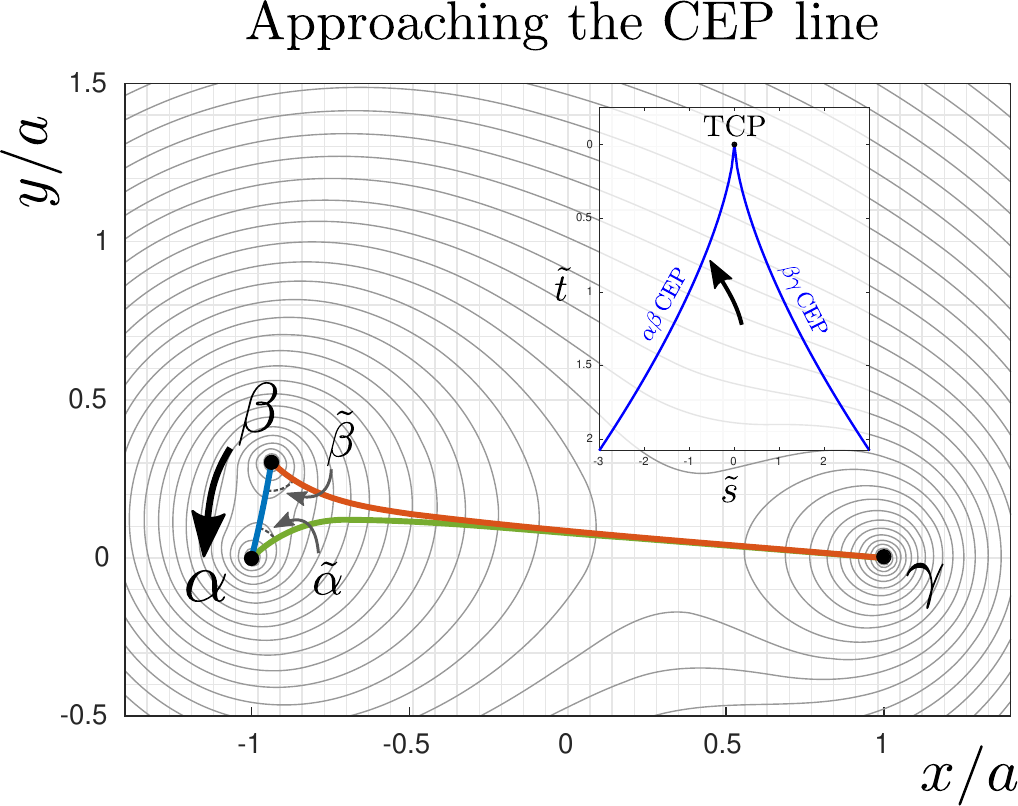}
\caption{\label{Fig19} As the $\alpha\beta$ CEP is approached from the three-phase coexistence region (see inset), the $\alpha$ and $\beta$ vertices in the 2D component plane merge, corresponding to the phases $\alpha$ and $\beta$ becoming identical. At the CEP, the tricuspid angle $\widetilde\gamma\to 0$, equivalent to $\theta_\gamma\to\pi$, leaving only the two angles $\widetilde\alpha$ and $\widetilde\beta$ satisfying $\widetilde\alpha+\widetilde\beta=\pi/2$. }
\end{figure}

Consider next that we vary the temperature and/or pressure to approach a point on the $\alpha\beta$ CEP line at $\tilde s\tilde t^{-3/2}=-1$, corresponding to the merging of the $\alpha$ and $\beta$ vertices -- see Fig.\ \ref{Fig19}. In order to understand the scaling properties of the tricuspid in this limit, we must rescale the density coordinates so that the tricuspid does not vanish. We do so by keeping the distance between $(x_\alpha,y_\alpha)$ and $(x_\beta,y_\beta)$ equal to unity (say). This rescaling does not affect the values of the tricuspid angles. In this limit, the rescaled path from $\alpha$ to $\beta $ becomes a straight line. This represents the universal scaling limit of the $\alpha\beta$ interface in the critical regime where each component of the density profile behaves as $x(t)=x_\alpha+(x_\beta-x_\alpha)\tanh \kappa t/2$, and $y(t)=y_\alpha+(y_\beta-y_\alpha)\tanh\kappa t/2$, where $\kappa=1/\xi$ is the inverse of the bulk correlation length $\xi$ of the $\alpha$ or $\beta$ phase, which diverges at the CEP. In the same limit, the rescaled $\gamma$ vertex is removed to infinity, corresponding to the tricuspid becoming infinitely long and thin with $\widetilde\gamma=0$ or, equivalently, to the contact angle $\theta_\gamma\to\pi$. This leaves just two angles $\widetilde\alpha$ and $\widetilde\beta$, satisfying
    \begin{equation}\widetilde\alpha+\widetilde\beta=\frac{\pi}{2}
    \label{alphabetacep}
    \end{equation}
provided we lie within the non-wetting gap, so the rescaled tricuspid exists. The angle $\widetilde\alpha$ is identified from $\widetilde\alpha=\tan^{-1}y'_{\alpha\beta}(x_\alpha)-\tan^{-1} y'_{\alpha\gamma}(x_\alpha)$. The $\alpha\beta$ path is a straight line of slope $ y'_{\alpha\beta}=x_\gamma$ as required by the Griffiths conditions, $y_\mu=-x_\mu^2$, and $x_\alpha+x_\beta+x_\gamma=0$. For the $\alpha\gamma$ path, we must use 
\begin{equation}
    \tan^{-1}y'(x)+\tan^{-1}\frac{y-y_\alpha}{x-x_\alpha}+\tan^{-1}\frac{y-y_\beta}{x-x_\beta}=\pi-3\tan^{-1}|x_\beta|
    \label{tanscep}
\end{equation}
which satisfies the required boundary condition that $y'\to x_\beta$ as also determined by the Griffiths constraints. It follows that the tricuspid angle satisfies $
    2\widetilde\alpha=3(\tan^{-1}x_\gamma+\tan^{-1}|x_\beta|)-\pi$
and, substituting $x_\gamma=2\sqrt{\tilde t}$ and $x_\beta=-\sqrt{\tilde t}$, we obtain, 
\begin{equation}
\tilde \alpha\;=\;\left\{\;
\begin{array}{l@{\hspace{1.5cm}}l@{\hspace{1.cm}}}
   \displaystyle\frac{3}{2}\,\tan^{-1}\left(\frac{3\sqrt{\tilde t}}{1-2\tilde t}\right)-\frac{\pi}{2}  &  \tilde t<\frac{1}{2}\\[.75cm]
   \!\!-\,\displaystyle\frac{3}{2}\,\tan^{-1}\left(\frac{3\sqrt{\tilde t}}{2\tilde t-1}\right)+\pi  & \tilde t>\frac{1}{2}
\end{array}\right.
\end{equation}

This is valid in the partial wetting regime, where the tricuspid has three sides. Thus, at the point of wetting neutrality ($\widetilde\alpha=\widetilde\beta$), we have $\tilde \alpha=\pi/4$, corresponding to $\tilde t=1/2$. Wetting takes  place when $\widetilde\alpha =0$, and hence $\tilde \beta=\pi/2$, corresponding to  $\tan (3\tilde t_-/(1-2\tilde t_-))=\pi/3$, which identifies that $\tilde t_-=(7-\sqrt{33}) /8$. This is the intercept of the line of wetting transitions by $\beta$ with the CEP line. Similarly, $\widetilde\alpha=\pi/2$ when $\tilde t_+=(7+\sqrt{33})/8$, identifying the intercept of wetting by $\alpha$ with the CEP line. Comparison with the Indekeu-Koga result (\ref{IKbeta}) for the contact angle, obtained from the surface tension formula, confirms that
\begin{equation}
    \theta_\beta=2\widetilde\alpha
    \label{CEPangles}
\end{equation}
which is precisely equivalent to $\widetilde\alpha=(\pi-\theta_\alpha)/2$ at the CEP. The scaling of the trajectories therefore reveals, consistently, the presence of a non-wetting gap between $\;\tilde t_- <\tilde t<\tilde t_+$.

\begin{figure}[h]
\includegraphics[width=.9\linewidth]{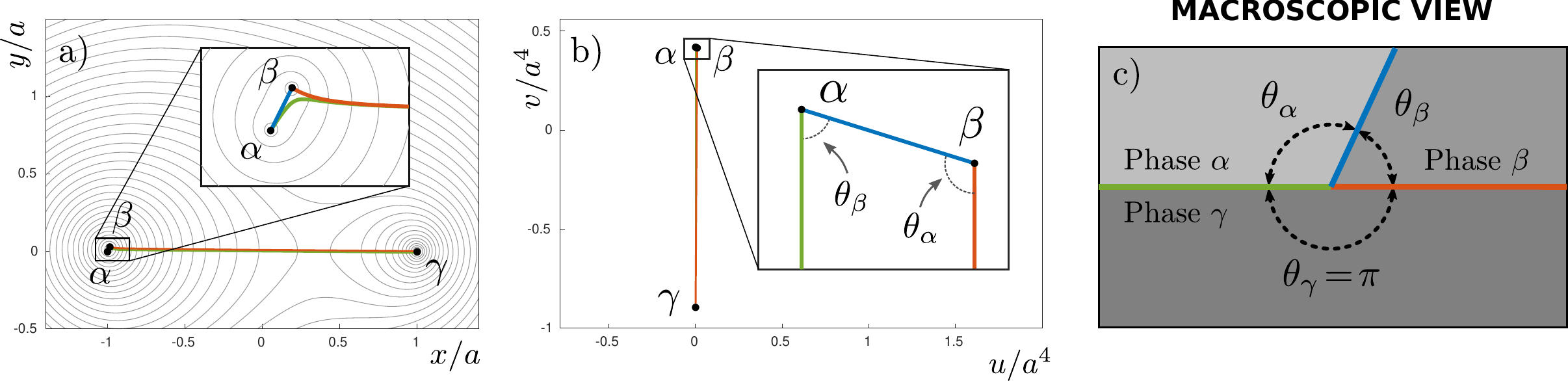}
\caption{\label{Fig20} a) The tricuspid of microscopic trajectories very close to the $\alpha\beta$ CEP with interior angles $\widetilde\alpha$ and $\widetilde\beta$ satisfying $\widetilde\alpha+\widetilde\beta=\pi/2$. b) The corresponding straight line trajectories after the conformal map, which meet at the $\alpha$ vertex with angle  $\theta_\beta=2\widetilde\alpha$, or equivalently at the $\beta$ vertex with angle $\theta_\alpha=\pi-2\widetilde\alpha$, forming the edge of an almost infinite parallelogram. c) The three-phase contact line, infinitesimally close to the $\alpha\beta$ CEP, where the $\alpha\beta$ interface meets a planar $\gamma$ interface at an angle $\theta_\beta$.}
\end{figure}

Fig.\ \ref{Fig20} shows the tricuspid and its conformal mapping close to the $\alpha\beta$ CEP, with the trajectories close to the $\alpha$ and $\beta$ vertices rescaled and shown inset. In the limit $\alpha\to\beta$, the re-scaled Neumann triangle is the end of an infinite parallelogram, whose interior angles are the contact angles of a macroscopic drop at a planar interface with the non-critical $\gamma$ phase. The rule for the tricuspid angles $\widetilde\alpha+\widetilde\beta=\pi/2$ is then equivalent to the simple geometrical condition $\theta_\alpha+\theta_\beta=\pi$, since $\widetilde\alpha=(\pi-\theta_\alpha)/2$ and $\widetilde\beta=(\pi-\theta_\beta)/2$.

\subsection{Conformal mapping at the phase boundary and for complete wetting.}

\begin{figure}[t]
\includegraphics[width=.9\linewidth]{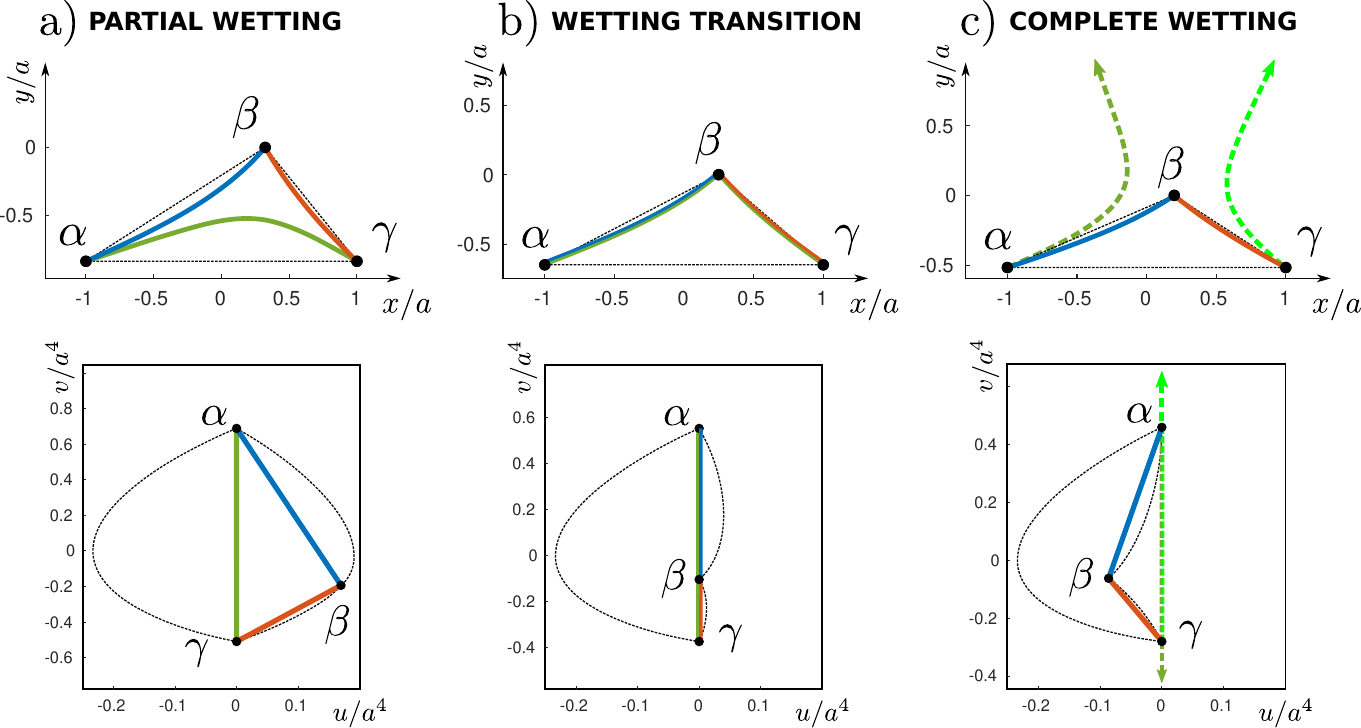}
\caption{\label{Fig21} The density paths representing the interfacial trajectories (top panels) and their conformal mapping $u+vi=f(x+yi)$ (bottom panels) for: a)  Partial wetting, b) At the wetting transition and c) For complete wetting. For partial wetting, and at the transition, the three trajectories map onto the sides of the Neumann triangle, whose lengths are the respective surface tensions. The triangle collapses at the wetting transition so that Antonov's rule is obeyed. For complete wetting, trajectories that connect two bulk phases exist only for the $\alpha\beta$ and $\beta\gamma$ interfaces. The branches of the $\alpha\gamma$ algebraic curve that pass through $\alpha$ and $\gamma$ are disconnected, do not pass through $\beta$, and instead fall to infinity (dashed dark and light green). These still map onto the vertical line $u=0$ (although they belong to different sheets of the mapping and  remain disconnected) but do not represent an interface profile. The mapping of the simple triangle connecting the bulk phases is also shown for each (dotted lines).}
\end{figure}

At the wetting phase boundary, the tricuspid disappears and the $\alpha\gamma$ trajectory is made of the separate $\alpha\beta$ and $\beta\gamma$ paths that meet at a right angle $\widetilde\beta=\pi/2$. The three trajectories are, therefore, described by the same algebraic curve $u(x,y)=0$, the branches of which intersect. For example, along the line of symmetry $\tilde s=0$, at the phase boundary $a=a_w=\sqrt{2\sqrt{3}-3}$, the algebraic curve is
\begin{equation}
x^4+y^4+\frac{8}{3}a_w^2y^3+2a_w^2(1+a_w^2)(y^2-x^2) -6x^2y^2-8a_w^2x^2y=0
\end{equation}
The transform $f(x+iy)=u+iv$ maps this, and hence all three trajectories, onto the same line $u=0$, equivalent to the collapse of the Neumann triangle -- see Fig.\ \ref{Fig21}b. The relation between the tricuspid angles and contact angles is maintained since $\pi-\theta_\beta=2\widetilde\beta=\pi$. 

In the complete wetting regime, the $\alpha\beta$ and $\beta\gamma$ paths are described by two different algebraic curves. For example, along the line of symmetry, when $a<a_w$, the $\alpha\beta$ path is described by the quartic curve eqn. (\ref{ualphabeta}) while the equation for the $\beta\gamma$ path has the sign of the first polynomial term reversed. These paths meet at an angle $\widetilde\beta >\pi/2$. There is no direct path from $\alpha$ to $\gamma$ that passes through $\beta$. The quartic curve (\ref{ualphagammasym}) still exists, but the branches that pass through $\alpha$ and $\gamma$ are disconnected, and do not pass through the $\beta$ vertex; instead, they escape to infinity -- see Fig.\ \ref{Fig21}c. This solution, therefore, does not describe an interfacial configuration. The transform $f(x+iy)=u+iv$ maps the $\alpha\beta$ and $\beta\gamma$ paths onto two straight lines of lengths $\sigma_{\alpha\beta}$ and $\sigma_{\alpha\gamma}$, respectively, that meet with an exterior opening angle $2\widetilde\beta$. The branches of the $\alpha\gamma$ quartic curve still map onto the straight line $u=0$, but these belong to different sheets of the mapping, and does not represent a path between $\alpha$ and $\gamma$ -- see Fig.\ \ref{Fig21}c.

\subsection{First-order dynamical equations}

Having determined the trajectories, surface tensions and phase diagram, we now turn our attention to the more difficult task of finding the actual density profiles, or equivalently, the full time dependence ${\bf r}(t)=(x(t),y(t))$ of the 2D motion. Progress can be made here because of a simplifying feature of the dynamics of the KWI model which, we will show later, generalizes to a much broader class of models. The dynamical motion is characterized by two conservation laws -- the conservation of energy, $(\dot x^2 +\dot y^2)/2=\omega(x,y)$, and the trajectory equation $u_x\dot x+u_y\dot y=0$. The potential $\omega(x,y)$ is a sixth-order polynomial while $u_x(x,y)$ and $u_y(x,y)$ are cubic polynomials, suggesting there is some connection between them. This must be the case since $|f'(z)|^2=2\omega(x,y)$ which is equivalent to 
\begin{equation}
    u_x^2+u_y^2=2\,\omega(x,y)
\end{equation}
 which may be checked explicitly using (\ref{omegagen}) with (\ref{ux}) and (\ref{uy}).  The conservation of energy is then the same as 
\begin{equation}
    \dot{x}^2+\dot{y}^2=u_x^2+u_y^2
\end{equation}
and together with,  $u_x\dot{x}+u_y\dot{y}=0$, it follows that the dynamics are described by the first-order equations
\begin{equation}
\dot x=\pm u_y,\hspace{1cm} \dot y =\mp u_x
\label{firstorder}
\end{equation}
where the signs depend simply on the bulk boundary conditions, i.e.\ whether $\alpha$ or $\gamma$ is at $t=\pm\infty$. These relations also make it clear that the singularities of the algebraic curves, where $u_x=u_y=0$, coincide with the bulk vertices, or stationary points of the classical motion, where $\dot x=\dot y=0$.  Equivalently, the velocity is specified by the conjugate harmonic function, so that $\dot{\bf{r}}=\pm\nabla v(x,y)$ i.e. the gradient flow equation of overdamped Langevin dynamics where the "potential" is the harmonic conjugate.

These equations of motion are particularly simple when written in terms of the complex variable $z=x+iy$ and its complex conjugate $\bar z=x-iy$. Using the Cauchy-Riemann equations, $u_x=u_y$ and $u_y=-v_x$, these read
\begin{equation}
    \dot{\bar z}=i f'(z),
    \label{dynz}
\end{equation}
and the equivalent complex conjugate result for $\dot{z}$.
Using this, we can prove the formula for the surface tension another way.  In terms of the complex variables, the expression for the surface tension (\ref{ST1}) is $
    \sigma_{\alpha\gamma}=\int_{-\infty}^{\infty}\!dt\, \dot z\,\dot{\bar z}$,
and substituting for $\dot{\bar z}$ gives $\sigma_{\alpha\gamma}=i(f(z_\gamma)-f(z_\alpha))$ and hence $\sigma_{\alpha\gamma}=(v_\alpha-v_\gamma)$ as earlier. 

The reduction in the order of the dynamical equations is a huge simplification allowing us to determine the trajectories, $x(t)$ and $y(t)$, by quadrature, similar to the familiar one-component theory. We illustrate this along the line of symmetry.

\subsubsection*{The wetting film thickness at $\tilde s=0$}

Recall that along the line of symmetry at $\tilde s=0$, the $\alpha$, $\beta$ and $\gamma$ bulk vertices lie at $(-a,-a^2)$, $(0,0)$ and $(a,-a^2)$ in the $(x,y)$ plane, with $x\equiv \rho_1$, $y\equiv \rho_2$ and $a=\sqrt{3\tilde t}$.  The $\alpha\gamma$ path is described by the quartic curve 
$u(x,y)=0$, where
\begin{equation}
\frac{u(x,y)}{\sqrt{2}}=x^4+y^4+\frac{8}{3}a^2y^3+2a^2(1+a^2)(y^2-x^2) -6x^2y^2-8a^2x^2y-\frac{a^4(a^4+6a^2-3)}{3}
\end{equation}
This is a only a quadratic equation in $x^2$, and has solution
\begin{equation}
    x^2(y)=a^2(1+a^2)+4a^2y+3y^2-\frac{2}{\sqrt3}(y+a^2)\sqrt{a^2(3+a^2)+4a^2y+6y^2}
    \label{xy}
\end{equation}
so the inverse $x(y)$ of the $\alpha\gamma$ curve is relatively simply. 

\begin{figure}[t]
\includegraphics[height=5.cm]{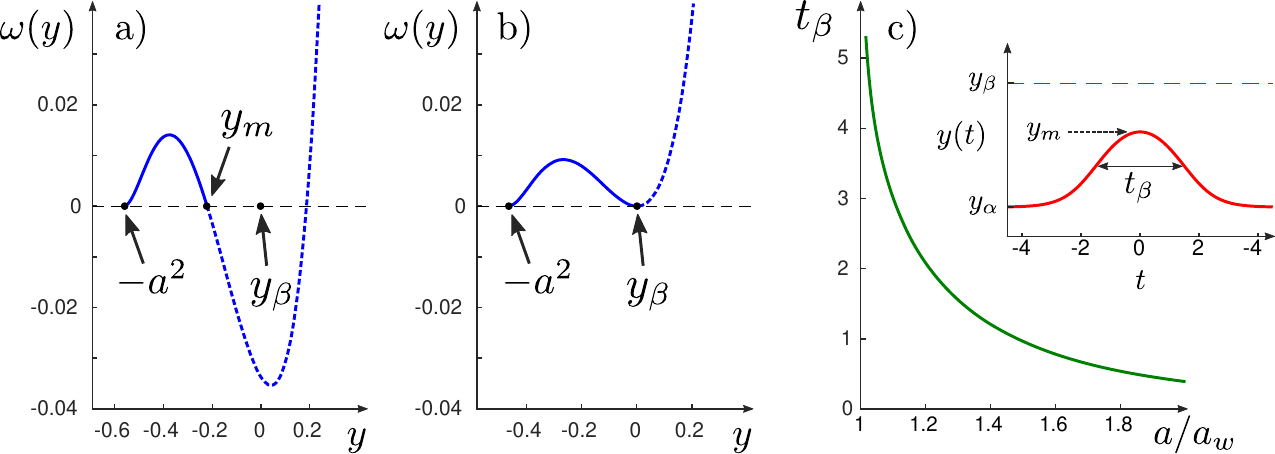}
\caption{\label{Fig22} The effective potential $\omega(y)$ determining $y(t)$ for the physically relevant range $[-a^2\!, y_m]$ (continuous) and unphysical range (dashed) is shown as a blue line in a) for partial wetting ($a/a_w=1.1$) and  b) at the wetting transition ($a_w=\sqrt{2\sqrt{3}-3}$), at the line of symmetry $\tilde s=0$.  Here, $y_\alpha=y_\gamma=-a^2$ is the bulk value where $\omega(-a^2)=\omega'(-a^2)=0$, while $y_m=y(0)$ is the midpoint value where $\omega(y_m)=0$. On approaching the transition, $y_m\to y_\beta\equiv0$ as $y_m\propto -\sqrt{a-a_w}$ and the potential is also a minimum at $y_\beta$.  c) The wetting film thickness $t_\beta$, defined in this case as the distance between inflection points of $y(t)$, as a function of $a$. As $a\to a_w$, the thickness diverges, corresponding to the wetting transition. Inset: The density $y(t)$ for $a/a_w=1.1$ (continuous red line) obtained using the quadrature (\ref{exactquad}). Other relevant magnitudes are also shown.}
 \end{figure}

Next, consider the first-order equations of motion (\ref{firstorder}). The derivatives $u_x$ and $u_y$ are easily determined so that the equations of motion read
\begin{equation}
    \frac{\dot x}{\sqrt2}=3x^2y-a^2(1+a^2)y+2a^2(x^2-y^2)-y^3
    \label{xdotsym}
\end{equation}
and 
\begin{equation}
    \frac{\dot y}{\sqrt2}=x(x^2-3y^2-4a^2y-a^2(1+a^2))
    \label{ydotsym}
\end{equation}
The equation for $\dot y$ is particularly simple because the dependence on $x$ can be eliminated using (\ref{xy}). A nice way of expressing this is as an effective one-dimensional conservation of energy 
\begin{equation}
    \frac{\dot y^2}{2}=\omega(y)
    \label{exactysym}
\end{equation}
where
\begin{equation}
 \omega(y)= \frac{4}{3}\,(y+a^2)^2\,\big(a^2(3+a^2)+4a^2y+6y^2\big)\,x^2(y)
 \label{eff}
 \end{equation}
is a one-component potential which determines, exactly, the dynamics of the $y$ coordinate. A similar effective potential exists for the $x$ coordinate but is intractably complicated and is not useful. The profile $y(t)$, which is symmetrical about $t=0$, is then determined by the usual quadrature
\begin{equation}
     t=\int_{y}^{y_m} \frac{dy}{\sqrt{2\;\omega(y)}}
    \label{exactquad}
\end{equation}
where we have assumed $t>0$. The effective potential (\ref{eff}) has a minimum at $y=y_\alpha=-a^2$, where $\omega(-a^2)=\omega'(-a^2)=0 $. The potential vanishes again at $y_m$, the maximum possible value of the $y$ coordinate, which occurs at $t=0$ where $x=0$, found from the solution of 
\begin{equation}
    y_m^4+\frac{8}{3}a^2y_m^3+2a^2(1+a^2)y_m^2=\frac{a^4(a^4+6a^2-3)}{3}
\end{equation}
For partial wetting ($a>a_w$), for which $y_m<0$, the potential is not a minimum at this point. However, at the wetting phase boundary ($a=a_w$), for which $y_m=y_\beta=0$, it is a minimum so that $\omega(0)=\omega'(0)=0$ -- see Fig.\ \ref{Fig22}. While rather complicated, the effective potential has the same qualitative features as the usual quartic Landau-like potential used in standard one-component square gradient theories of wetting. Not surprisingly, the integral in (\ref{exactquad}) cannot be evaluated in terms of elementary functions. However, expanding $\omega(y)$ about the minimum occurring for $y>y_m$, and noting that $y_m\propto \sqrt{a-a_w}$, determines that the wetting film thickness $t_\beta$ of the adsorbed $\beta$ phase diverges continuously on approaching the wetting phase boundary, as
 \begin{equation}
     t_\beta\sim -\xi_\beta\ln (a-a_w)
     \label{film}
 \end{equation}
 where  $\xi_\beta=(9+5\sqrt3)/12\sqrt2\approx 1.04064$ is the bulk correlation length of the $\beta$ phase at the wetting transition, identified using $\xi_\beta=\omega_{xx}(x_\beta,y_\beta)^{-1/2}$. The logarithmic divergence of the film thickness is the expected result for mean-field descriptions of wetting in systems with short-ranged forces \cite{Sullivan1986}. 

\subsection{Exact density profiles at wetting from the algebraic curve}

\begin{figure}[t]
\includegraphics[height=4.cm]{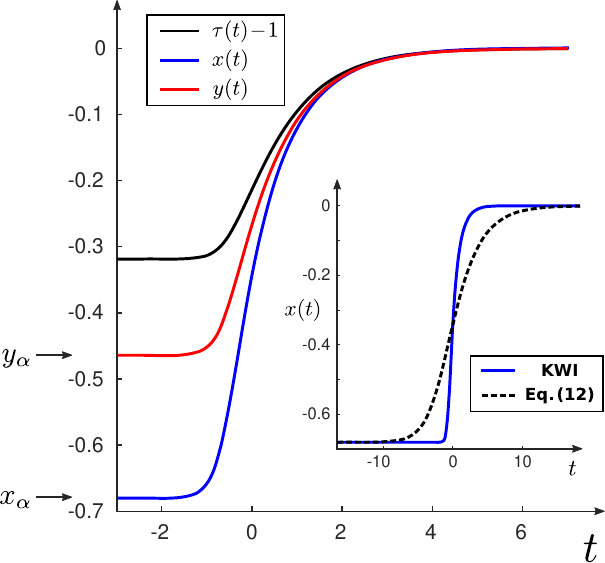}
\caption{\label{Fig23} Density component profiles $x(t)$ and $y(t)$, together with that of the parameter $\tau(t)-1$, for the $\alpha\beta$ interface at the wetting transition,  $a=\sqrt{2\sqrt{3}-3}$. Inset: Comparison of the primary density component profile $x(t)$ of the KWI model  with the one-component result, eq.\ (\ref{onecompkwi}). The origin of $t$ has been chosen so that $x(0)=x_\alpha/2$, midway between the density of the bulk $\alpha$ phase, $x_\alpha=-a$, and bulk $\beta$ phase, $x_\beta=0$. }
 \end{figure}

By specializing to the wetting phase boundary, we can determine both coordinates $x(t)$ and $y(t)$ exactly. This can be done by combining the first-order equations of motion with the rational parameterization of the trajectory curves, allowing the coordinates to be determined implicitly by a transcendental equation. This is by far the most involved part of the solution to the KWI model, and we only quote the final result for the component density profiles along the line of symmetry at $\tilde s=0$. The spatial dependence of the component density profiles $\rho_1(z)$ and $\rho_2(z)$, or equivalently, the coordinate time dependence $x(t)$ and $y(t)$, are expressed as a function of a function. The parametric equation for the curve is
\begin{equation}
    x=x(\tau),\hspace{1cm} y= \tau x(\tau)
\end{equation}
where 
\begin{equation}
x(\tau)=\frac{6(9-5\sqrt{3})(\tau^2-1)}{2a_w^2\tau(3-\tau^2)+3c(\tau^2-a_w^2)\sqrt{1-\tau^2/b^2}}
\end{equation}
which is equivalent to (\ref{xtau}), and we have specialized to the $\alpha\beta$ section of the trajectory running from $\tau=a_w$ at $\alpha$ to $\tau=1$ at $\beta$. Recall that $b^2=2\sqrt{3}+3$ and $c=\sqrt{1+1/\sqrt{3}}$. The time dependence of the parameter $\tau$ is then specified by the transcendental equation
\begin{equation}
    \Big(\frac{\sqrt{b^2-1}+\sqrt{b^2-\tau^2}}{1-\tau}\Big)^{\xi_\beta}\Big(\frac{\tau^2-a_w^2}{(\sqrt{b^2-a_w^2}+\sqrt{b^2-\tau^2})^2}\Big)^{\xi_\alpha}=e^{t-t_0}
    \end{equation}
where $t_0$ is a constant specifying the arbitrary location of the interface and
\begin{equation}
 \xi_\alpha=\frac{(2\sqrt{3}+3)\sqrt{\sqrt{3}+1}}{12\sqrt{2}},\hspace{1cm}\xi_\beta=\frac{9+5\sqrt{3}}{12\sqrt{2}}   
\end{equation}
are the (true) bulk correlation lengths of the $\alpha$ and $\beta$ phases, i.e.\ governing the exponential decay of correlation functions. This solution is much more complicated than that for the one-component model (\ref{onecompkwi}), but displays the same structure with each component decaying exponentially into the bulk phase, each side of the interface, with the appropriate correlation length. Since $\,\xi_\beta\approx 1.04064\,$ is larger than $\,\xi_\alpha\approx 0.629587$, the interface decays more sharply into the bulk $\alpha$ phase. The exact profiles for $\tau(t)-1$, $x(t)$ and $y(t)$ are shown in Fig.\ \ref{Fig23}, together with a comparison
of the primary density $x(t)$ with the one-component result, eq.\ (\ref{onecompkwi}). The interface is markedly broader in the one-component description. This simply reflects that the numerical values of both bulk correlation lengths $\xi_\alpha$ and $\xi_\beta$ are larger. Full details of this calculation and, for the general trajectory at wetting, away from the line of symmetry are provided in the Appendix. Simplicity arises from scaling behaviour as we approach the lines of critical end points, which we shall discuss in detail later.

\section{Two-component square-gradient models with local XY symmetry}

The methods and results of the last section show how the KWI model may be solved exactly. This integrability was, of course, strongly hinted at by the numerical observations of Koga and Widom, particularly regarding the properties of the tricuspid of the density profile paths. We have explained this by showing that the tricuspid can be conformally mapped onto the Neumann triangle determining the surface tensions. However, there are several aspects of the solution which suggest that a simpler mathematical approach is possible. For example, the following questions remain unanswered: What is the origin of the  conformal invariance and the first-order dynamical equations? What property of the KWI model makes it integrable, and does this apply to other models? Finally, what property of the KWI  model leads to non-wetting gaps?

In this section, we show that these questions may be answered when we reformulate the variational problem using complex analysis from the very beginning. Combining the real density variables $\rho_1$ and $\rho_2$, or equivalently $x$ and $y$, into the complex variable $z=x+iy$ and its conjugate $\bar{z}=x-iy$, will allow us to spot, and better exploit, the underlying component isotropy in a broad class of models, and arrive very quickly at the trajectories of the (real) density components, and the surface tensions, without doing anything more than a single integration of a simple polynomial.

\subsection{Using complex variables}

 Let us consider a general class of two-component models for which the excess square-gradient grand potential functional is 
 \begin{equation}
 \Omega[x,y]=\int_{-\infty}^\infty\!\!\!dt\;\left(\,\frac{1}{2}(\dot x^2+\dot y^2)+\omega(x,y)\right)
\end{equation}
and where the potential $\omega(x,y)$ vanishes at three bulk densities $\alpha$, $\beta$ and $\gamma$ situated at $(x_\alpha,y_\alpha)$,  $(x_\beta,y_\beta)$ and $(x_\gamma,y_\gamma)$. We keep the convention for the ordering $x_\alpha\le x_\beta\le x_\gamma$. The two Euler-Lagrange equations are
\begin{equation}
\ddot x=\omega_x, \hspace{1cm} \ddot y=\omega_y
\end{equation}
and have a first-integral, equivalent to the equation for the conservation of energy $E$, with $E=0$, 
\begin{equation}
\frac{1}{2}(\dot x^2+\dot y^2)=\omega(x,y)
\end{equation}
for profiles that end at the bulk densities. Using this conservation law the surface tension of the $\alpha\gamma$ interface (say) is 
\begin{equation}
\sigma_{\alpha\gamma}=\int_{-\infty}^\infty\!\!\! dt\;(\dot x^2+\dot y^2)
\end{equation}
 We now re-write these equations by introducing a complex variable $z$ and its complex conjugate $\bar{z}$:
 \begin{equation}
 z=x+iy,\hspace{1cm}\bar{z}=x-iy
 \end{equation}
 Using these, the grand potential functional is
\begin{equation}
    \Omega[z,\bar{z}]=\int_{-\infty}^\infty\!\!\! dt\; \left(\frac{\dot{z}\dot{\bar{z}}}{2}+w(z,\bar{z})\right)
\end{equation}
where 
\begin{equation}
w(z,\bar{z})=\omega(x,y)
\end{equation}
 The potential vanishes at the three bulk vertices in the complex plane, $z_\mu=x_\mu+iy_\mu$, where $w(z_\mu,\overline{z_\mu})=0$, for $\mu=\alpha,\beta,\gamma$, where it attains a minimum. The Euler-Lagrange equations, found from minimizing the grand potential, are
\begin{equation}
\ddot{z}=2\:\frac{\partial w(z,\bar{z})}{\partial\bar{z}},\hspace{1cm}\ddot{\bar{z}}=2\:\frac{\partial w(z,\bar{z})}{\partial z}\
\end{equation}
and the equation for the conservation of energy becomes
\begin{equation}
    \frac{\dot{z}\dot{\bar{z}}}{2}=w(z,\bar{z})
\end{equation}
Finally, the expression for the surface tension of the $\alpha\gamma$ interface (say) is
\begin{equation}
 \sigma_{\alpha\gamma}=\int_{-\infty}^{\infty}\!\!\! dt\;\,    \dot{z}\dot{\bar{z}}
 \label{STcomplex}
\end{equation}
which, of course, can also be applied to the other interfaces.

\subsection{Models with local XY symmetry}

The advantage of this reformulation emerges if we now specialize to a class of real potentials of the form
\begin{equation}
    w(z,\bar{z})=w(z)\overline{w(z)}
    \label{wclass}
\end{equation}
involving the product of a function $w(z)$, with its conjugate $\overline{w(z)}$, requiring that it satisfies $w(z_\mu)=0$. Clearly, for this class of potentials the Euler-Lagrange equations read 
\begin{equation}
\ddot{z}=2\; w (z)\overline{w'(z)},\hspace{1cm}\ddot{\bar{z}}=2\;w'(z)\overline{w(z)}
\end{equation}
For example, if we suppose that
\begin{equation}
w(z)=\prod_\mu (z-z_\mu)^{n_\mu}
\end{equation}
then the original bulk potential, in component form, is
\begin{equation}
\omega(x,y)=\big((x-x_\alpha)^2+(y-y_\alpha)^2\big)^{n_\alpha}\big((x-x_\beta)^2+(y-y_\beta)^2\big)^{n_\beta}\big((x-x_\gamma)^2+(y-y_\gamma)^2\big)^{n_\gamma}
\label{KWIgen}
\end{equation}
which is clearly a generalization of the KWI model, which is recovered on setting $n_\alpha=n_\beta=n_\gamma=1$. More generally, the index $n_\beta=1$ means the $\beta$ phase is non-critical, while $n_\beta=2$ means it is critical, etc.

\subsection*{First-order equations}

While the potentials $\omega(x,y)=w(z)\overline {w(z)}$ are not central potentials, familiar from orbital mechanics, they all have a local $O(2)$ or XY symmetry at each bulk vertex at {\it{and}} away from bulk criticality, where the potential is {\it{locally}} radial. This property, which reflects directly the component isotropy, makes this class of models integrable, i.e.\ there are two conservation laws matching the number of components. These are most simply expressed as the complex conjugate pair of first-order gradient flow equations
\begin{equation}
     \dot{\bar{z}}=i f'(z), \hspace{1cm} \dot{z}=-i\overline{f'(z)}
     \label{ccpair}
\end{equation}
where $f(z)$ is given by
\begin{equation}
    f'(z)=\sqrt{2}\;e^{i\psi} w(z)
    \label{fprime}
\end{equation}
and, hereafter, referred to as the trajectory function. Here, $\psi$ is similar to a gauge factor, not apparent in $\omega(x,y)=|f'(z)|^2/2$. The proof of this is straightforward: taking the derivative of $\dot{\bar{z}}$ (say) gives $\ddot{\bar{z}}=if''(z)\dot z$ and substituting for $f''(z)=\sqrt{2}e^{i\psi} w'(z)$ and $\dot{z}=-i\overline{f'(z)}$ removes the gauge factor so that $\ddot{\bar z}=2w'(z)\overline{w(z)}$, recovering the Euler-Lagrange equation.
Similarly, the product of the two first-order equations (\ref{ccpair}) gives, directly,
\begin{equation}
\frac{\dot{z}\dot{\bar{z}}}{2}=w(z)\overline{w(z)}
\end{equation}
recovering the conservation of energy. For potentials with a local XY symmetry, the two conservation laws (\ref{ccpair}) are, therefore, a consistent factorization of the equation for the conservation of energy in the complex coordinate plane when $E=0$. 

\subsection*{The trajectory and surface tension}

Next, from the trajectory function $f(z)$ we define functions $u(x,y)$ and $v(x,y)$ from
\begin{equation}
f(z)=u(x,y)+iv(x,y)
\end{equation}
or, equivalently, new coordinates $u$ and $v$ via the transform $f(x+iy)=u+iv$. Since $f(z)$ is a function of a complex variable, these functions are harmonic, and must satisfy the Laplace equation
\begin{equation}
    \nabla^2u(x,y)=0,\hspace{1cm} \nabla^2v(x,y)=0
\end{equation}
with $v(x,y)$ the conjugate of $u(x,y)$, obeying the Cauchy-Riemann conditions $u_x=v_y$ and $u_y=-v_x$. The derivative of the function $f(z)$ with respect to "time" $t$ is clearly $df(z)/dt=f'(z) \dot{z}$ and using the dynamical equation for $\dot{z}$ we obtain
\begin{equation}
    \frac{d f(z)}{dt}=-i|f'(z)|^2
\end{equation}
Thus, the dynamical equations of motions for the coordinates $u$ and $v$ are
\begin{equation}
    \dot{u}=0,\hspace{1cm} \dot{v}=-|f'(z)|^2
    \label{udotvdot}
\end{equation}
implying that $u(x,y)$ is a constant. Both the gauge $\psi$ and constant of integration in $f(z)$ are, so far, arbitrary. We choose the gauge $\psi=\psi_{\alpha\gamma}$ so that $\Re{(f(z_\alpha))}=\Re(f(z_\gamma))=u_{\alpha\gamma}$
and the constant of integration so that $u_{\alpha\gamma}=0$. This determines the trajectory of the $\alpha\gamma$ interface, provided it exists, i.e.\ in the partial wetting regime, as
\begin{equation}
   \Re({f(z)})\equiv u(x,y)=0
\end{equation}
This may be regarded as an alternative expression of the second conservation law for these models, in addition to the conservation of energy. Finally, substituting $\dot{\bar{z}}=if'(z)$ into the expression (\ref{STcomplex}) immediately determines the surface tension as
 \begin{equation}
 \sigma_{\alpha\gamma}=\Im{(f(z_\alpha)-f(z_\gamma))}
 \end{equation}
or equivalently
 \begin{equation}
 \sigma_{\alpha\gamma}=|f(z_\alpha)-f(z_\gamma)|
 \label{sigmacomplex}
 \end{equation}
 which is independent of the gauge $\psi$ and the constant of integration in $f(z)$.

 \subsection*{The general tricuspid mapping}

All these models display the same remarkable property linking directly microscopic and macroscopic physics. As before, the $\alpha\beta$, $\beta\gamma$ and $\alpha\gamma$ trajectories outline a tricuspid with internal angles $\widetilde\alpha$, $\widetilde\beta$ and $\widetilde\gamma$. The conformal map $f(x+iy)=u+iv$ straightens all three paths, since the function $f(z)$ is identical for all of them, up to a rotation determined by the appropriate value of $\psi$. The mapping with $\psi=\psi_{\alpha\gamma}$ maps them onto a triangle, with the $\alpha\gamma$ trajectory along the vertical line $u=0$, whose three side lengths are the corresponding surface tensions. Therefore, in all these models, the tricuspid defined by the microscopic density profile paths maps onto the Neumann triangle for the macroscopic contact angles, and we can identify that $\psi_{\alpha\gamma}-\psi_{\alpha\beta}=\pi-\theta_\alpha$, etc.  The conformal map preserves angles except at the singularities, residing at the three bulk vertices, where $f'(z_\mu)=0$. At these points, the angles between lines are multiplied by the index describing the singularity, which follows from de Moivre's theorem. Therefore, the generalized relation between the tricuspid and contact angles is
\begin{equation}
    \widetilde\alpha=\frac{\pi-\theta_\alpha}{n_\alpha+1},\hspace{1cm}
    \tilde \beta=\frac{\pi-\theta_\beta}{n_\beta+1},\hspace{1cm}
 \widetilde\gamma=\frac{\pi-\theta_\gamma}{n_\gamma+1}
\end{equation}
so the rule for the sum of the tricuspid angles becomes
\begin{equation}(n_\alpha+1)\widetilde\alpha+(n_\beta+1)\widetilde\beta+(n_\gamma+1)\widetilde\gamma=\pi
    \label{tricuspidgen}
\end{equation}
At a wetting transition, where $\beta$ wets the $\alpha\gamma$ interface, the tricuspid disappears, leaving the outline provided by the paths of the $\alpha\beta$ and $\beta\gamma$ trajectories. In this limit, $\widetilde\alpha=\widetilde\gamma=0$, while
\begin{equation}
    \widetilde\beta=\frac{\pi}{n_\beta+1}
\end{equation}
 which is the same as the value of the scattering angle $\Delta\theta_{scat}(n_\beta)$, eq.\ (\ref{tscat}).  Reformulating the initial variational problem using complex analysis is therefore a much more direct route to determining the trajectories from the bulk potential, without having to integrate the differential equation (\ref{pathODE1}) for the path $y(x)$. This approach reveals the origin of the conformal invariance, leading to the map between the tricuspid and the Neumann triangle, which applies to a wider class of models, each exhibiting a local XY symmetry. Two examples of this are given below, before we apply the analysis to a binary mixture at a wall, equivalent to the properties of the KWI model near the CEP.

 \subsection{Surface tension formulae}.

 Here, we derive surface tension formulae for the generalized KWI model defined by the potential (\ref{KWIgen}) using the same convenient choice of coordinates with $\alpha$ at $(-a,0)$, $\beta$ at $\ell(\cos\phi,\sin\phi)$, and $\gamma$ at $(0,a)$ in the $(x,y)$ plane. Three examples are given: \\
 
 a) Suppose $\alpha$ and $\gamma$ are multi-critical, with index $n_\alpha=n_\gamma=n$ while $\beta$ is non-critical with $n_\beta=1$. Thus, $w(z)=(z^2-a^2)^n(z-\ell e^{i\phi})$, and from (\ref{sigmacomplex}) the surface tension of the $\alpha\gamma$ interface follows as $\sigma_{\alpha\gamma}=\sqrt{2}|\int_{-a}^adz\, w(z)|$ giving
 \begin{equation}
 \sigma_{\alpha\gamma}=\frac{\sqrt{2}\;2^{n+1}\;n! \;a^{2n+1}\;\ell} {(2n+1)!!}
 \end{equation}
 Intriguingly, this is independent of the median angle $\phi$, recovering, for example, the KWI result, $\sigma_{\alpha\gamma}=4\sqrt{2}a^3\ell/3$, when $n=1$, while $\sigma_{\alpha\gamma}=16\sqrt{2}a^5\ell/15$ and $\sigma_{\alpha\gamma}=32\sqrt{2}a^7\ell/35$, when $n=2$ and $n=3$, respectively.\\

 b) Next, suppose that $\alpha$ and $\gamma$ are non-critical, with $n_\alpha=n_\gamma=1$, while $\beta$ is critical with $n_\beta=2$. Then, $w(z)=(z^2-a^2)(z-\ell e^{i\phi})^2$ and the integral, $\sigma_{\alpha\gamma}=\sqrt{2}|\int_{-a}^adz\, w(z)|$, determines the tension as
 \begin{equation}
 \sigma_{\alpha\gamma}=\frac{4\sqrt{2}}{15}\; a^3\sqrt{a^4+10a^2\ell^2\cos2\phi+25\ell^4}
 \label{121}
 \end{equation}
which now depends on $a$, $\ell$ and $\phi$, and hence the full shape of the triangle defined by the bulk vertices. We shall use this result shortly to illustrate a wetting transition involving a critical phase with XY symmetry.\\

 c) Finally, we suppose that $\alpha$, $\beta$ and $\gamma$ are all critical, with $n_\alpha=n_\beta=n_\gamma=2$. Then, $w(z)=(z^2-a^2)^2(z-\ell e^{i\phi})^2$ and the integral, $\sigma_{\alpha\gamma}=\sqrt{2}|\int_{-a}^adz\, w(z)|$, determines that
 \begin{equation}
 \sigma_{\alpha\gamma}=\frac{16\sqrt{2}}{15}\; a^5\sqrt{a^4+14a^2\ell^2\cos2\phi+49\ell^4}
 \end{equation}
similarly depending on $a$, $\ell$ and $\phi$.\\

\subsection{Wetting transition by a critical layer with XY symmetry}

Building on these results, we now explore an example of wetting when a critical phase $\beta$ intrudes between two non-critical phases $\alpha$ and $\gamma$. Because the wetting layer is critical, its thickness will be determined by long-ranged thermal Casimir-like forces arising from the confinement of the critical layer, which attract or repel the $\alpha\beta$ and $\beta\gamma$ interfaces. If these forces are net repulsive, the critical phase completely wets the $\alpha\gamma$ interface. This example is useful because a continuous change from attraction to repulsion is only possible because of the component isotropy, allowing us to connect with previous results for the Casimir effect in the 3D XY model.

\begin{figure}[t]
\includegraphics[height=7.cm]{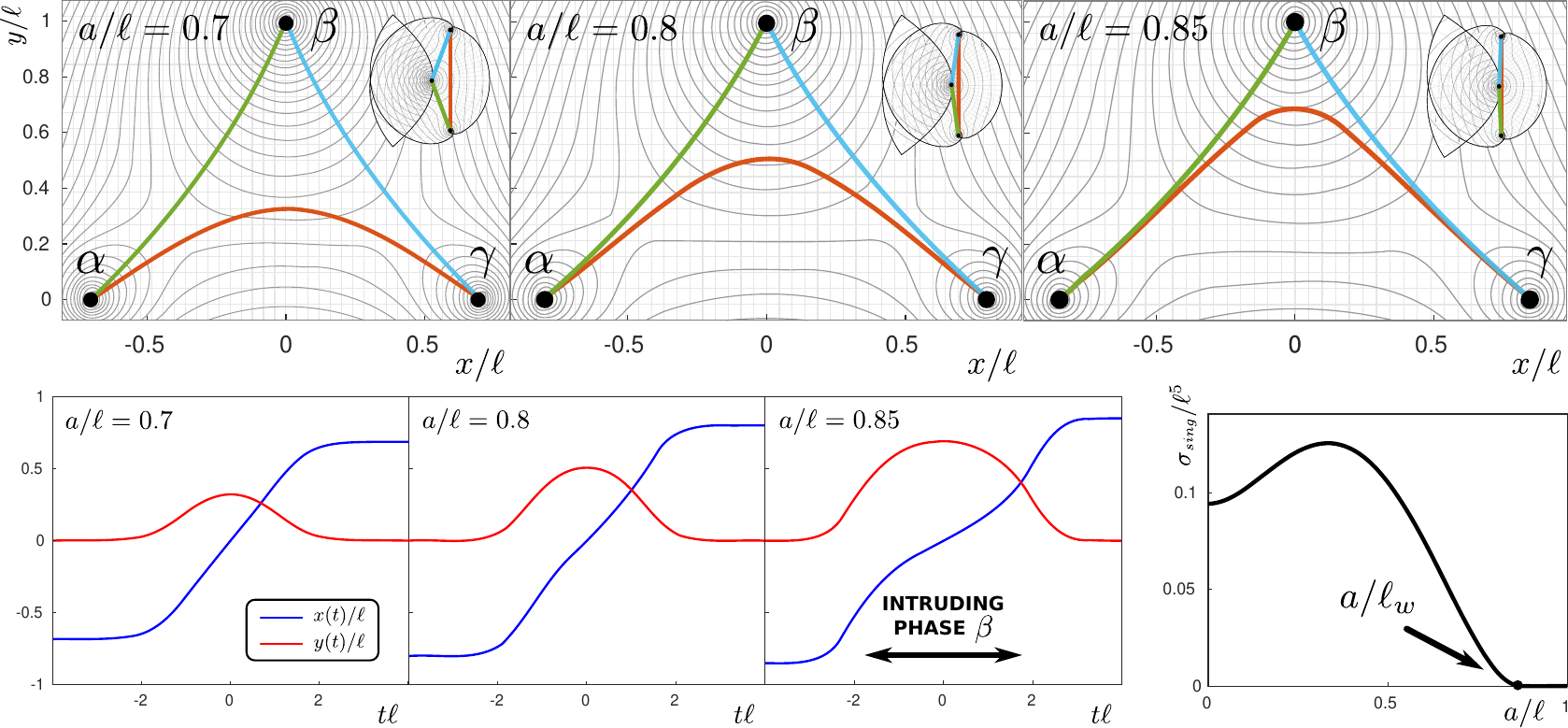}
\caption{\label{Fig24} Upper panels: Density profile trajectories when a critical XY phase $\beta$ is adsorbed between non-critical phases $\alpha$ and $\gamma$, together with the contour lines of the potential, as the wetting transition is approached. The conformal mapping of the trajectories onto the Neumann triangle is shown inset. Lower panels: Component profiles $x(t)$ and $y(t)$ for the $\alpha\gamma$ interface showing the growth of the wetting layer of $\beta$, and the singular contribution to the surface tension. A second-order phase transition occurs when $\ell_w/a=\sqrt{2\sqrt{10}-5}$ where $\widetilde\beta=\theta_{scat}(2)=\pi/3$, and the Casimir force between the non-critical interfaces vanishes.}
 \end{figure}

To study this, we use a similar coordinate system to that used for the  KWI model along the line of symmetry with the bulk vertices $(x_\mu,y_\mu)$ for $\mu=\alpha$, $\beta$ and $\gamma$ located at $(-a,0)$, $(0,\ell)$ and $(a,0)$, respectively. For this choice of vertices, the $\alpha\beta$ and $\beta\gamma$ trajectories are simply reflections of each other and $\sigma_{\alpha\beta}=\sigma_{\alpha\gamma}$. The bulk potential $\omega(x,y)$ is given by
\begin{equation}
\omega(x,y)=\big((x+a)^2+y^2\big)\,\big(x^2+(y-\ell)^2\big)^2\,\big((x-a)^2+y^2\big)
\end{equation}
similar to the KWI model, but with the contribution from $\beta$ phase quartic rather than quadratic. As noted above, the isotropy means that this description of criticality belongs to the universality class of an $n=2$-vector model. The trajectories may be determined in two ways. For example, the first integral of the path equation (\ref{pathODE}) gives
\begin{equation}
    \tan^{-1}y'(x)+\tan^{-1}\frac{y}{x+a}+\tan^{-1}\frac{y}{x-a}+2\tan^{-1}\frac{y-\ell}{x}=K
\end{equation}
where the constant $K$ is different for the $\alpha\beta$ and $\alpha\gamma$ trajectories. Cyclically applying this at the ends of the trajectories already confirms that the internal angles of the tricuspid paths, for partial wetting, must obey
\begin{equation}
    \widetilde\alpha+\frac{3}{2}\,\widetilde\beta+\widetilde\gamma=\frac{\pi}{2}
    \label{sumangles}
\end{equation}
in accordance with (\ref{tricuspidgen}). We focus primarily on the $\alpha\gamma$ trajectory, which will reveal the wetting transition. In that case $K_{\alpha\gamma}=\pi$, which follows from the symmetry condition that $y'(0)=0$. One way of obtaining the path is combining the inverse tangents. This leads to an exact derivative, $u_xdx+u_ydy=0$, which then integrates further to give the trajectory. It is far easier, however, to use the complex analysis and identify that the
\begin{equation}
    w(z)=(z^2 -a^2)(z-\ell)^2
\end{equation}
 corresponding to $n_\alpha=n_\gamma=1$, and $n_\beta=2$. This means that the algebraic curve describing the trajectory has two double points at $\alpha$ and $\gamma$, and a triple point at $\beta$, if the trajectory passes through these singularities. Integration of (\ref{fprime}) determines the trajectory function as
\begin{equation}
    f(z)=-i\sqrt{2}\,\left(\frac{z^5}{5}-i\ell\frac{z^4}{2}-(\ell^2 +a^2)\frac{z^3}{3}+ia^2\ell z^2+a^2\ell^2 z\right)
\end{equation}
In this expression, the value of the gauge is $\psi_{\alpha\gamma}=-\pi/2$, chosen so that the $\Re(f(-a))=\Re(f(a))\equiv u_{\alpha\gamma}$, where $u_{\alpha\gamma}=a^2\ell/2$. Both methods lead to the same equation for the trajectory, 
\begin{equation}
   \frac{y^5}{5}+x^4y-2x^2y^3+3\ell x^2y^2-(2\ell+a^2)x^2y-\frac{\ell}{2}(x^4+y^4)+a^2\ell(x^2-y^2)+\frac{(a^2+\ell^2)}{3}y^3+a^2\ell^2y=\frac{a^2\ell}{2}
\end{equation}
which is an algebraic curve of degree $d=5$. The curve has genus $g=4$ for partial wetting, but $g=2$ at the wetting phase boundary when it passes through all three singularities. The complex analysis also determines the surface tension $\sigma_{\alpha\gamma}=|f(a)-f(-a)|$, which evaluates as
\begin{equation}
    \sigma_{\alpha\gamma}=\frac{4\sqrt{2}a^3}{15}\,(5\,\ell^2-a^2)
\end{equation}
which is the same as (\ref{121}) on setting $\phi=\pi/2$. As $\ell$ is lowered, the trajectory gets closer to the $\beta$ vertex (see Fig.\ \ref{Fig24}), signaling that a wetting layer of $\beta$ intervenes between the $\alpha$ and $\gamma$ phases, the thickness of which diverges continuously as $\ell\to\ell_w$. The value of $\ell_w$ follows from requiring that the trajectory just touches the $\beta$ vertex at $(0,\ell)$, which reduces to $\ell_w^4+10\,a^2\ell_w^2-15\,a^2=0$, identifying that
\begin{equation}
\frac{\ell_w}{a}=\sqrt{2\sqrt{10}-5}
\end{equation}

This is larger that the corresponding value $\ell_w/a=\sqrt{2\sqrt{3}-3}$ for the KWI model for which the $\beta$ vertex is non-critical. The reason for this is that the required value of the scattering angle at the wetting transition is $\widetilde\beta=\pi/3$, when the $\beta$ vertex is a triple point,  which is smaller than the requirement $\widetilde\beta=\pi/2$, when it is a double point. The conformal transform $f(x+iy)=u+iv$ maps the trajectories onto the Neumann triangle for the tensions and contact angles, determining that the tricuspid angles are
\begin{equation}
    \widetilde\alpha=\frac{\pi-\theta_\alpha}{2},\hspace{1cm}\widetilde\beta=\frac{\pi-\theta_\beta}{3},\hspace{1cm}\widetilde\gamma=\frac{\pi-\theta\gamma}{2}
\end{equation}
which, of course, leads to the summation rule (\ref{sumangles}). This confirms that, at the wetting transition, the only non-vanishing tricuspid angle $\widetilde\beta$ takes the value of the quartic scattering angle $\widetilde\beta=\Delta\theta_{scat}(2)=\pi/3$. This is the same as the angle between surface vectors on opposing walls, for which the long-ranged Casimir contribution to the free-energy of a confined XY model at its critical point vanishes -- a result first derived by Krech \cite{Krech1997}.

Similar methods determine the trajectory and surface tensions for the $\alpha\beta$ (and symmetric $\beta\gamma$) interfaces. For example, the surface tension follows from $\sigma_{\alpha\beta}=\sigma_{\beta\gamma}=|f(i\ell)-f(-a)|$, where we may use the same function $f(z)$, since the phase factor $\psi_{\alpha\beta}$ is only needed for the orientation of the trajectory, yielding
\begin{equation}
 \sigma_{\alpha\beta}=\frac{\sqrt{2}\,}{30}\,(a^2+\ell^2)^2\,\sqrt{16a^2+\ell^2}  
\end{equation}
Application of Antonov's rule, $\sigma_{\alpha\gamma}=\sigma_{\alpha\beta}+\sigma_{\beta\gamma}$, then recovers consistently the location of the wetting phase boundary, $\ell_w/a=\sqrt{2\sqrt{10}-5}$, as found from the trajectory equation. The singular contribution to the free-energy, $\sigma_{sing}\equiv2\sigma_{\alpha\beta}-\sigma_{\alpha\gamma}$, is shown in Fig.\ \ref{Fig24} and vanishes as 
\begin{equation}
    \sigma_{sing}\propto (\ell-\ell_w)^2
\end{equation}
as $\ell\to\ell_w^+$, so that, as for the KWI model, the wetting transition is second-order with exponent $\alpha_s=0$. The divergence of the thickness of the wetting layer can also be determined analytically. Integrating the trajectory in the vicinity of the $\beta$ vertex, where it is described by (\ref{radialtraj}), yields
\begin{equation}
t_\beta\propto (\ell-\ell_w)^{-\beta_s}
\end{equation}
identifying the critical exponent for the adsorption as $\beta_s=1/3$. The standard critical exponent relation for wetting, $2-\alpha_s=2\nu_\parallel-2\beta_s$, then implies that the parallel correlation length $\xi_\parallel\propto (\ell-\ell_w)^{-\nu_\parallel}$, describing the growth of capillary-wave-like fluctuations due to interfacial wandering, diverges with exponent $\nu_\parallel=4/3$. The values of the exponents,
\begin{equation}
\alpha_s=0,\hspace{1cm} \beta_s=\frac{1}{3},\hspace{1cm} \nu_\parallel=\frac{4}{3}
\end{equation}
are different from those found for wetting transitions involving a critical layer in one-component systems, which also require 
the presence of long-ranged dispersion-like intermolecular forces to balance the Casimir interaction \cite{Nightingale1985b}. These differences are due to the XY symmetry of the adsorbed critical phase.

\section{A model of wetting by a binary mixture near a wall}

We now return to the most striking prediction of the KWI model -- the presence of a non-wetting gap along the CEP lines. The origin of this becomes far more transparent when we consider a related two-component isotropic square-gradient model of wetting in a binary fluid mixture near a planar wall. This model may also be solved using the techniques described above, which allows the trajectories, surface tensions and component density profiles to determined exactly. These, we shall show, describe the scaling behaviour that emerges in the KWI model near the CEP lines. To begin, however, we recall the results of the very simple one-component theory of wetting at a planar wall, which displays critical point wetting, but will fit consistently within the two-component description.

\subsection{One-component theory: The Nakanishi-Fisher phase diagram}

 Consider a fluid, for which two bulk phases $\alpha$ and $\beta$ coexist  below a critical temperature $T_c$, in contact with a planar wall. We keep with the same notation and use $x(t)$ to denote the density profile with the wall placed in the $t=0$ plane, with the density/concentration taking the bulk values $x_\alpha=-a$ or $x_\beta=a$. We suppose the value of the density at the wall is fixed, $x(0)=x_0$ -- this is all we will need to see the connection with the KWI model. Within the standard square gradient description of wetting, the equilibrium density profile is obtained from minimization of the grand potential functional per unit area
\begin{equation}
    \Omega[x]=\int_0^\infty \!\!\!dt\;\left(\frac{\dot x^2}{2}+\omega_0(x)\right)
\end{equation}
where $\omega_0(x)=(x^2-a^2)^2$
is a quartic potential modeling phase coexistence below $T_c$. Thus, identifying $a\propto (T_c-T)^{1/2}$ describes the usual mean-field singularity of the bulk order-parameter, while the curvature of the potential $\omega''(a)=\kappa^2$ identifies the inverse of bulk correlation length, $\kappa=1/\xi$, as $\kappa=2\sqrt{2}\,a$. Minimization of $\Omega[x]$ leads to the Euler-Lagrange equation $\ddot x=\omega_0'(x)$, which has a first integral $\dot x^2/2=\omega_0(x)$, equivalent to the conservation of energy, where again the total energy is $E=0$. Specializing to the wall-$\alpha$ interface, with $0<x_0<a$, further integration determines the equilibrium density profile, which is just a portion of the free $\alpha\beta$ profile, as $
    x(t)=a\tanh(\kappa(t_\beta-t)/2)$
where $t_\beta$ denotes the thickness of the wetting layer of $\beta$, defined here to be where $x(t_\beta)=0$. It follows that the thickness of the wetting layer is
\begin{equation}
    t_\beta=\xi\,\ln\left(\frac{a-x_0}{a+x_0}\right)
    \end{equation}
which diverges continuously as $x_0\to a$, corresponding to a second-order (critical) wetting transition. The wall is completely wet by $\beta$ when $x_0>a$ . Similarly, the wall-$\beta$ interface is completely wet by $\alpha$ when $x_0<-a$. The surface tensions of the wall-$\alpha$ and wall-$\beta$ interfaces are the values of the grand potential for the respective equilibrium profiles and reduce to integrals that can be easily evaluated. For example $\sigma_{w\alpha}=\int_{-a}^{x_0}dx\sqrt{2\,\omega_0(x)}$, yielding
\begin{equation}
    \sigma_{w\alpha}=|f(x_0)-f(-a)|
\end{equation}
where, by simple integration of $\sqrt{2\omega_0(x)}$, the function $f(x)$ is given by 
\begin{equation}
    f(x)=\sqrt{2}\,\left(\frac{x^3}{3}-a^2x\right)
    \label{STone}
\end{equation}
Similarly, $\sigma_{w\beta}=|f(x_0)-f(a)|$ and  $\sigma_{\alpha\beta}=|f(a)-f(-a)|$ where the latter recovers the very well known mean-field expression $\sigma_{\alpha\beta}=4\sqrt{2}a^3/3$.

\begin{figure}[h]
\includegraphics[height=5.cm]{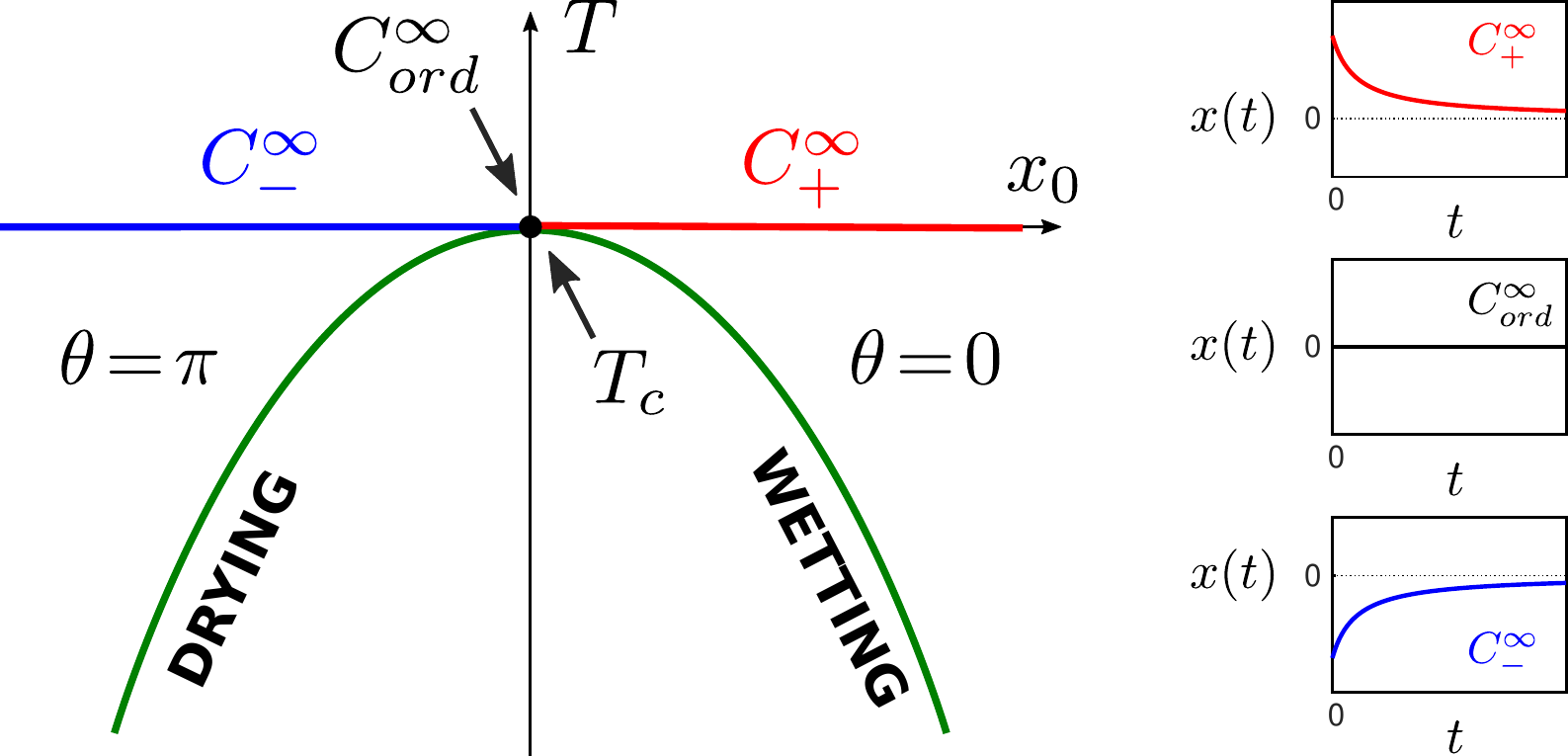}
\caption{\label{Fig25} Section of the surface phase diagram at bulk coexistence illustrating the necessity of critical point wetting for the one-component square-gradient theory with fixed boundary condition $x(0)=x_0$ at the wall. Wetting and drying phase boundaries occur along $|x_0|=a$, with $a\propto (T_c-T)^{1/2}$ and terminate at an ordinary surface phase transition $C_{ord}^\infty$, where the profile is exactly flat, $x(t)=0$. Inset: Profiles along the critical line at $T=T_c$. The ordinary surface transition separates lines of critical adsorption, $C_{+}^\infty$, and critical desorption, $C_{-}^\infty$, where the profile decays as $x(t)\propto \pm 1/t$.}
 \end{figure}

Using the tensions, we can identify the contact angle from Young's equation, $\sigma_{w\alpha}=\sigma_{w\beta}+\sigma_{\alpha\beta}\cos\theta$,
which varies between $\theta=\pi$ and $\theta=0$ as $x_0$ is increased from $-a$ to $a$. This confirms that the wetting transitions at these phase boundaries are second-order since,  $\sigma_{sing}\propto(|x_0|-a)^2$, as $|x_0|\to 0$.  A section of the Nakanishi-Fisher surface phase diagram in the $(x_0,T)$ plane is shown in Fig.\ \ref{Fig25}, showing lines of wetting and drying that converge to the same a point on the critical line $T=T_c$, which corresponds to an ordinary surface phase transition $C_{ord}^\infty$. This separates lines of critical adsorption $C_+^\infty$ and critical desorption $C_-^\infty$, where the order-parameter profile exhibits a slow algebraic decay $x(T)\sim\pm t^{-\beta/\nu}$, where $\beta=\nu=1/2$ are the mean-field values of the critical exponents for the bulk order-parameter and correlation length. The same consistent connection between the lines of wetting, drying and surface criticality applies for a uniaxial ferromagnet with a surface field (and negative surface enhancement parameter) and also in realistic molecular models of simple fluids with short-ranged forces, where the coordinate $x_0$ is replaced by the strength of the short-ranged wall-fluid potential \cite{Nakanishi1982,Evans2019,Parry2023b}.

\subsection{Two-component theory}

We now generalize to a model of wetting near wall by a binary fluid mixture with two density components $x$ and $y$. We use the same isotropic square-gradient theory as per the KWI model, but with only two bulk phases $\alpha$ and $\beta$. The grand potential per unit area is written
\begin{equation}
    \Omega[x,y]=\int_0^\infty\!\!\! dt \left(\frac{1}{2}(\dot x^2+\dot y^2)+\omega(x,y)\right)
\end{equation}
where
\begin{equation}
    \omega(x,y)=\big((x-a)^2+y^2\big)\,\big((x+a)^2+y^2\big)
    \label{binomega}
\end{equation}
so that the bulk $\alpha$ and $\beta$ phases sit at coordinates $(-a,0)$ and $(a,0)$ respectively, and $a\propto (T_c-T)^{1/2}$ measures the temperature deviation from the critical/consolute temperature of demixing. The isotropy means both components are characterized by the same inverse bulk correlation length $\kappa=2\sqrt{2}\,a$, identified from $\kappa^2=\omega_{xx}(\pm a,0)=\omega_{yy}(\pm a,0)$. The wall is again placed in the $t=0$ plane where the densities take fixed values $(x_0,y_0)$. If we set $y_0=0$, the theory reduces to the one-component model, since $\omega(x,0)=\omega_0(x)$. When the wall boundary condition is infinitely far away from $\alpha$ and $\beta$, however, the trajectories are same as the scaling limit of the KWI model near the CEP lines. Minimization of the grand potential leads, of course, to the two coupled Euler-Lagrange equations $\ddot x=\omega_x(x,y)$ and $\ddot y=\omega_y(x,y)$
with the conservation of energy relation, $(\dot x^2+\dot y^2)/2=\omega(x,y)$.
The condition that the mechanical energy is $E=0$ means, once again, that the particle takes infinite time to fall from its initial condition $(x_0,y_0)$ to a minimum at either $\alpha$ or $\beta$. These represent the two possible wall-$\alpha$ and wall-$\beta$ interfaces. Only the trajectories for which the initial condition is placed along the line $y_0=0$ are straight lines, in which case $y(t)=0$ for all $t$, and the $x(t)$ path is the same as for the one-component theory.
Here, we show how the model can be solved exactly for the trajectories, surface tensions and time dependent coordinates $(x(t),y(t))$. This can be done either by integration of the differential equation for the path or using the complex analysis. Both are explained here for completeness.

\subsection{The trajectory paths and surface phase diagram}

For the binary mixture, the differential equation for the path (\ref{pathODE1}) gives us
\begin{equation}
    \frac{y''}{1+y'^2}=\frac{y-(x+a)y'}{(x+a)^2+y^2}+\frac{y-(x-a)y'}{(x-a)^2+y^2}
\end{equation}
This can be integrated to give
\begin{equation}
    \tan^{-1}y'+\tan^{-1}\frac{y}{x+a}+\tan^{-1}\frac{y}{x-a}=\tan^{-1}c
    \label{tanswall}
\end{equation}
implying, as before, a conservation of the sum of angles made between the tangent of the path and the straight lines to the bulk vertices. The constant $c$ depends on the boundary conditions at the wall $(x_0,y_0)$ and in the bulk, and is therefore different for the wall-$\alpha$ and wall-$\beta$ trajectories, taking  values $c_{w\alpha}$ and $c_{w\beta}$. 

\begin{figure}[t]
\includegraphics[height=5.cm]{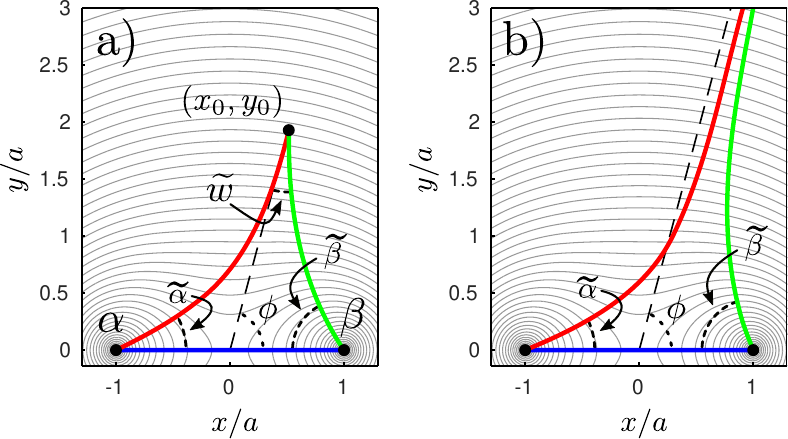}
\caption{\label{Fig26} a) A tricuspid formed from the trajectories of the wall-$\alpha$, wall-$\beta$ and (straight line) $\alpha\beta$ interfaces, with internal angles $\widetilde\alpha$, $\widetilde\beta$ and $\widetilde w$, satisfying $\widetilde\alpha+\widetilde\beta+\widetilde w/2=\pi/2$. The (fixed) density at the wall corresponds to the point $(x_0,y_0)$. b) A tricuspid when the particle falls from infinity along a straight line with angle $\phi$. In this case, we have $\widetilde w=0$, reproducing the scaling behaviour for trajectories in the KWI model asymptotically close to a CEP.}
\end{figure}

Combining the tangents we obtain
\begin{equation}
    \big(2xy+c(y^2+a^2-x^2)\big)\,dx+\big(x^2-y^2-a^2+2cxy\big)\,dy=0
    \label{binexact}
\end{equation}
which is an exact differential. Integrating determines that the wall-$\alpha$ path is
\begin{equation}
  x^2y-a^2y-\frac{y^3}{3}+c_{w\alpha}\left(xy^2+a^2x-\frac{x^3}{3}\right) =-\frac{2}{3}c_{w\alpha}a^3
  \label{trajwallalpha}
\end{equation}
where, the R.H.S. is $2c_{w\beta}a^3/3$ for the wall-$\beta$ path. The values of the constants are then given by
\begin{equation}
c_{w\alpha}(x_0,y_0)=\frac{y_0^3+3y_0(a^2-x_0^2)}{x_0(3y_0^2+3a^2-x_0^2)+2a^3},\hspace{1cm}c_{w\beta}(x_0.y_0)=\frac{y_0^3+3y_0(a^2-x_0^2)}{x_0(3y_0^2+3a^2-x_0^2)-2a^3}
\label{calphabeta}
\end{equation}
which differ only in the sign of $a^3$ in the denominator.
In the partial wetting regime, the two trajectories, together with the straight line $y_{\alpha\beta}=0$ representing the free $\alpha\beta$ interface, form a tricuspid in the density plane, with internal angles $\widetilde\alpha$, $\widetilde\beta$ and $\widetilde w$ -- see Fig.\ \ref{Fig26}. Cyclically applying the rule (\ref{tanswall}) for the derivatives at the ends of the paths, we observe that these angles always obey
\begin{equation}
    \widetilde\alpha+\widetilde\beta+\frac{\widetilde w}{2}=\frac{\pi}{2}
    \label{tribin}
\end{equation}
similar to the KWI model, but with $\widetilde w/2$ replacing $\widetilde\gamma$. 
In the limit when the initial coordinates $x_0\to\infty$ and $y_0\to\infty$, the tricuspid becomes infinitely long and thin so that $\widetilde w\to 0$, and trajectories fall along asymptotes that make an angle $\phi$ to the $x$ axis (say), i.e.\ $x_0\to\infty,y_0\to\infty$ with $y_0/x_0=\tan\phi$. In this case, the tricuspid angles obey the same geometrical constraint as for the KWI model, 
\begin{equation}\widetilde\alpha+\widetilde\beta=\frac{\pi}{2}
\end{equation}
when the temperature and pressure are asymptotically close to the line of critical end points -- see Fig.\ \ref{Fig26}b. For such trajectories, that fall from infinity, the constants $c_{w\alpha}$ and $c_{w\beta}$ take the same value $c$ which depends only on $\phi$. Substituting for $y_0=x_0\tan\phi$ into (\ref{calphabeta}), and taking the limit $x_0\to\infty$, gives $c=\tan\phi\,(\tan^2\phi-3)/(3\tan^2\phi-1)$,
which simplifies to
\begin{equation}
c=\tan(3\phi-\pi)
\end{equation}
We will soon show that this simplifies even further in the partial wetting regime.

\begin{figure}[t]
\includegraphics[height=5.cm]{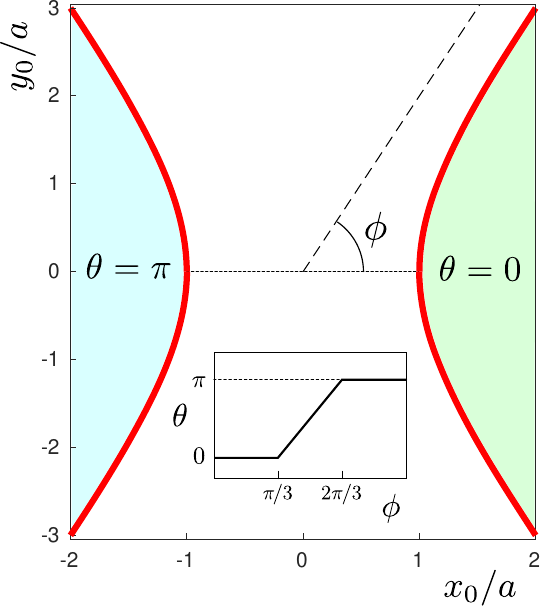}
\caption{\label{Fig27} Surface phase diagram at bulk coexistence ($T<T_c$) for a binary mixture with local XY symmetry, at a wall with fixed boundary conditions. The wetting phase boundaries lie along $y_0^2= 3(x_0^2-a^2)$, separating regions of complete wetting ($\theta=0$) and drying ($\theta=\pi$). The white area between them corresponds to a non-wetting gap that persists up to $T_c$. Inset: The contact angle as a function of the median angle $\phi$ characterizing the asymptotic direction of motion for trajectories that fall from infinity.}
\end{figure}

Returning to the general case, we note that the trajectories are unique so that any wetting transition must be continuous. The same requirement on the local scattering angle applies so that $\tilde \beta=\pi/2$ at wetting of the wall-$\alpha$ interface by $\beta$, and $\widetilde\alpha=\pi/2$ at wetting by $\alpha$ at the wall-$\beta$ interface. At these phase boundaries, the tricuspid disappears so that, for example, the wall-$\alpha$ trajectory decomposes into those for the wall-$\beta$ and $\alpha\beta$ paths, which meet at a right angle. The requirement that the path passes through both bulk vertices implies that at the phase boundary the constant $c=0$. In this case, the algebraic curve $u(x,y)=0$ is reducible and simplifies to the straight line $y=0$, representing the $\alpha\beta$ path, and the explicit equation
\begin{equation}
y=\pm\sqrt{3(x^2-a^2)}
\end{equation}
representing the path of the wall-$\alpha$ and wall-$\beta$ interfaces for $|x|>a$. Since the path at wetting includes the initial condition, $(x_0,y_0)$, it follows that the wetting phase boundary is the hyperbola
\begin{equation}
    y_0^2=3(x_0^2-a^2)
    \label{phaseboundary}
\end{equation}
which divides the density plane into regions of complete wetting by $\alpha$ ($\theta=\pi$), complete wetting by $\beta$ ($\theta=0$) and partial wetting -- see Fig.\ \ref{Fig27}. Thus, $y_0=0$ corresponds to $x_0=\pm a$, recovering the wetting phase boundary of the one-component theory. On the other hand, as $x_0\to\pm\infty$, the phase boundary has straight line asymptotes $y_0=\pm\sqrt{3}x_0$, which make an angle of $\pi/3$ with the horizontal -- the scattering angle $\Delta\theta_{scat}(2)$ for a quartic potential. In the mechanical analogy, the wetting phase boundary is the separatrix between the basins of attraction for the two fixed points at $(\pm a,0)$

We can already see that a non-wetting gap must occur in this model of a binary mixture at a wall, since, as $T\to T_c$, the wetting phase boundaries tend to the same straight lines $y_0=\pm\sqrt{3}\,x_0$. This leaves the region, $2\pi/3>\phi>\pi/3$, where the limiting value of the contact angle must lie in the range $\pi>\theta>0$. The identification of a phase boundary at $\phi=\pi/3$ is then identical to the condition, $y'_\beta(-1)=\sqrt{3}$, we used to determine the intercept of the lines of wetting with the CEP which gives $\tilde t_-$ in the KWI model. The second phase boundary at $\phi=2\pi/3$, similarly determines the value of $\tilde t_+$. This connection will be made more explicit when we determine the contact angle.

\subsection{The surface tension, contact angle and non-wetting gap}

The same expression for the trajectory follows from the complex analysis since the potential $\omega(x,y)$ belongs to the class of models with local XY symmetry, with $w(z)=(z^2-a^2)$. The trajectory function is therefore found from integration of 
\begin{equation}
f'(z)=\sqrt{2}\,e^{i\psi}\big(z^2 -a^2\big)
\end{equation}
giving
\begin{equation}
f(z)=\sqrt{2}\,e^{i\psi}\left(\frac{z^3}{3} -a^2z\right)
\end{equation}
where we can identify the gauge consistently from $c=\cot\psi$. The trajectory of the wall-$\alpha$ interface (\ref{trajwallalpha}) then follows directly from $\Re(f(z))=\Re(f(-a))$. The surface tension of the wall-$\alpha$ interface, given  by eq.\ (\ref{STcomplex}), is
\begin{equation}
\sigma_{w\alpha}=\int_0 ^\infty \!\!\!dt\; \dot z \dot{\bar{z}}
\end{equation}
and substitution of the first-order dynamical equation, $\dot{\bar{z}}=if'(z)$, gives
\begin{equation}
\sigma_{w\alpha}=|f(z_0)-f(-a)|
\end{equation}
where $z_0=x_0+iy_0$. More explicitly, the surface tensions are given by 
\begin{equation}
     \sigma_{w\alpha}=\sqrt{2}\,\left|\,\frac{z_0^3}{3}-a^2z_0-\frac{2a^3}{3}\right|, \hspace{1cm} \sigma_{w\beta}=\sqrt{2}\,\left|\,\frac{z_0^3}{3}-a^2 z_0+\frac{2a^3}{3}\right|
     \label{STbin}
 \end{equation}
 which are the natural generalization of the results of the one-component theory (\ref{STone}). Using Antonov's rule, $\sigma_{w\alpha}=\sigma_{w\beta}+\sigma_{\alpha\beta}$, for complete wetting by $\beta$, it is straightforward to recover the wetting phase boundary (\ref{phaseboundary}), derived from the trajectories, and also to show that the wetting transitions are always second-order. We will do this more elegantly below. From the surface tensions, we can determine the contact angle from Young's equation
 \begin{equation}    
     \sigma_{w\alpha}=\sigma_{w\beta}+\sigma_{\alpha\beta}\cos\theta
     \end{equation}
 where it is apparent that $\theta$ depends on two scaling variables, so that 
 \begin{equation}
     \theta=\theta\big(y_0/x_0,x_0/a\big)
 \end{equation}
 Thus, on approaching the critical point, $a\to 0^+$, we obtain the limiting value of the contact angle $\theta$, which depends only on $y_0/x_0$. In this case, the wetting phase boundaries tend to the straight lines $y_0=\pm\sqrt{3}x_0$ leaving a wetting gap where partial wetting has persisted up to the critical point. The contact angle within this gap satisfies
 \begin{equation}
     \sec\theta=\pm\sqrt{1+\frac{y_0^2(y_0^2-3x_0^2)^2}{x_0^2(x_0^2-3y_0^2)^2}}
     \label{thetawall1}
 \end{equation}
where the $\pm$ is simply the sign of $x_0$. This result makes it clear that critical point wetting only occurs when we set $y_0=0$. In this case, which is equivalent to the one-component theory, a wetting transition by $\alpha$ or $\beta$ must occur as $a\to 0$, unless $x_0=0$, corresponding to the line of wetting neutrality where $\theta=\pi/2$. However, when $y_0\ne 0$, the surface phase diagram displays a non-wetting gap, the same as the KWI model. 

 Exactly the same scaling expression for $\theta$ applies if we keep the value of $a$ fixed, that is $T<T_c$, and consider that the starting point of the trajectory is infinitely far away. As discussed earlier, in this case the tricuspid becomes infinitely long and thin so that the tricuspid angle $\widetilde w=0$ and emulates, precisely, the scaling behaviour found near the CEP lines in the KWI model. Substituting $y_0=x_0\tan\phi$ into (\ref{thetawall1}), gives $\tan\theta=\tan\phi \tan(\phi-\pi/3)\tan(\phi+\pi/3)$, reducing to
 \begin{equation}
   \theta=3\phi-\pi
 \end{equation}
 so that
\begin{equation}
 c=\tan\theta
\end{equation}
 in the partial wetting regime $\frac{2\pi}{3}\ge\phi\ge\frac{\pi}{3}$. Thus, the contact angle varies linearly between $0$ and $\pi$ as the tilt angle $\phi$ sweeps between the values at wetting by $\alpha$ and $\beta$, which are determined by the scattering angle $\Delta\theta_{scat}(2)=\pi/3$. We note that the linearity near the wetting phase boundaries at $\phi=\pi/3$ and $\phi=2\pi/3$, proves that the wetting transitions are second-order.
 
\begin{figure}[h]
\includegraphics[height=7cm]{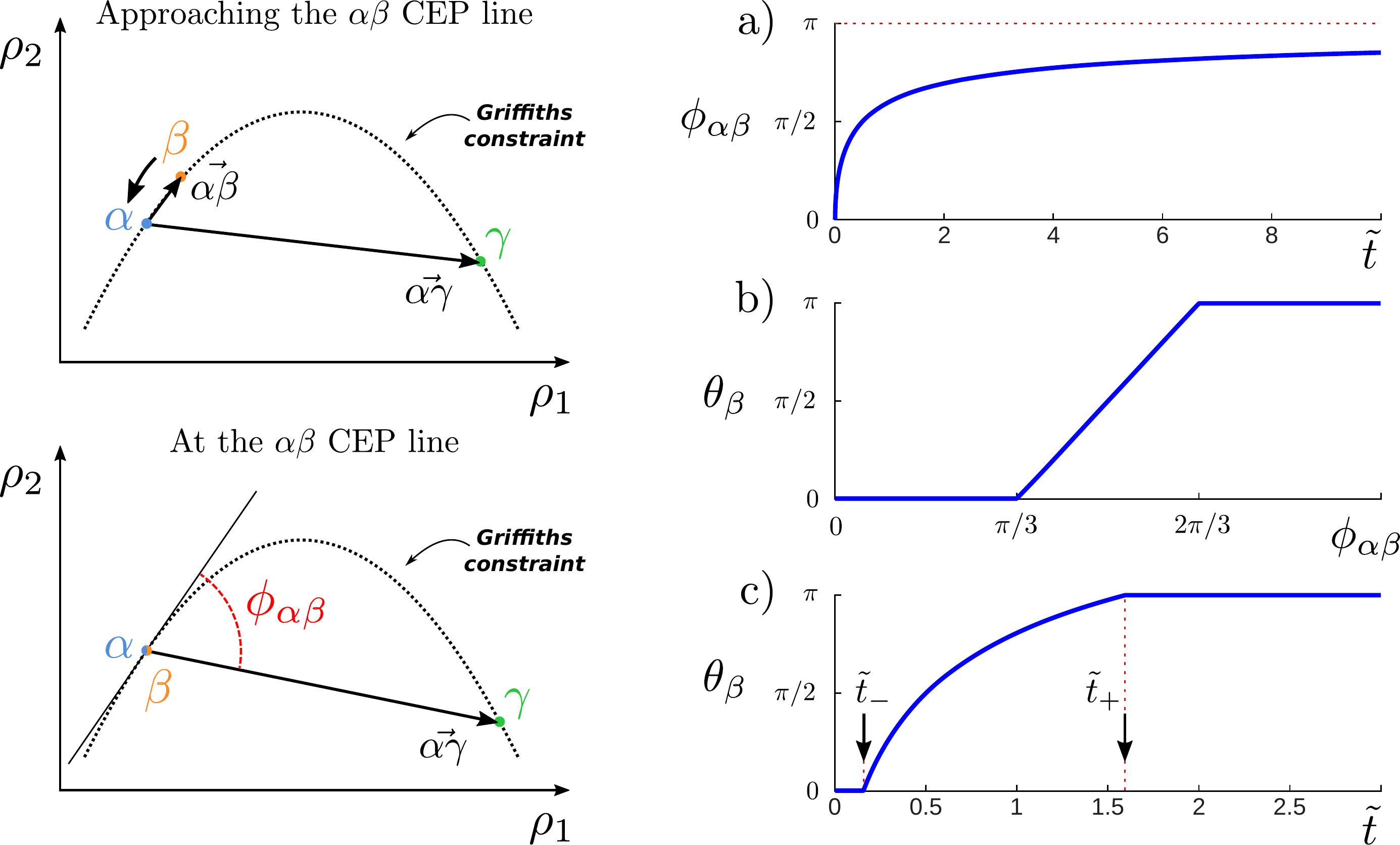}
\caption{\label{Fig28} LHS upper panel: The location of the component densities of the three bulk phases, following the Griffiths constraint $\rho_2=-(\rho_1)^2$, on approaching a point on the $\alpha\beta$ CEP (corresponding to the merging of $\alpha$ and $\beta$), and the vectors $\vec{\alpha\beta}$ and $\vec{\alpha\gamma}$. LHS lower panel: The median angle $\phi_{\alpha\beta}$ defined between the two vectors at this point on the CEP. RHS: a) The value of $\phi_{\alpha\beta}$ as a function of the scaling temperature-like field $\tilde t$ along the CEP, b) The contact angle $\theta_\beta$ as a function of $\phi_{\alpha\beta}$, c) The contact angle $\theta_\beta$ as a function of $\tilde t$ along the CEP showing the non-wetting gap between $\tilde t_-$ and $\tilde t_+$, where $\phi_{\alpha\beta}=\pi/3$ and $\phi_{\alpha\beta}=2\pi/3$, respectively. }
\end{figure}

For the KWI model, the value of $\phi$ we must use along the $\alpha\beta$ CEP is $\phi_{\alpha\beta}$. This measures the limiting value of the angle between the vector $\vec{\alpha\beta}$, from $\alpha$ to $\beta$, with $\vec{\alpha\gamma}$ between $\alpha$ and $\gamma$ in the 2D density plane -- see Fig.\ \ref{Fig28}. These are determined by the Griffiths values of the bulk densities which, recall, satisfy $\rho_1^\alpha+\rho_1^\beta+\rho_1^\gamma=0$ and $\rho_2^\mu= - (\rho_1^\mu)^2$. It follows that $\vec{\alpha\beta}$, which is now just the direction of the tangent to the Griffiths constraint curve $\rho_2=-(\rho_1)^2$  at $\alpha$, satisfies $\vec{\alpha\beta}\propto(1,\rho_1^\gamma)$, while  $\vec{\alpha\gamma}\propto(1,\rho_1^\alpha)$. Their scalar product then determines, that, when expressed in terms of the temperature scaling variable $\tilde t$, of the bulk phase diagram,
\begin{equation}
\cos\phi_{\alpha\beta}=\frac{1-2\tilde t}{\sqrt{(1+\tilde t)(1+4\tilde t)}}
\label{phiCEP}
\end{equation}
 where we have substituted $\rho_1^\gamma=2\tilde t^{\,1/2}$ and $\rho_1^\alpha=\rho_1^\beta=-\tilde t^{\,1/2}$. Using this value in $\theta_\beta=3\phi_{\alpha\beta}-\pi$, recovers the expression (\ref{IKbeta}) for the contact angle $\theta_\beta$ in the non-wetting gap, conjectured by Indekeu and Koga -- see Fig.\ \ref{Fig28}. \\ 
 
 \subsection{Conformal map of the paths onto a droplet}

\begin{figure}[t]
\includegraphics[height=4.cm]{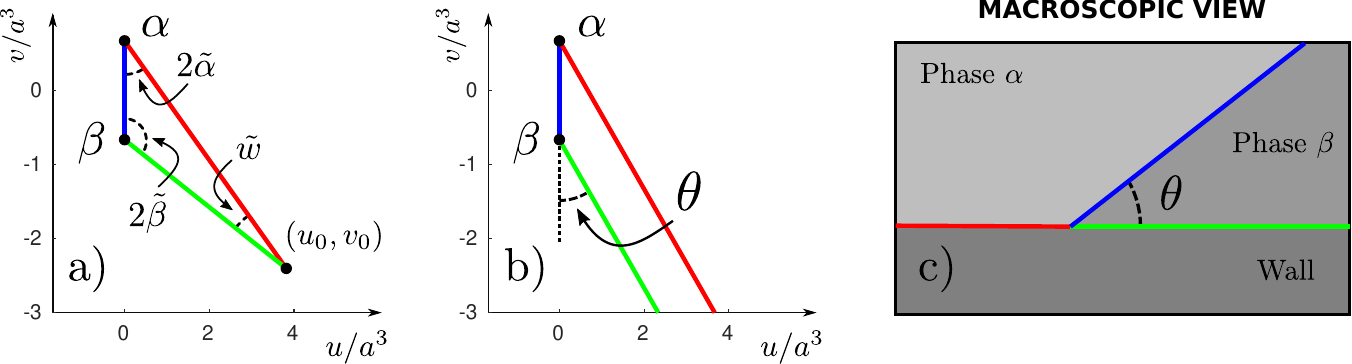}
\caption{\label{Fig29} a) The conformal map, $f(x+iy)=u+iv$, of the wall-$\alpha$, wall-$\beta$ and $\alpha\beta$ trajectories, as shown in Fig.\ \ref{Fig26}a, onto a triangle with interior angles $2\widetilde\alpha,\;2\widetilde\beta$ and $\widetilde w$ at $\alpha$, $\beta$ and $(u_0,v_0)$, respectively. b) For trajectories that fall from infinity, along the asymptote $y/x\to\tan\phi$ (as shown Fig. 26b) the tricuspid maps onto the sides of an infinite parallelogram, tilted at the contact angle, i.e.\ $2\widetilde\alpha=\theta$ with $\theta=3\phi-\pi$. c) The parallelogram outlines the shape of a macroscopic drop at a wall, with contact angle $\theta$.  }
\end{figure}

 We now use the function $f(z)$ to define new coordinates $(u,v)$ using $f(x+iy)=u+iv$. This maps the algebraic curve of the wall-$\alpha$ interface onto a section of the vertical straight line $u=0$ , which is of length $\sigma_{w\alpha}$. Since the mapping function $f(z)$ is the same for the wall-$\beta$ and $\alpha\beta$ paths, up to a rotation, these are also mapped onto straight lines, whose lengths are $\sigma_{w\beta}$ and $\sigma_{\alpha\beta}$, respectively. The tricuspid is therefore mapped onto a triangle whose lengths are the three surface tensions -- see Fig.\ \ref{Fig29}a, which we have oriented so that the $\alpha\beta$ path is vertical with the initial position $(x_0,y_0)$ mapping to $(u_0,v_0)$. There is no analogue of the Neumann triangle for the present wall-fluid system, since there is only one contact angle $\theta$, but nevertheless the angles of the triangle are uniquely defined by the values of $\sigma_{w\alpha}$, $\sigma_{w\beta}$ and $\sigma_{\alpha\beta}$, as specified in (\ref{STbin}). The conformal map preserves the value of $\widetilde w$, since $f'(z_0)\ne 0$, but, by de Moivre's theorem, doubles the angles subtended at $\alpha$ and $\beta$, since $f'(z_\mu)=0$ while $f''(z_\mu)\ne 0$, at these bulk vertices, with $\mu=\alpha,\beta$ . The angles of the triangle are, therefore, $2\widetilde\alpha$, $2\widetilde\beta$ and $\widetilde w$, that sum to $\pi$ consistent with the rule for the tricuspid $\widetilde\alpha+\widetilde\beta+\widetilde w/2=\pi/2$. Using Young's equation, it follows that the macroscopic contact angle $\theta$ is, in general, related to the tricuspid angles via
 \begin{equation}
     \cos\theta=\frac{\sin(\widetilde\beta-\widetilde\alpha)}{\sin{(\widetilde\beta+\widetilde\alpha)}}
     \label{cosrulebib}
\end{equation}
Therefore, when the paths start from a point $(0,y_0)$ on the line of symmetry, for which the tricuspid angles are equal by symmetry ($\widetilde\alpha=\tilde \beta$), the contact angle is $\theta=\pi/2$, corresponding to perfect wetting neutrality. The wetting transitions correspond to the limiting cases where the triangle collapses to a line, reproducing Antonov's rule. Then, $\theta=\pi$ (wetting by $\alpha$) is equivalent to $\widetilde\alpha=\pi/2$ and $\tilde \beta=\widetilde w= 0$, while $\theta=0$ (wetting by $\beta$) is equivalent to $\widetilde\beta=\pi/2$ and $\widetilde\alpha=\widetilde w=0$.

 Again, extra simplicity arises from the scaling limit if we suppose  that the starting point of the trajectory is infinitely far away, $x_0\to \infty$, $y_0\to\infty$ with $y_0/x_0=\tan\phi$ fixed, so that $\widetilde w=0$. In this case, the relation with the contact angle, (\ref{cosrulebib}),  simplifies to 
 \begin{equation}
     \theta=2\widetilde\alpha\hspace{1cm}
     \label{dropmap}
 \end{equation}
 or, equivalently, $\theta=\pi-2\widetilde\beta$. The remarkable implication of the relation (\ref{dropmap}) is that the conformal mapping of the tricuspid shape straightens the profile paths and outlines the shape of a macroscopic drop on a wall characterized by the contact angle -- see Fig.\ \ref{Fig29}b,c. There is, therefore, a direct conformal mapping between microscopic and macroscopic interfacial properties and emulates precisely the behaviour of the KWI model near the CEP, whose mapping was shown in Fig.\ \ref{Fig20}. 

 \subsection{The exact density profiles}

A final simplicity of the binary model is that the actual component density profiles themselves, $x(t)$ and $y(t)$, can be determined exactly. We start again from the first-order dynamical equation
\begin{equation}
\dot {\bar z}=if'(z)
\end{equation}
and its complex conjugate, which solve the Euler-Lagrange equations. In component form, these are the coupled equations
 of motion
\begin{equation}
    \frac{\dot x}{\sqrt{2}}=\frac{1}{\sqrt{1+c^2}}\,\big(x^2-y^2-a^2+2cxy\big)
    \label{xdotbin}
\end{equation}
and
\begin{equation}
\frac{\dot y}{\sqrt{2}}=\frac{c}{\sqrt{1+c^2}}\,\left(x^2-y^2-a^2-\frac{2}{c}\,xy\right)
\label{ydotbin}
\end{equation}
These maybe solved for both the wall-$\alpha$ and wall-$\beta$ interfaces for the corresponding values of $c$, eq.\ (\ref{calphabeta}). We illustrate this first for the wetting phase boundary and for a line of neutrality, which are useful checks on the general solution.

\subsubsection*{At the wetting phase boundary, $\theta=0$.}

At the wetting transition, the trajectory of the wall-$\alpha$ interface decomposes into those of the wall-$\beta$ and free $\alpha\beta$ interfaces. The $\alpha\beta$ interface has the standard solution, $x(t)=a\tanh\kappa(t_\beta-t)/2$ and $y(t)=0$, of the one-component theory, where the interface location $t_\beta$ is infinitely far from the wall, and recall $\kappa=2\sqrt{2}\,a$. The first-order dynamical equations describing the wall-$\beta$ interface simplify, since $c=0$, and substituting for the trajectory $y=\sqrt{3(x^2-a^2)}$, we obtain
$\dot x=2\sqrt{2} (a^2-x^2)$. Integrating then determines the solution
\begin{equation}
    x(t)=a\coth \kappa(t+t_w),\hspace{1cm}y(t)=\frac{\sqrt{3}a} {\sinh \kappa(t+t_w)}
    \label{xwet}
\end{equation}
where the constant $t_w$ determines the boundary value at the wall, $x_0=a\coth\kappa t_w$. For particles that fall from infinity, these simplify even further since $t_w=0$ -- see Fig.\ \ref{Fig30}. We note that at, this wetting phase boundary, the $x$ component decay towards its bulk $\alpha$ value, as $x(t)-a\propto e^{-2\kappa t}$. This  higher-order exponential decay is consistent with the general mean-field expectation that, at a continuous wetting transition, the local density profile flattens \cite{Dietrich1991,Parry2023b}.

\subsubsection*{ Along the line of neutrality, $\theta=\pi/2$}

\begin{figure}[h]
\includegraphics[height=5.cm]{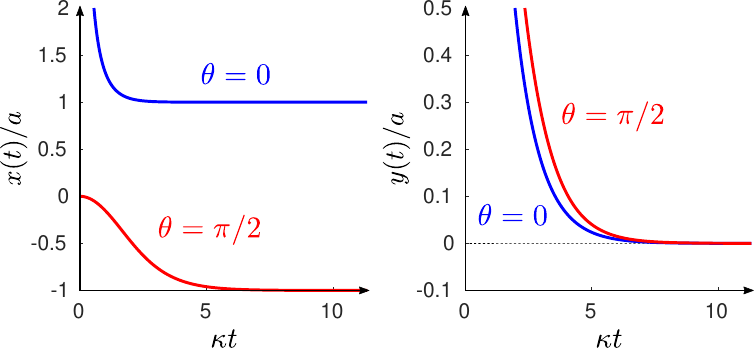}
\caption{\label{Fig30} The primary component profile, $x(t)$, and secondary component profile, $y(t)$, for  the wall-$\alpha$ interface at wetting ($\theta=0$), and at wetting neutrality ($\theta=\pi/2$), illustrated here for paths that fall from infinity. These paths represent the scaling limit in the critical region, $a\propto \sqrt{T-T_c}\to 0$, with $x/a$ and $\kappa t$, fixed. }
\end{figure}

A second simplification occurs if we take the limit $c\to \infty$, in which case the cubic algebraic curve reduces to
\begin{equation}
    y^2=\frac{(x+1)^2(x-2)}{3x}
\end{equation}
where we have set $a=1$ since this can be scaled out. This describes a trajectory of the wall-$\alpha$ interface starting from $x_0=0$ and $y_0=\infty$, i.e.\ from a point infinitely distant on the line of neutrality, so that by symmetry $\theta=\pi/2$. Notice that the trajectory ends at the bulk vertex with slope $y'(-1)=1$, identifying $\widetilde\alpha=\pi/4$, consistent with the conformal mapping result $\theta=2\widetilde\alpha=\pi/2$. The dynamical equations also simplify in this limit. Substituting  $y(x)$ into the equation for $\dot x$ integrates to gives
\begin{equation}
x(t)=a\;\frac{2\;\mathcal{T}(t)^2}{\mathcal{T}(t)^2-3},\hspace{1cm} y(t)=a\;\frac{3(1-\mathcal{T}(t)^2)}{\mathcal{T}(t)(3-\mathcal{T}(t)^2)}
\label{xbin}
\end{equation}
where
\begin{equation}
    \mathcal{T}(t)=\tanh\frac{\kappa t}{2}
\end{equation}
and we have reinstated the factor of $a$ and the inverse bulk correlation length $\kappa=2\sqrt{2}\,a$. Both components now decay towards their bulk values as $x(t)-a\propto e^{-\kappa t}$ and $y(t)\propto e^{-\kappa t}$, and do not exhibit the flattening which occurs at the wetting phase boundary -- see Fig \ref{Fig30}. \\

\subsection{The general profile from the algebraic curve}

To obtain the solutions $x(t)$ and $y(t)$ for the general trajectory, with an arbitrary value of $c$, we now use the theory of algebraic curves. In general, the profile path $u(x,y)=0$ is a cubic (i.e.\ degree $d=3$) with one double point singularity, where $u_x=u_y=0$, at the bulk vertex $\alpha$ (i.e $\delta_{tot}$ =1). The degree-genus theorem, eq.\ (\ref{genus}), identifies the genus as $g=0$, implying that a rational parameterization is available even for partial wetting. To find this, we shift the coordinate, defining $x'=x+1$, so that the $\alpha$ vertex lies at the origin (and we have set $a=1$). With this shift, the equation for the curve becomes
\begin{equation}\label{180}
    x'^2y-2x'y-\frac{y^3}{3}+c\Big(x'y^2-y^2+x'^2-\frac{x'^3}{3}\Big)=0
\end{equation}
and allows us to find the rational parameterization by considering lines through the origin, $x'=x'(\tau)$ and $y=\tau x'(\tau)$. Substitution of these expressions in (\ref{180}) identifies the function $x'(\tau)$ and yields
\begin{equation}
    x(\tau)=\frac{3\tau+\tau^3-2c}{3\tau-\tau^3+c(3\tau^2-1)}
    \label{ratbin}
\end{equation}
with $y(\tau)=\tau(1+x(\tau))$, which we express here shifting back to the original coordinates. The parameter $\tau$ decreases from an initial value $\tau_0$, determined by $(x_0,y_0)$, to the limiting value $\tau_\alpha=(\sqrt{1+c^2}-1)/c$ in the bulk $\alpha$ phase.
Substituting $x(\tau)$ and $y(\tau)$ into the equations for $\dot x$ and $\dot y$ determines the dynamical equation for the rational parameter itself
\begin{equation}
    \frac{\dot\tau}{\sqrt{2}}=-\frac{c(1-\tau^2)-2\tau}{\sqrt{1+c^2}}
    \label{taudotbin}
\end{equation}
This consistently identifies the value of $\tau_\alpha$ at the bulk $\alpha$ vertex, where $\dot\tau=0$. This first-order equation has the solution 
\begin{equation}
\tau(t)=\frac{1}{c}\left(\sqrt{1+c^2}\coth\frac{\kappa (t+t_0)}{2}-1\right)
\label{taubin}
\end{equation}
where $t_0$ is determined by the boundary condition at the wall, and we have assumed $x(0)>0$ so that the contact angle is in the range $0<\theta<\pi/2$. Substitution of $\tau(t)$ into $x(\tau)$ and $y(\tau)$ then determines the exact coordinates $x(t)$ and $y(t)$. Here, we focus on the coordinate $x(t)$ which reveals all the wetting behaviour. The result is expressed most clearly if we take the scaling limit in which the particle falls from infinity along the asymptotic line, $y_0/x_0\to\tan\phi$, characterized by the median angle $\phi$ for which $\theta=3\phi-\pi$ and $
 c= \tan \theta$. The trajectory starts infinitely far away, at $\tau_0=\tan\phi$, and finishes at $(-a,0)$, for which $\tau_\alpha=\tan(\theta/2)$. The condition $\tau_0=\tan\phi$ determines that the constant $t_0$ must satisfy
 \begin{equation}
     \coth\frac{\kappa t_0}{2}=-\frac{\cos2\phi}{\cos\phi}
     \label{cotht0}
 \end{equation}
 Therefore, $t_0$ increases from $0$ at $\phi=\pi/2$ (wetting neutrality) and diverges on approaching the wetting phase boundary as
 \begin{equation}
 t_0=\xi\ln\frac{2}{\sqrt{3}\;\theta}+\cdots
     \end{equation}
 where the higher-order terms vanish as $\theta\to 0$. The physical meaning of this length will become apparent shortly. The coordinate $x(t)$ is then exactly given by
\begin{equation}
\frac{x_{w\alpha}(t)}{a}=\frac{(\cos^2\theta-2)\;\mathcal{T_+}(t)^3 +3\cos\theta \;\mathcal{T_+}(t)^2-3\cos^2\theta \;\mathcal{T_+}(t)+\cos\theta}{(2\cos^2\theta-1) \;\mathcal{T_+}(t)^3-3\cos\theta \;\mathcal{T_+}(t)^2+3\mathcal{T_+}(t)-\cos\theta}
\label{binxwalpha}
\end{equation}
where
\begin{equation}
    \mathcal{T_+}(t)=\tanh \frac{\kappa(t+t_0)}{2}
\end{equation}
satisfying the required boundary conditions at the wall, $x_{w\alpha}(0)=\infty$, and in the bulk, $x_{w\alpha}(\infty)=-a$. Setting $\theta=\pi/2$ reproduces our earlier result (\ref{xbin}) for wetting neutrality.

\begin{figure}[h]
\includegraphics[height=5.cm]{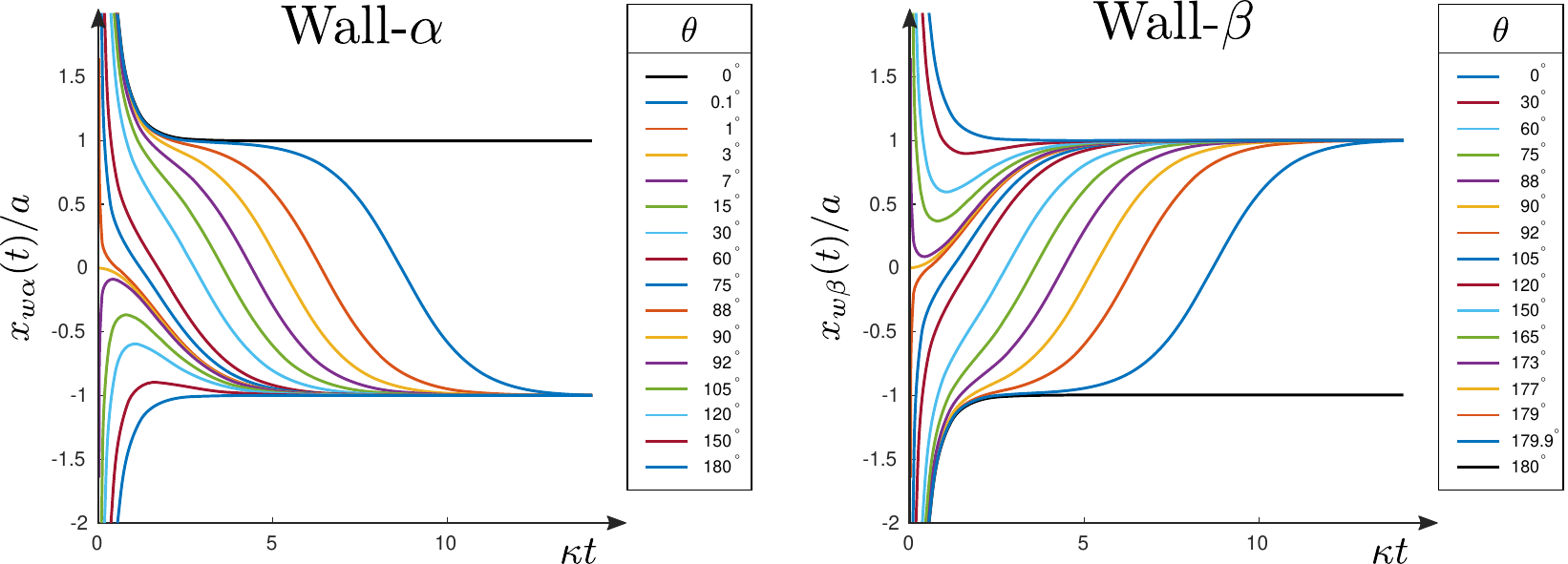}
\caption{\label{Fig31} Density component profiles, $x_{w\alpha}(t)$ and $x_{w\beta}(t)$, for the wall-$\alpha$ and wall-$\beta$ interfaces, for different contact angles $\theta=3\phi-\pi$. The divergence of the wetting film thickness for the wall-$\alpha$ interface as $\theta\to 0$ coincides with a qualitative change from a non-monotonic to a monotonic decay of the profile for the wall-$\beta$ interface. The profiles show the expected component symmetry $x_{w\beta}(t,\theta)=-x_{w\alpha}(t,\pi-\theta)$. }
\end{figure}

 A similar calculation determines the density profile for the wall-$\beta$ interface. In this case, the rational parameterization of the primary profile $x=x(\tau)$ is given by 
\begin{equation}
    x(\tau)=-\frac{3\tau+\tau^3-2c}{3\tau-\tau^3+c(3\tau^2-1)}
    \label{ratbin2}
\end{equation}
which is the negative of the expression (\ref{ratbin}) for the wall-$\alpha$ interface. This is also the case for the secondary profile, since $y(\tau)=\tau(x(\tau)-1)$. The value of $\tau$ now varies from $\tau_0$ at the wall, which for particles that fall from infinity is again $\tau_0=\tan\phi$, to $\tau_\beta=-(\sqrt{1+c^2}+c)/c$ in the bulk. This means that the solution for $\tau(t)$ is simply obtained from (\ref{taubin}) replacing $t$ with $-t$, so that 
\begin{equation}
\tau(t)=-\frac{\sqrt{1+c^2}\coth\frac{\kappa (t-t_0)}{2}+1}{c}
\label{taubin2}
\end{equation}
where $t_0$ is unchanged from (\ref{cotht0}). Thus, for particles that fall from infinity, the exact expression for $x(t)$ for the wall-$\beta$ interface is given by 
\begin{equation}
\frac{x_{w\beta}(t)}{a}=\frac{(2-\cos^2\theta)\;\mathcal{T_-}(t)^3 +3\cos\theta \;\mathcal{T_-}(t)^2+3\cos^2\theta \;\mathcal{T_-}(t)+\cos\theta}{(2\cos^2\theta-1) \;\mathcal{T_-}(t)^3+3\cos\theta \;\mathcal{T_-}(t)^2+3\mathcal{T_-}(t)+\cos\theta}
\label{xexactwbeta}
\end{equation}
where 
\begin{equation}
    \mathcal{T_-}(t)=\tanh \frac{\kappa(t-t_0)}{2}
\end{equation}
This solution satisfies the boundary conditions at the wall, and in the bulk, $x_{w\beta}(\infty)=a$, and is the negative of the result for wall-$\alpha$ interface when $\theta=\pi/2$, as required by symmetry. Both component density profiles, $x_{w\alpha}(t)$ and $x_{w\beta}(t)$, are shown in Fig.\ \ref{Fig31}, over the range of contact angles $\pi\ge\theta\ge 0$. From these exact results, we can now determine three properties of physical interest.

\subsection*{The growth of the wetting layer}

At the wall-$\alpha$ interface, a wetting layer of $\beta$ grows continuously as $\theta\to 0$ (i.e. $\phi\to \pi/3$), while at the wall-$\beta$ interface, a wetting layer of $\alpha$ grows as $\theta\to \pi$ (i.e. $\phi\to 2\pi/3$). We can determine the divergence of the wetting film thickness analytically, illustrated here for wetting by $\beta$ at the wall-$\alpha$ interface. We identify the wetting layer thickness $t_\beta$ as the distance from the wall at which $x=0$. In the mechanical analogy, this is the time taken to fall from infinity and cross the $x$ axis. From (\ref{ratbin}), it follows that this happens when the rational parameter satisfies 
$\tau^3+3\tau-2c=0$, which for small $c$ has solution $\tau\approx 2c/3$. Then, from (\ref{taubin}) and (\ref{cotht0}), it follows that as $\phi\to \pi/3$, the film thickness diverges as
\begin{equation}
     t_\beta=-\xi\ln(3\phi-\pi)+\cdots
\end{equation}
where $\xi=1/\kappa$ is the bulk correlation length. The logarithmic divergence is the expected mean-field result for systems with short-ranged forces. This also describes the wetting layer thickness in the KWI model, in the asymptotic critical region close to the $\alpha\beta$ CEP line, as $\tilde t$ is reduced to $\tilde t_-$. The coordinates $x$ and $y$ are then just a linear combination of the primary and secondary densities -- as in equations (\ref{rot1}) and (\ref{rot2}), but with the index $\gamma$ replaced with $\beta$.\\

\begin{figure}[h]
\includegraphics[height=6.cm]{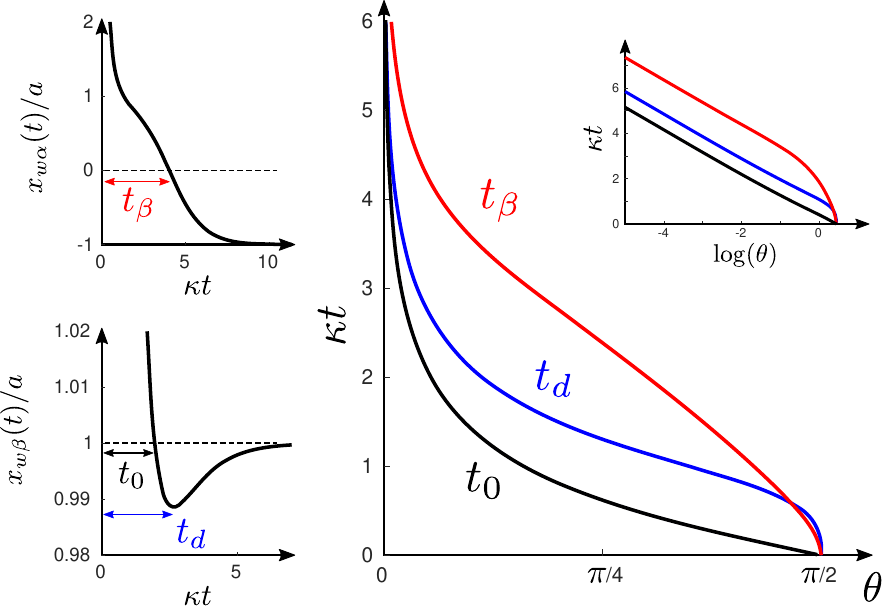}
\caption{\label{Fig32} Component density profiles, $x(t)$, for the (critical) wall-$\alpha$ interface (left upper) and the (non-critical) wall-$\beta$ interface (left lower) when $\theta=10^\circ$, illustrating the length scales, $t_\beta$, $t_d$ and $t_0$. Their dependence on the contact angle $\theta$ is shown right, together with their logarithmic divergence at wetting (inset). As $\theta\to 0$, we find $\kappa (t_\beta-t_d)=\ln 9/2$ while $\kappa( t_d- t_0)=\ln 2$.}
\end{figure}

\subsection*{A qualitative change in the wall-$\beta$ profile}

The expression for $x_{w\beta}(t)$, often referred to as the non-critical interface, since no wetting film is present, reveals a qualitative change in the density profile of the wall-$\beta$ interface as $\theta\to 0$. In the partial wetting regime, $\phi>\pi/3$, this profile has a non-monotonic decay. This is apparent since the condition $x_{w\beta}(t)=a$ has two solutions -- at $t=\infty$ and $t=t_0$, i.e.\ in the mechanical analogy, it takes time $t_0$ to cross the line $x=a$. Lying very close to this is a minimum at $t=t_d$, where $\dot x_{w\beta}(t_d)=0$. By contrast, in the completely wet regime, the profile exhibits no such minimum and has a simple monotonic decay to the bulk. Therefore, there is a qualitative change from non-monotonic to monotonic behaviour at the wetting transition -- a feature which has also been reported in models of continuous (critical) wetting transitions in systems with long-ranged forces \cite{Dietrich1991,Parry2023b}.

To quantify this more precisely, we note that, for partial wetting, the minimum corresponds to the point on the trajectory curve where $y'(x)=\infty$ and represents a region of slight local depletion. Since the minimum in $x_{w\beta}(t)$ lies further away from the wall than $t_0$, the region of depletion must move away from the wall as $\theta\to 0$. This is the reason why the profile at the transition is flattened, decaying as $e^{-2\kappa t}$. This change from non-monotonic to monotonic decay constitutes a non-thermodynamic singularity, and is not associated with critical behaviour in the surface tension $\sigma_{w\beta}$.  Fig.\ \ref{Fig32} shows representative density profiles for the wall-$\alpha$ and wall-$\beta$ interfaces, together with the growth of $t_\beta$, $t_0$ and $t_d$ approaching the wetting transition. It is straight forward to determine the divergence of $t_d$ analytically, similar to the determination of $t_\beta$. From the rational parameterization (\ref{ratbin2}), it follows that $\dot x=0$ also requires that $dx/d\tau=0$, which for small $c$ identifies that $\tau\approx-4/c$. Then, from 
(\ref{taubin2}) we have $\tanh \kappa(t_d-t_0)/2=1/3$ and hence $t_d-t_0=\xi\ln 2$ so that 
\begin{equation}
    t_d=-\xi\ln (\phi-\pi/3)+\cdots
\end{equation}
This diverges in the same manner as the wetting layer thickness $t_\beta$ and $t_0$. All of these lengths lie within a bulk correlation length of each other. 

\begin{figure}[h]
\includegraphics[height=6.cm]{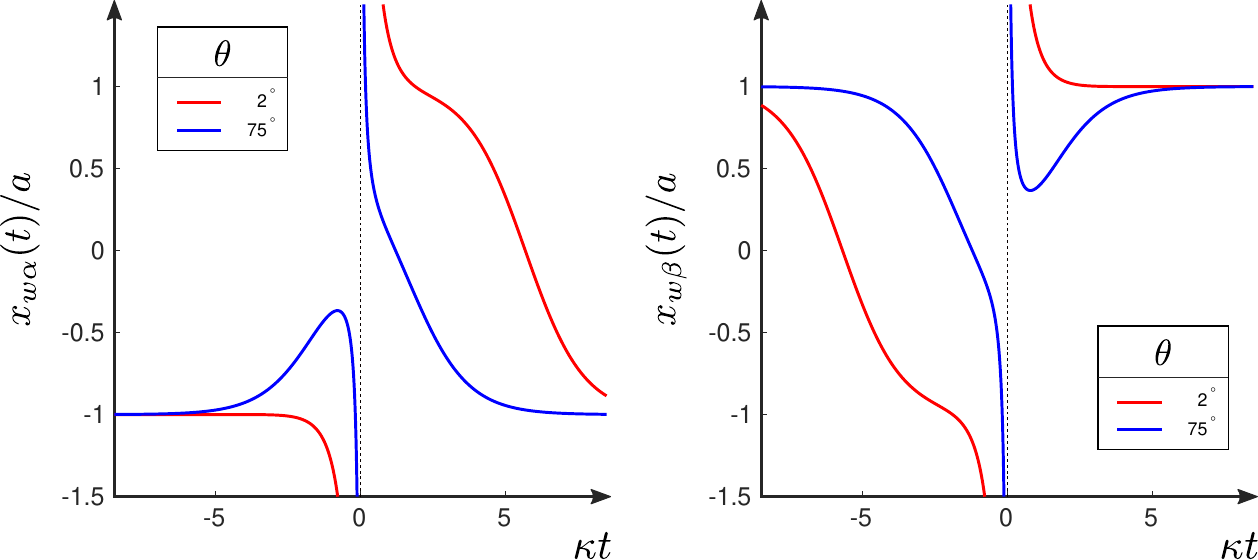}
\caption{\label{Fig33} The symmetries between the expressions for the component densities profiles, for the wall-$\alpha$ and wall-$\beta$  interfaces, when the coordinate $t$ is continued through the singularity at the wall, illustrated here for contact angles $\theta=2^\circ$ and $75^\circ$. 
Note that the wetting layers and regions of depletion are interchanged when $t$ is replaced with $-t$, following eq.\ (\ref{wbwa}). }
\end{figure}

 The density profiles of the wall-$\alpha$ and wall-$\beta$ interfaces are related by two symmetries. To appreciate this, we write them as $x_{w\alpha}(t,\theta)$ and $x_{w\beta}(t,\theta)$, including the dependence on the contact angle. The first symmetry is straightforward and reflects the equivalence between wetting by $\beta$ at the wall-$\alpha$ interface and wetting by $\alpha$ at the wall-$\beta$ interface, i.e.
\begin{equation}
x_{w\beta}(t,\theta)=-x_{w\alpha}(t,\pi-\theta)
\end{equation}
However, there is a second symmetry which can be seen by allowing $t$ to explore the whole range $-\infty<t<\infty$, i.e.\ when continued through the singularity at the wall. For example, using only the expression for $x_{w\alpha}(t,\theta)$ for the wall-$\alpha$ interface, the continuation into the region $t<0$ is the simply the negative of the profile representing the wall-$\beta$ interface, for the same contact angle $\theta$, so that
\begin{equation}\label{wbwa}
x_{w\beta}(t,\theta)=-x_{w\alpha}(-t,\theta)
\end{equation}
and similarly $x_{w\alpha}(t,\theta)=-x_{w\beta}(-t,\theta)$. This is illustrated in Fig.\ \ref{Fig33}, again showing the locations of the wetting layer and local depletion region for both interfaces.\\

\subsection*{The relation to the disjoining pressure} 

The qualitative change in the density profile of the non-critical interface bares a striking resemblance to that of the binding potential at a second-order wetting transition. Here, we show that these two different quantities, each defined for different interfaces, are intimately related and have the same decay from the wall.

The binding potential is the essential ingredient in mesoscopic interfacial Hamiltonian descriptions of wetting \cite{Forgacs1991}, and corresponds to the surface free-energy of a wetting film that is fixed to be a given thickness. Usually this constrained thickness is written as the variable $\ell$ but here we shall write it as $t$ to allow comparison with the density profile. We shall also use a subscript to emphasize that the potential is defined for the wall-$\alpha$ interface. Thus, $W_{w\alpha}(t)$ is a surface free-energy found by partially minimizing the  grand potential functional under a constraint that fixes the wetting film thickness to $t$ \cite{Fisher1991}. The derivative of (minus) the binding potential with respect to the film thickness is then the disjoining pressure, which vanishes at equilibrium. For systems with short-ranged forces, the binding potential describing second-order wetting transitions has the well known form \cite{Brezin1983,Fisher1991,Squarcini2022}
\begin{equation}
W_{w\alpha}(t)\;\approx\; a_0\,(T-T_w)\, e^{-\kappa t}+b_0\,e^{-2\kappa t}+\cdots
\end{equation}
where $T_w$ denotes the wetting temperature, $\kappa\equiv1/\xi$, and $a_0,b_0>0$ are positive constants. The minimum value of the binding potential lies at the mean-field wetting layer thickness, $W_{\alpha\beta}'(t_\beta)=0$, where $W_{w\alpha}(t_\beta)=\sigma_{\alpha\beta}(1-\cos\theta)$ is the singular contribution to the surface tension, which, close to the wetting transition, can be expanded as $\cos\theta\approx 1-\theta^2/2$. The curvature of the potential then determines the mean-field parallel correlation length, $\xi_\parallel=\sqrt{\sigma_{\alpha\beta}/W_{w\alpha}''(t_\beta)}$, describing the build up of capillary-wave fluctuations at the unbinding $\alpha\beta$ interface. This approach, very simply, recovers the standard mean-field critical singularities $t_\beta\approx -\xi\ln (T_w-T)$, $\sigma_{sing}=-a_o^2 (T_w-T)^2/4b_o$ and $\xi_\parallel\propto (T_w-T)^{-1}$, for short-ranged continuous (critical) wetting.

\begin{figure*}[t]
\includegraphics[height=5.cm]{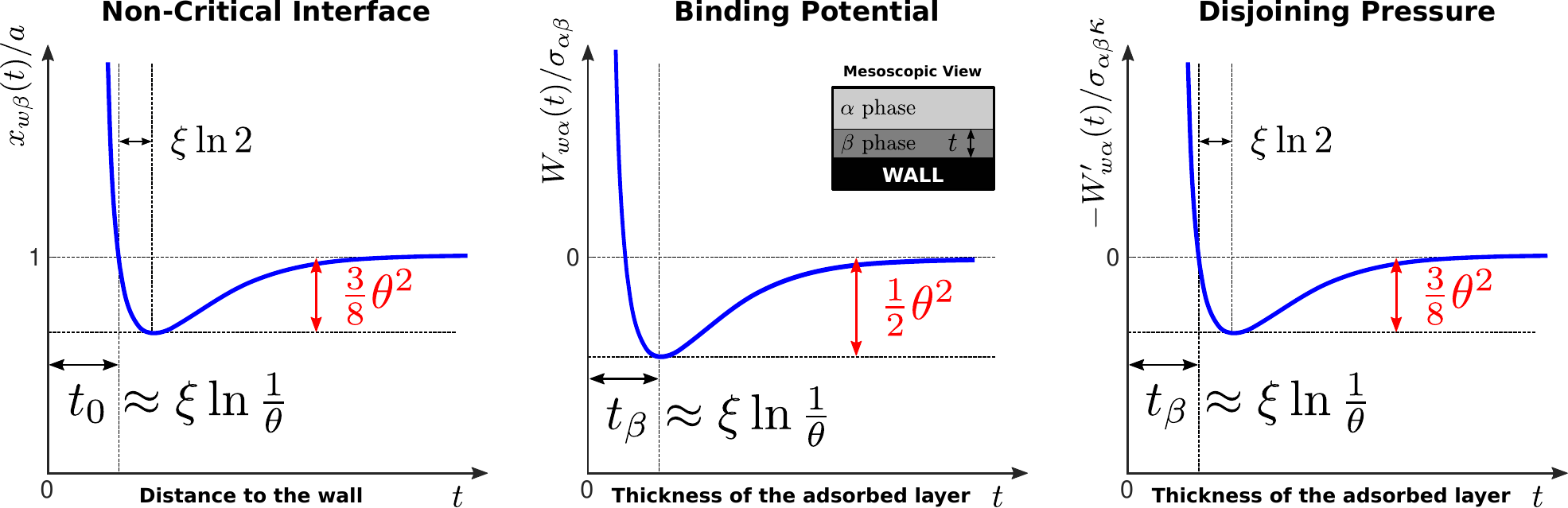}
\caption{\label{Fig34} 
The decays of the component profile of the non-critical wall-$\beta$ interface $x_{w\beta}(t)$, the binding potential $W_{w\alpha}(t)$ of the wall-$\beta$ interface (Inset: A wetting layer of constrained thickness $t$), and the disjoining pressure $-W_{w\alpha}'(t)$ of the wall-$\beta$ interface. \\
}
\end{figure*}

 The values of the constants $a_0$ and $b_0$ depend on the precise definition of the interface position . For example, we could use a crossing criteria which fixes the value of the $x$ component at the interface location, or alternatively fix the adsorption. The standard crossing criterion fixes the value of the component $x=0$ at the interface which is consistent with how we have identified $t_\beta$. These definitions of interface location give slightly different values of $a_0$, $b_0$ but do not affect the critical singularities.  It is possible to rewrite the binding potential, and its derivatives, in such a way that this dependence gets adsorbed into a length scale of order the microscopic bulk correlation length. This can be done by simply using $a_0(T_w-T)=\sqrt{2\sigma_{\alpha\beta}b_0}\;\theta$ to re-express  $t_\beta$ in terms of the contact angle. In this case, the derivative of the binding potential is given by 
\begin{equation}
\frac{ W_{w\alpha}'(t)}{\sigma_{\alpha\beta}\kappa}\;\approx\;\frac{3}{2}\,\theta^2\,e^{-\kappa ( t-t_\beta)}(1-e^{-\kappa( t-t_\beta)})+\cdots
\end{equation}
This way of re-writing the binding potential exposes the relation with the density profile $x_{w\beta}(t)$. Close to the wetting phase boundary, we again expand $\cos\theta\approx 1-\theta^2/2$, and working to quadratic order in the contact angle, the expression (\ref{xexactwbeta}) reduces to 
\begin{equation}
\frac{x_{w\beta}(t)}{a}\approx 1+3\theta^2\;\frac{\mathcal{T}_-(t)(\mathcal{T}_-(t)-1)}{(1+\mathcal{T}_-(t))^2}+\cdots
\end{equation}
valid provided $\theta/(\mathcal{T}_-+1)<<1$, equivalent to $\kappa t>>1$. Substituting for $\mathcal{T}_-=\tanh\kappa(t-t_0)/2$ then gives
\begin{equation}
\frac{x_{w\beta}(t)}{a}\approx 1-\frac{3}{2} \,\theta^2\,e^{-\kappa (t-t_0)}(1-e^{-\kappa (t-t_0)})+\cdots
\end{equation}
where no further expansion has been used. In this way, we can equate
\begin{equation}
\frac{x_{w\beta}(t)}{a}\approx1-\frac{W'_{w\alpha}(t-\Delta)}{\kappa\;\sigma_{\alpha\beta}}
\end{equation}
where $\Delta\equiv t_\beta-t_0 \approx \xi$ is a microscopic length that depends on what definition of the interface location is used, e.g.\ a crossing criterion or constrained adsorption, etc. This length remains microscopic because $t_\beta$ and $t_0$  exhibit precisely the same critical singularity at the wetting transition. We use the standard crossing criterion definition that $x_{w\alpha}(t_\beta)=0$, natural by symmetry, which gives $\Delta=\xi\ln9$. Such an "uncertainty" between microscopic and mesoscopic descriptions must be present since the interface location, and hence the  binding potential itself, is only meaningfully defined on length scales greater than the bulk correlation length. Thus,  apart from the near vicinity of the wall, $\kappa t\lesssim 1$, the decay of the microscopic density profile and the disjoining pressure, each defined for different interfaces,  are the same -- see Fig.\ \ref{Fig34}. This is the reason why there is a qualitative change in the shape of the non-critical profile at the wetting phase boundary, and why $t_0$, or equivalently $t_d$,
has the same divergence as the wetting film thickness $t_\beta$.\\

 \section{Summary and conclusions}

Firstly, great credit must be given to Koga and Widom for introducing a (deceptively simple) square-gradient model showing wetting transitions. And also for their numerical analysis of the profile paths and conjectures for the tensions, generalized by Koga and Indekeu, that suggested that an exact solution was possible. There are three parts to the exact solution:
\begin{itemize}
    \item[a)] The derivation of the density profile paths, $u(x,y)=0$, which are algebraic curves that outline a tricuspid in the density component plane.
\item[b)] The derivation of the Koga-Indekeu conjecture for the surface tension by conformal mapping the tricuspid of microscopic paths onto the Neumann triangle.
\item[c)] The determination of the component density profiles, $x(t),y(t)$, which satisfy first-order dynamical equations, which may be obtained analytically at the wetting phase boundary, and even more generally near the CEPs, using the rational parameterization of algebraic curves. 
\end{itemize}

We remark that the conformal mapping, $f(x+iy)=u+iv$ with $f'(z)=\sqrt{2}e^{i\psi} w(z)$, used to straighten the trajectories for models with a local XY symmetry, is a generalization of the mapping $(x+iy)=(u+iv)^2$, used by Levi-Civita to regularize orbital motion 
\cite{Arnold1989}. In this case the gravitational potential, i.e. the potential of an inverse-square law force, corresponds to $w(z)=z^{-1/2}$, and the trajectory function identifies the Levi-Civita square mapping. The method presented here generalizes this to a broader class of classical potentials, but restricted to the case when the energy $E=0$.  We remark, of course, that within the mechanical analogy there is no analogue of the Neumann triangle for the contact angles. Further details of this are provided in Appendix B.

To finish our article, we now focus on the broader physical implications stemming from the analytical solution of the KWI model, with the aim of understanding if critical point wetting, or a non-wetting gap, occurs for ordinary fluid mixtures.

 \subsection{Non-wetting gaps and XY surface criticality}
 
 The exact solution of the KWI model confirms that wetting transitions occur in the two-component description that are not present within the one-component theory. It also confirms that non-wetting gaps appear along the CEP lines  in the surface phase diagram. It is this, rather surprising, prediction that we focus on now. Our first remark is that the absence of critical point wetting is broader than in the KWI model. The same non-wetting gap also occurs in a model of a binary mixture at a wall with fixed surface boundary conditions, where the profile trajectories are identical to those of the KWI model close to the CEP. This means that the absence of critical point wetting {\it{does not}} arise, as had been speculated, because of a difference between wetting at walls and wetting involving fluid interfaces only. These mean-field predictions are unaltered, in three dimensions, when interfacial fluctuation are included, since these do not change the location of the (continuous) wetting phase boundaries \cite{Brezin1983,Squarcini2022}. 
 
 Here we argue that the reason why critical point wetting is suppressed is because, at $T_c$ (or along the CEPs), these models have an XY symmetry which means the surface criticality is different to that for fluids belonging to the bulk Ising universality class. This symmetry is immediately evident in the KWI model at the TCP, where the potential is radial so $\omega(\rho_1,\rho_2)=(\rho_1^2+\rho_2^2)^3$. The same local radial symmetry applies along the CEP lines, or equivalently for the model of a binary mixture at a wall where, at $T_c$, the potential is $\omega(x,y)=(x^2+y^2)^2$. This XY symmetry is also the origin of the wetting transition when a critical phase is adsorbed between two non-critical phases. Here, we show the deep connection between the presence or absence of critical point wetting and the universality class describing surface criticality. 
 
 Having established that critical point wetting is absent for the binary mixture with local XY symmetry, let us break the symmetry so that each component has a different bulk correlation length, by adding an anisotropic term, so that
\begin{equation}
\omega(x,y)=\big((x-a)^2+y^2\big)\big((x+a)^2+y^2\big)+\frac{\lambda y^2}{2}
\label{binlambda}
\end{equation}
with $\lambda>0$. For this potential, the bulk correlation lengths, given by $1/\xi_x^2=\omega_{xx}(-a,0)$ and $1/\xi_y^2=\omega_{yy}(-a,0)$, are
\begin{equation}
\xi_x=\frac{1}{2\sqrt{2}a}, \hspace{1cm }\xi_y=\frac{1}{\sqrt{8a^2+\lambda}}
\end{equation}
with $\xi_y$ {\it{remaining finite}} at $T_c$. This way of introducing anisotropy is pertinent to ordinary fluid mixtures and identifies $x$ as the scalar order-parameter exhibiting the critical singularities -- something we shall elaborate upon later in connection with the Griffiths constraints. If we were to set $\lambda=\infty$, so that $\xi_y=0$, the particle would fall instantly to the line $y=0$, from its initial position $(x_0,y_0)$. Then $x(t)$ would evolve exactly as within the one-component theory.

\begin{figure*}[t]
\includegraphics[height=5.cm]{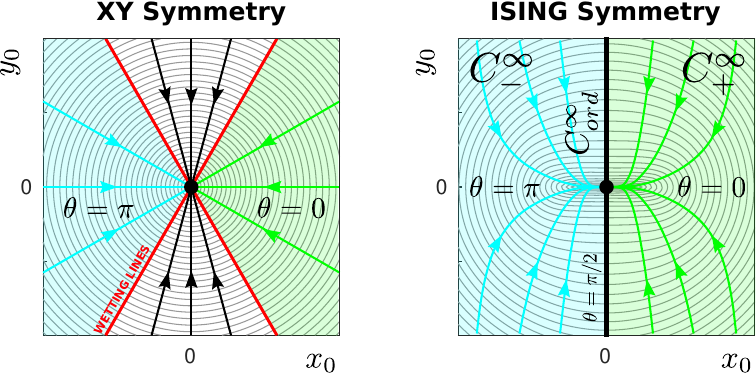}
\caption{\label{Fig35} 
Surface phase diagrams and density profile trajectories (arrows) for a two-component mixture near a wall at bulk criticality $T_c^-$, showing the limiting values of the contact angle, $\theta=0$ (green) and $\theta=\pi$ (blue):  a) Component isotropy/XY symmetry ($\lambda=0$) with a remnant phase boundary (red) at $\phi=\pi/3$ and $2\pi/3$ and non-wetting gap between them (white), where the contact angle $0<\theta<\pi$ at bulk criticality. b) Component anisotropy/Ising symmetry ($\lambda>0)$, with critical point wetting and regions of critical adsorption, $C_+^\infty$, and desorption $C_-^\infty$. In this case, the non-wetting gap is absent. \\
}
\end{figure*}

Let us now define scaling variables which adsorb the dependence on $a\propto \sqrt{T_c-T}$. Defining  $\tilde x= x/a$, $\tilde y= y/a$ and $ t'=ta$, the equation for the conservation of energy, reads
\begin{equation}
\frac{1}{2}\big(\,\dot{\tilde x}^2+\dot{\tilde y}^2\big)=\big((\tilde x-1)^2+\tilde y^2\big)\big((\tilde x+1)^2+\tilde y^2\big)+\frac{1}{2}\Big(\frac{\lambda}{\,a^2}\Big) \tilde y^2
\end{equation}
showing that the anisotropy rescales as $\lambda/a^2$ and is, therefore, a {\it{relevant field}}. Thus, the wetting phase boundary is $|x_0|=\sqrt{a^2+y_0^2/3}$ for $\lambda=0$ giving the gap, while $\lambda=\infty$ is the one-component theory, for which the phase boundary is $|x_0|=a$, giving critical point wetting \cite{Nakanishi1982}. Both phase boundaries are the same for large $a$, but since $\lambda$ is a relevant variable, the presence of any anisotropy drives the trajectories to those of the one-component theory leading to critical point wetting as $a\to 0$. More specifically, 
as $\lambda/a^2\to \infty$, the rescaled trajectories become vertical lines and $\tilde x(t')$ reduces to the result of the one-component theory with wetting phase boundaries at $|\tilde x|=1$. Equivalently, at $T_c$, the true unscaled trajectories approach the origin as straight lines $y\propto x$,  when $\lambda=0$, corresponding to a star node representing XY surface criticality, where both components decay as a power-law 
\begin{equation}
(x(t),y(t))\sim(\cos\phi,\sin\phi)/\sqrt{2}t
\end{equation}
These trajectories do not cross the limit of the wetting phase boundaries $\phi=\pi/3$ and $2\pi/3$, allowing for a non-wetting gap -- see Fig.\ \ref{Fig35}. However, for $\lambda>0$, the trajectories approach the origin as $y\propto \exp(-\sqrt{\lambda/2x^2})$ and form a stable node, as pertinent to Ising surface criticality. In this case, all trajectories, except those at $x_0=0$ where $\theta=\pi/2$ by symmetry, would cross any wetting phase boundary, so critical point wetting must be restored. At $T_c$, a line of ordinary surface transitions $C_{ord}^\infty$, at wetting neutrality, separates regions of critical adsorption and desorption $C_\pm^{\infty}$, where {\it{only}} the $x$ component decays as a power-law  $x\sim\pm (\sqrt{2}t)^{-1}$, and the limiting values of the contact angle are $\theta=0$ and $\theta=\pi$, respectively -- see Fig.\ \ref{Fig35}. For the ternary mixture, component anisotropy near the $\alpha\beta$ CEP means that the intercept with the lines of wetting by $\alpha$ or $\beta$ occurs at $\tilde t=1/2$, which is the point of wetting neutrality, leading to Fig.\ \ref{Fig03}b. The two different scenarios represented in Fig.\ \ref{Fig03},  therefore, reflect the local symmetry of the order-parameter near criticality where, in each case, the contact angles are continuous functions of $T$ and $p$ within the three-phase region. Therefore, the presence or absence or critical point wetting is determined by a universality principle and simply reflects the universality class of the surface critical behaviour. We believe this is robust, and is unaltered by the inclusion of interfacial fluctuations occurring beyond mean-field, or if long-ranged dispersion forces are present, since these do not alter wetting phase boundaries \cite{Brezin1983,Squarcini2022,Evans2019}. 

\subsection{Ordinary fluids: Square-gradient theory, the Sullivan model, and Griffiths revisited}

 In ordinary binary and ternary fluid mixtures, anisotropy is synonymous with the Ising universality class describing bulk critical behaviour, and, for example, is the reason why the structure factor associated with the concentration, but not the number density, is singular at $T_c$ in the Bhatia-Thornton theory of resistivity in binary alloys \cite{Bhatia1970}. In microscopic approaches, based on density functional models,  bulk criticality only requires that the second derivatives of the grand potential per unit volume, $\omega(x,y)$, where here we may suppose $x=n_1$ and $y=n_2$ are the homogeneous component densities, or linear combinations of them expanded about the critical point, satisfies
\begin{equation}
\omega_{xx}\omega_{yy}-\omega_{xy}^2=0
\label{det}
\end{equation}
 whereas in the KWI model all the second derivatives vanish, i.e. $\omega_{xx}=\omega_{yy}=\omega_{xy}=0$. 
 
 A rather general square-gradient theory of interfaces in binary mixtures, showing both consolute and critical end points, was developed by Telo da Gama and Evans \cite{TeloDaGama1983a,TeloDaGama1983b}. In their approach the bulk potential $\omega(n_1,n_2)$, referred as $f_0(\{n_i\})$, is constructed by integrating compressibility relations, similar to the structure factors in the Bhatia-Thornton theory, involving the zeroth Fourier coefficients, $c_{ij}(q=0)$, of the three Ornstein-Zernike direct correlation functions for the two component system. Similarly, the coefficients of the gradient terms, which depend on the densities and include a cross term, involve the second-moments (the coefficient of $q^2$) in the expansions of $c_{ij}(q)$. The condition for criticality (\ref{det}) is fulfilled without a local XY symmetry. This can be demonstrated very clearly by specializing to a particular type of local density functional theory, where the connection with square-gradient theory can be precised further. This was done by Ding and Hauge, who showed how a square-gradient model, with isotropic gradient terms, as in the KWI model, may be derived from an $N$ component mean-field local density functional theory -- specifically, a generalized Sullivan model of fluid adsorption near a wall, situated in the $t=0$ plane (say), which has Yukawa intermolecular interactions \cite{Ding1989}. In this case the scaling densities $x$ and $y$ are linear combinations of the two local hard-sphere chemical potentials. For this Sullivan model, the bulk potential $\omega(x,y)$ of the equivalent square-gradient theory, with a precise mechanical analogy, can be constructed exactly. Full details of this are provided in Appendix C. Here, we need only note that, at the critical point where coexistence between a low density gas and a higher density liquid ends, the potential has the expansion
 \begin{equation}
     \omega(x,y)=\omega_{40}(x-1)^4+\omega_{20} y^2+\cdots
     \label{Ding}
 \end{equation}
 where $(1,0)$ is the location of the critical density in component space and $\omega_{20}$ and $\omega_{40}$ are both positive coefficients. This analysis of a microscopic density functional model therefore confirms the presence of anisotropy. This potential does not have a local XY symmetry and the surface criticality is the same as for the Ising universality class, where only the $x$ component displays critical behaviour. The same conclusions can be drawn for three component systems along the whole CEP line - only one component of the order-parameter, describing the "easy" direction of fluctuations in $(n_1,n_2,n_3)$ space displays critical behaviour. This is the reason why critical point wetting was still found in the numerical study by Leermakers and Ergorov, whose model is essentially the same as the Sullivan theory \cite{Leermakers2025}.

\begin{figure*}[t]
\includegraphics[height=5.cm]{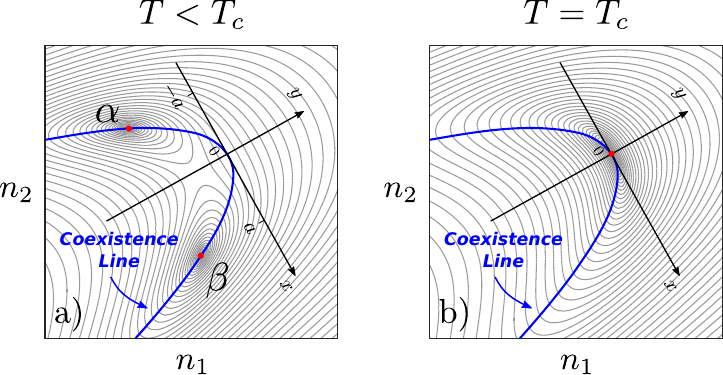}
\caption{\label{Fig36} 
Contours of the potential $\omega(x,y)=(x^2+y)^2+(y+a^2)^2$, and the orientation of the axes of the primary and secondary scaling densities, in the microscopic density plane ($n_1,n_2$) for: a) $T<T_c$, b) $T=T_c$. At mean-field level, the bulk values of the coexisting microscopic components follow a parabola, leading to the Griffiths constraint $y=-x^2$, with $x=\pm a$ for the corresponding values in the bulk $\alpha$ and $\beta$ phases. In both cases, the anisotropy of the potential is apparent. \\
}
\end{figure*}

Our final remark concerns the origin of the Griffiths constraints themselves. The location of the non-wetting gap ($\tilde t_-\le \tilde t\le \tilde t_+ $) within the KWI model comes from applying the Griffiths constraints for the three densities, $(\rho_1^\mu,\rho_2^\mu)$ in the 2D component plane, to the general expressions for the trajectories and the surface tensions. Specifically, it is the values of the bulk densities that specifies how the median angle $\phi_{\alpha\beta}$ (the angle between the $\alpha\beta$ trajectory and the bulk $\gamma$ vertex in the 2D component plane) depends on $\tilde t$ along the CEP -- recall eqn.\ (\ref{phiCEP}). However, the Griffiths constraint $\rho_2^\mu= -(\rho_1^\mu)^2$ on the secondary density is itself a consequence of an assumed Ising universality class near a consolute or critical end point, which requires an anisotropy. This was discussed by Rowlinson and Widom for the case of a binary mixture near its consolute point, where precisely the same constraint applies \cite{Rowlinson1982}.  Fig.\ \ref{Fig36} shows the coexistence curve for the microscopic component densities $n_1$ and $n_2$, and how the primary, $x$, and secondary, $y$, bulk scaling densities are defined in terms of them. The primary density  $x$ describes a coordinate along the tie lines of the coexisting microscopic densities at $(n_1^\alpha,n_2^\alpha)$ and $(n_1^\beta,n_2^\beta)$. The direction of the primary density is therefore fixed and is used to model the bulk potential $\omega(x)= (x^2-a^2)^2$, with $a\propto (T_c-T)^{1/2}$, in one-component descriptions - this is the "easy" direction for fluctuations. In the two-component description the secondary density $y$ is chosen to span the same space reflecting the {\it{parabolic}} shape of the mean-field coexistence curve. Ignoring unimportant constants, the two-component potential that models this is
\begin{equation}
\omega(x,y)\approx (x^2+y)^2+(y+a^2)^2
\end{equation}
The minima of this potential occur at $x=\pm a$ and $y=-a^2$, leading to the Griffiths constraint that the bulk densities must satisfy $y=-x^2$. The bulk critical singularities are determined by the primary density $x$, which is the only density for which the associated bulk correlation length diverges. The Griffiths constraint on the secondary density therefore implicitly assumes an underlying anisotropy, and the resulting surface criticality modeled by these potentials belongs to the Ising universality class as in Fig.\ \ref{Fig35}. Note that at the critical point, $a=0$, the potential is the same as (\ref{Ding}), as obtained from a fully microscopic density functional theory.  \\

\subsection{The KWI model its implications}

We end our paper with a summary of the implications and insights into interfacial phenomena and wetting transitions, which arise from the exact solutions of the square-gradient models described here: \\

I) Non-wetting gaps are present in systems with short-ranged forces if there is a local XY symmetry arising from component isotropy. This occurs both for wetting involving only fluid-fluid interfaces, as for the KWI model of a ternary mixture, and for wall-fluid interfaces, as shown for the model of a binary mixture. The existence of non-wetting gaps reflects the change in surface critical behaviour when there is an XY symmetry at a critical point or critical end point.\\

II) The local XY symmetries that lead to a non-wetting gap are not present for ordinary fluid mixtures, where criticality belongs to the Ising universality class and leads to "normal" surface critical behaviour \cite{Fisher1978,Fisher1990a}. The solution of the KWI model and of the binary mixture, therefore, provides strong support that critical point wetting is present for ordinary fluid mixtures, and certainly microscopic models of fluids with short-ranged forces, reflecting the universality class of surface phase transitions. \\

III) Wetting transitions involving critical layers that have an XY symmetry, for example a superfluid along the $\lambda$ line, would have different critical singularities to those in ordinary fluids. This reflects the change in the Casimir interaction, between the unbinding $\alpha\beta$ and $\beta\gamma$ interfaces, mediated by the critical $\beta$ phase.\\

IV) Tricritical point wetting is more general than critical point wetting in ternary mixtures. Even when non-wetting gaps exist at the CEP lines, the phase of intermediate density $\beta$ wets the $\alpha\gamma$ interface near the TCP.\\

There are two other very surprising features of square-gradient models which have a local $XY$ symmetry, which relate microscopic, mesoscopic and macroscopic interfacial properties: \\

V) The density profile $x_{w\beta}(t)$, of the non-critical wall-$\beta$ interface, is dual to the disjoining pressure for the wetting transition at the wall-$\alpha$ interface. These exhibit the same qualitative and quantitative properties near the transition.  \\

\begin{figure*}[h]
\includegraphics[height=5.cm]{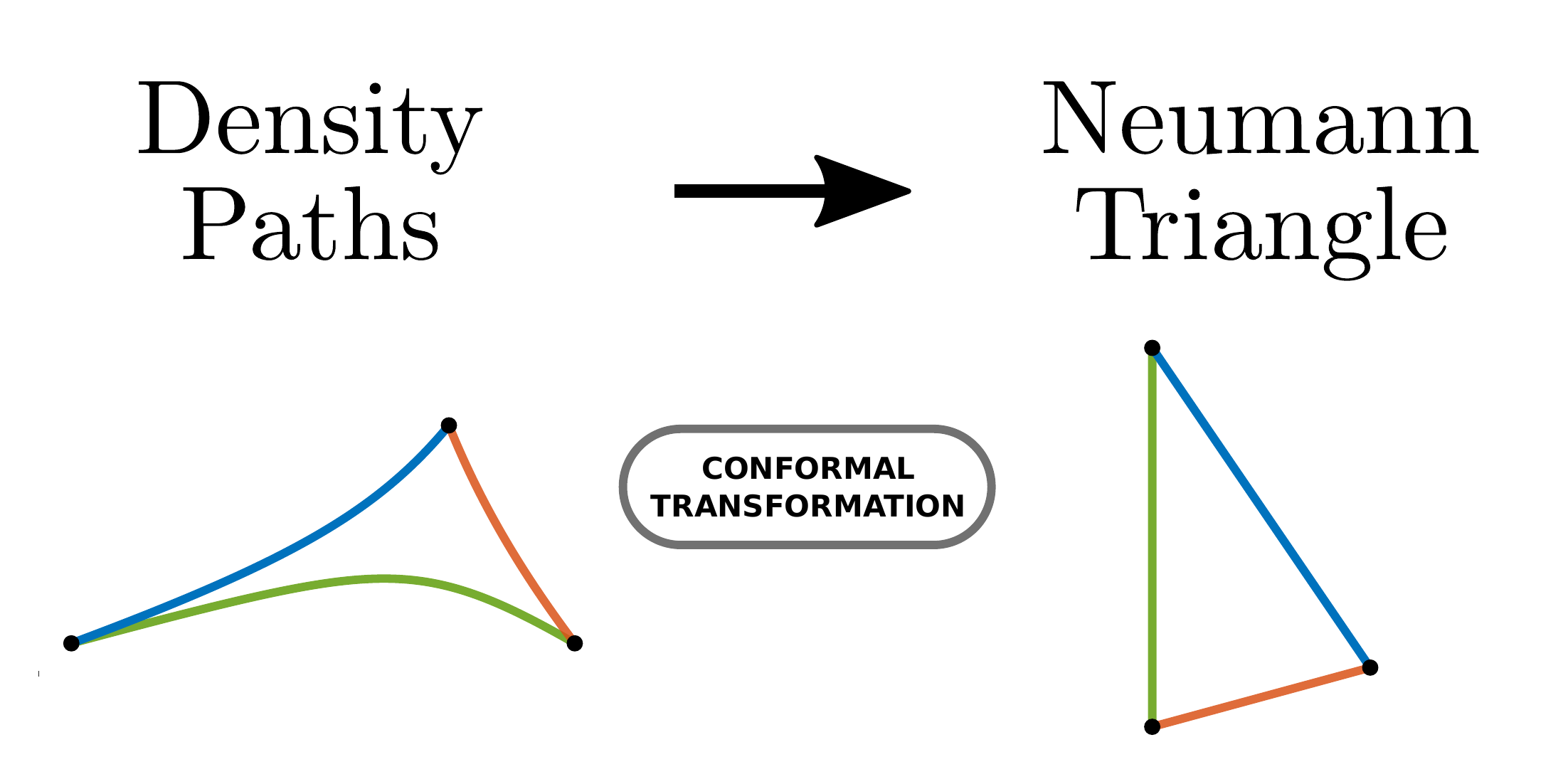}
\caption{\label{Fig37} The straightening of the microscopic density profile paths under a conformal transform which maps the tricuspid onto the Neumann triangle for the surface tensions and contact angles.
 \\
}
\end{figure*}

VI) There is a direct mapping between microscopic density profile paths and the Neumann triangle specifying the contact angles of a macroscopic drop near the contact line -- see Fig.\ \ref{Fig37}.\\ 

These predictions can certainly be tested using more microscopic models. We mention in particular that considerable advances have been made recently combining  machine learning methods within the formalism of density functional theory \cite{Sammuller2025,Robitschko2025}. This has already been used, very successfully, to study the coexistence curve, surface tension and complete drying at hard-walls in simple fluids \cite{Henderson1985}. Applying these methods to binary mixtures near walls, and the free interfaces of ternary mixtures, would test whether critical point wetting occurs for ordinary fluid mixtures always, as our analysis suggests. Studies of ternary mixtures could also test whether a transition from partial to complete wetting occurs on approaching the TCP.

Finally, of course, further experimental studies of wetting in fluid mixtures would be very welcome. Wetting in ternary mixtures has an advantage over wetting at solid-fluid interfaces because the location of the non-critical phase $\gamma$ (say), in component space relative to $\alpha$ and $\beta$, changes as we move along the CEP. As shown here this emulates the variation of the surface-field strength determining the fluid density at the wall in binary mixtures, which is of course much harder to directly manipulate. This is one of the reasons why wetting transitions at solid-fluid interfaces, and in particular continuous (critical) wetting transitions, have proved so elusive experimentally. Wetting transitions in three-phase fluid mixtures do not suffer that problem, where the distance along the CEP replaces the surface field strength. Indeed several previous studies have observed both first-order and continuous wetting transitions. For example, mixtures of alkanes of different lengths show first-order, continuous and tricritical wetting transitions \cite{Ross2001}, whose surface phase diagram is  in accord with the square-gradient theory of Nakanishi and Fisher theory, and hence the presence of critical point wetting. Some studies of water, oil and non-ionic amphiphile mixtures have reported that partial wetting may persist very close to the critical end points, although issues with reaching equilibrium have also been noted \cite{Kahlweit1993}. Revisiting these would be extremely useful in clarifying if partial wetting persists, or if a wetting transition occurs very close to the CEP.

\begin{acknowledgments}
We are deeply indebted to J.O.\ Indekeu and K.\ Koga for  sharing numerical results for trajectories near the critical end points. This was invaluable in leading to the analytical solution. We are particularly grateful also to R.Evans, for a detailed reading of our manuscript, and to R.P.W.\ Thomas for patiently explaining the theory of algebraic curves. We have also benefited from many discussions T.\ Bertrand, Y.\ Mart\'inez-Rat\'on and A.\ Malijevsk\'y. CR acknowledges the support  of grant PID2021-126307NB-C21(MCIN/AEI/10.13039/501100011033/FEDER,UE).
\end{acknowledgments}

\appendix

\section*{Appendix A: Exact profiles at wetting in the KWI model}

We start by considering the wetting transition occurring along the line of symmetry with $\tilde s=0$. With bulk vertices at $(-a_w,-a_w^2)$, $(0,0)$ and $(a_w,-a_w^2)$, the equation for all three trajectories, at the wetting transition, is
\begin{equation}
 x^4+y^4+\frac{8}{3}a_w^2y^3+2a_w^2(1+a_w^2)(x^2-y^2)-6x^2y^2-8a_w^2x^2y=0
 \label{trajsym}
\end{equation}
Since a rational parameterization is available, we write
\begin{equation}
    x=x(\tau),\hspace{1cm} y(\tau)=\tau x(\tau)
\end{equation}
and substitution into (\ref{trajsym}) gives
\begin{equation}
  (\tau^4-6\tau^2+1)x^2+\frac{8}{3}a_w^2\tau(\tau^2-3)x+2a_w^2(1+a_w^2) (\tau^2-1)=0 
  \label{xquad}
\end{equation}
This has the solution
\begin{equation}
x(\tau)=\frac{3a_w^2(1+a_w^2)(\tau^2-1)}{2a_w^2\tau(3-\tau^2)+3c(\tau^2-a_w^2)\sqrt{1-\tau^2/b^2}}
\label{xtausym}
\end{equation}
which is equivalent to (\ref{xtau}), and we have specialized to the $\alpha\beta$ section of the trajectory running from $\tau=a$ at $\alpha$ to $\tau=1$ at $\beta$. This partial form of the rational parameterization will be sufficient to determine the profile exact and we will not need to specify $\tau$ as a function of some rational parameter $s$ as in (\ref{taus}). Instead we shall determine the dependence of $\tau$ on "time" $t$ i.e. the original spatial coordinate.  

Substituting $y=\tau x$ into the first-order equation (\ref{xdotsym}) for $\dot x$ gives
\begin{equation}
    \frac{\dot x}{\sqrt2}=\tau(3-\tau^2)x^3+2a_w^2(1-\tau^2)x^2-a_w^2(1+a_w^2)\tau x
\end{equation}
and similarly substituting into (\ref{ydotsym}) for $\dot y$ gives
\begin{equation}
    \frac{\dot y}{\sqrt2}=(1-3\tau^2)x^3-4a_w^2\tau x^2-a_w^2(1+a_w^2)x
\end{equation}
But $\dot y=\dot\tau x+\dot x\tau$ and after combining the above equations we arrive at the dynamical equation for the parameter $\tau(t)$
\begin{equation}
\frac{\dot\tau}{\sqrt2}=\frac{3a_w^2(1+a_w^2)c(1-\tau^2)(\tau^2-a_w^2)\sqrt{1-\tau^2/b^2}}{2a_w^2\tau(3-\tau^2)+3c(\tau^2-a_w^2)\sqrt{1-\tau^2/b^2}}
\label{taudotsym}
\end{equation}
This first-order equation can be integrated giving a transcendental equation for $\tau$ as a function of $t$,
\begin{equation}
    \Big(\frac{\sqrt{b^2-1}+\sqrt{b^2-\tau^2}}{1-\tau}\Big)^{\xi_\beta}\Big(\frac{\tau^2-a_w^2}{(\sqrt{b^2-a_w^2}+\sqrt{b^2-\tau^2})^2}\Big)^{\xi_\alpha}=e^{t-t_0}
    \label{ttau}
\end{equation}
where $t_0$ is an arbitrary constant. Here $\xi_\alpha=1/\kappa_\alpha$ and $\xi_\beta=1/\kappa_\beta$ are length scales which determine the decay of $\tau(t)$, and hence $x(t)$ and $y(t)$, each side of the $\alpha\beta$ interface: thus $\tau(t)=1+\mathcal{O}(e^{-\kappa_\beta t})$, as $t\to\infty$, while $\tau(t)=a_w+\mathcal{O}(e^{\kappa_\alpha t})$ as $t\to-\infty$. These lengths are identical to the two bulk correlations identified form the curvature of the KWI potential, $\kappa_\mu^2=\omega_{xx}(x_\mu,y_\mu)$, which along the line symmetry take the values 
\begin{equation}
    \kappa_\alpha=2\sqrt2a_w^2\sqrt{1+a_w^2}=4(2\sqrt3-3)\sqrt{\sqrt3-1}
\end{equation}
and
\begin{equation}
\kappa_\beta=\sqrt2 a_w^2(1+a_w^2)=2\sqrt2(9-5\sqrt3)
\end{equation}
Thus $\tau(t)$ itself has an interfacial structure controlling both density profiles, $x(t)$ and $y(t)$, of the full two-component KWI model.

 The above analysis can be generalized to determine the  density profiles for the general $\alpha\gamma$ path at wetting or equivalently the separate $\alpha\beta$ and $\beta\gamma$ paths that meet at a right angle. This is, by far, the most algebraically demanding part of the exact solution of the KWI model, and is handled most efficiently using the complex analysis. The first step is to find the rational parameterization of the curve. To do this we translate the locations of the bulk vertices so that $\beta$ lies at the origin. Thus
\begin{equation}
 z_\alpha=-a-\ell e^{i\phi},\hspace{1cm}z_\beta=0,\hspace{1cm}z_\gamma=a-\ell e^{i\phi}
\end{equation}
The equation for the $\alpha\gamma$ curve at wetting is
\begin{equation}
    \Re{(f(z))}=0
\end{equation}
where
\begin{equation}
    f(z)=-ie^{-i\phi}z^2\Big(z^2+\frac{8}{3}z\ell e^{i\phi}+2(\ell^2e^{2i\phi}-a^2)\Big)
\end{equation}
and the R.H.S. is zero since, at the phase boundary the trajectory passes through the $\beta$ vertex at the origin. We now consider the same parameterization 
\begin{equation}
    x=x(\tau), \hspace{1cm} y(\tau)=\tau x(\tau)
\end{equation}
so that the complex variable is
\begin{equation}
    z(\tau)= (1+i\tau)x(\tau)
\end{equation}
Substitution then reduces the quartic in $z$ to only a quadratic equation in $x$, given by
\begin{equation}
    \Re \Big(i(1+i\tau^2)e^{-i\phi}\big((1+i\tau)^2x^2+\frac{8}{3}\ell(1+i\tau)e^{i\phi}x+2(\ell^2e^{2i\phi}-a^2)\big)\Big)=0
\end{equation}
With some patience this can be solved giving the two roots that represents the parameterized $\alpha\beta$ and $\beta\gamma$ parts of the curve. For the $\alpha\beta$ interface we find
\begin{equation}
   \frac{x}{a}=\frac{6\sin\phi}{\sqrt{1+\frac{2}{\sqrt3}\sin\phi}}\frac{(\tau-\tau_\beta^-)(\tau-\tau_\beta^+)}{(2\tau(3-\tau^2)+(\tau-\tau_\alpha)(\tau-\tau_\gamma)\sqrt{(1-\sqrt3\sin\phi)\tau^2-2\sqrt3\tau\cos\phi+\sqrt3(\sqrt3+\sin\phi)})}
   \label{xtaugen}
\end{equation}
where the $\beta\gamma$ part has a negative sign in front of the final square root term in the denominator. In this expression
\begin{equation}
    \tau_\beta^\pm=-\frac{\cos\phi}{\sqrt3+\sin\phi}\mp\sqrt{\frac{\cos^2\phi}{(\sqrt3+\sin\phi)^2}+1}
\end{equation}
are the values of $\tau$ at $x=0^\pm$. Note that the product $\tau_\beta^+\tau_\beta^-=-1$, which is equivalent to the condition, $\tilde \beta=\pi/2$, that the $\alpha\beta$ and $\beta\gamma$ paths meet at a right angle. Similarly
\begin{equation}
    \tau_{\alpha,\gamma}=\frac{\sin\phi}{\cos\phi\pm\sqrt{1+\frac{2}{\sqrt3}}\sin\phi}
\end{equation}
are the values of the parameter $\tau$ at $\alpha$ ($+$ sign) and $\gamma$ ($-$ sign). Thus, for example, the parameterization of the $\alpha\beta$ curve runs from $\tau=\tau_\alpha$ to $\tau=\tau_\beta^-$, and similarly between $\tau_\beta^+$ and $\tau_\gamma$ for the $\beta\gamma$ section. This reduces to (\ref{xtausym}) when $\phi=\pi/2$, corresponding to the line of symmetry where the bulk vertices form an isosceles triangle.

We now use either of the dynamical equation of motions (\ref{dynz}), which for the complex conjugate gives
\begin{equation}
    \frac{\dot{\bar z}}{\sqrt2}=-e^{-i\phi}(z^3+2\ell e^{i\phi}z^2+(\ell^2e^{2i\phi}-a^2)z)
\end{equation}
together with the rational parameterization of the conjugate, $\bar{z}=(1-i\tau)x(\tau)$. Explicitly, we have
\begin{equation}
    \frac{d}{dt} (1-i\tau)x=\sqrt{2}e^{-i\phi}x((1+i\tau)^3x^2+2(1+i\tau)^2\ell e^{i\phi}x+(1+i\tau)(\ell^2e^{2i\phi}-a^2))
\end{equation}
and comparing the $\Re$ and $\Im$ parts allows us to determine that
\begin{equation}
    \frac{\dot\tau}{\sqrt2}=\frac{2}{3}\ell\tau(3-\tau^2)x+(\ell^2+a^2)(1-\tau^2)\sin\phi+2(\ell^2-a^2)\tau\cos\phi
\end{equation}
Then, using (\ref{xtaugen}), for the rational parameterization of $x(\tau)$, we arrive, finally at the first-order equation  
\begin{equation}
    \dot\tau=-d(\phi)\Theta(\tau)
    \end{equation}
where 
\begin{equation}
            \Theta(\tau)=\frac{(\tau-\tau_\beta^-)(\tau-\tau_\beta^+)(\tau-\tau_\alpha)(\tau-\tau_\gamma)\sqrt{(1-\sqrt{3}\sin\phi)\tau^2-2\sqrt{3}\tau\cos\phi+\sqrt{3}(\sqrt{3}+\sin\phi)}}{2\tau(3-\tau^2)+(2+\sqrt{3}\sin\phi)(\tau-\tau_\alpha)(\tau-\tau_\gamma)\sqrt{(1-\sqrt{3}\sin\phi)\tau^2-2\sqrt{3}\tau\cos\phi+\sqrt{3}(\sqrt{3}+\sin\phi)}}
            \label{Theta}
\end{equation}
and $d(\phi)=2\sqrt{2}a^2\sin\phi (\sqrt{3}+\sin\phi)$. Notice that the function $\Theta(\tau)$ vanishes, as required, at the three bulk vertices $\alpha$, $\beta$ and $\gamma$, which are the stationary points of the motion and recovers (\ref{taudotsym}) when $\phi=\pi/2$. The first-order equation for $\dot\tau$ can be integrated to obtain the inverse function, $t(\tau)$ which then implicitly determines the coordinates/ component profiles from the parameterization $x(\tau)$ and $y=\tau x(\tau)$. We note that at $\phi= 0,\pi$ the discriminant is a perfect square. This corresponds to the scaling limit, near a CEP, where the analysis simplifies and is equivalent to that of a binary mixture at a wall.

\section*{Appendix B: The Levi-Civita map in orbital mechanics and Bogomolny equations}

The solution to the KWI model is equivalent to solving a problem in 2D classical mechanics for a potential that does not correspond to a central force, i.e.\ where there is no conservation of angular momentum. The solution is restricted to energy $E=0$, describing free-interfaces or semi-infinite systems. In this case, for potentials that factorize as $\omega(x,y)=w(z)\overline{w(z)}$, the conservation of energy 
also factorizes into the first-order equation, $
\dot{\overline{z}} = if '(z)$,
together with its complex conjugate, where $f'(z)=\sqrt{2}e^{i\psi}w(z)$ defines the complex trajectory function $f(z)$. It is this factorization that makes the model integrable. This can also  be viewed as part of a suitable transformation to straight line, inertial, motion. In the original Cartesian coordinates, the Euler-Lagrange equations are equivalent to Newton's law $ \ddot{\bf{x}}=\nabla\omega(x,y)$, where ${\bf{x}}=(x,y)$. We now transform to new spatial coordinates ${\bf{u}}=(u,v)$, using $f(z)$ as a conformal map, and also a new time variable $dt'=|f'(z)|^2\;dt\,$:
\begin{equation}
    f(x+iy)=u+iv,\hspace{1cm}dt'=2\;\omega(x,y)\;dt
\end{equation}
where, in the transformation of the time variable, the dependence of $x$ and $y$ on $t$ is implicit. Then, for $E=0$, the acceleration and velocity in the new coordinates,   corresponds to the simple linear motion of a free particle, i.e.
\begin{equation}
\ddot{\bf{u}}=0,\hspace{1cm}\dot{{\bf{u}}}=-1
\label{inertial}
\end{equation}
which follow from (\ref{udotvdot}), and where now the dot denotes differentiation with respect to $t'$. The linear motion in the $(u,v)$ plane, reduces the dimensionality of the problem, thereby determining the path in the original $(x,y)$ coordinates. The transformation of the time variable means that there is also conservation of linear momentum in the new coordinates.

We can apply this method to the familiar 2D Kepler problem since $\omega(x,y)=(x^2+y^2)^{-1/2}\equiv 1/r$ is the potential of the attractive inverse square force law. Then $w(z)=z^{-1/2}$ so the trajectory function is simply $f(z)\propto\sqrt{z}$. The equation for the path, $\Re( \sqrt{z})=\sqrt{r_0}$, with $z=r e^{i\theta}$, then immediately recovers the usual polar equation, $r(\theta)=2 r_0/(1+\cos\theta)$, for the parabolic orbit of a body with energy $E=0$. In this particular Kepler problem, the transformation of the coordinates, $x+iy=(u+iv)^2$, and $d t'\propto dt/r$, are the same as the Levi-Civita mapping used to regularize orbital motion \cite{Arnold1989}. This mapping is, of course, not required to solve the equations of motion but is introduced to remove singularities in them when the distance between the orbiting bodies is small. When applied to elliptical orbits, corresponding to $E<0$, this maps the equation of motion onto that of a harmonic oscillator in the $(u,v)$ plane with an angular frequency $\Omega=\sqrt{-E}$. When $E=0$, this simplifies further to the equations of motion, eqn.(\ref{inertial}), of a free particle. The method we have used here can therefore be viewed as a generalization of the Levi-Civita mapping to a broader class of potentials, for which $\omega(x,y)=w(z)\overline{w(z)}$, but restricted to the specific case of $E=0$ which always reduces the problem to the linear motion of a free particle. This is no longer the case when $E \ne 0$ and the motion remains two dimensional in the transformed $(u,v)$ plane. In this case, it is not difficult to show that the equation for the conservation of energy in the transformed coordinates reads
\begin{equation}
    \frac{1}{2}\dot{\bf{u}}^2=\frac{1}{2}+\frac{E}{2\;\omega(x,y)}
\end{equation}
where on the R.H.S. we must substitute for $x=x(u,v)$ and $y=y(u,v)$ into $\omega(x,y)$ thus generating an effective potential for the motion in the $(u,v)$ plane. For the Kepler problem, when $E<0$, for example, this is just the mechanical potential of a harmonic oscillator, $\Omega^2(u^2+v^2)/2$, where $\Omega=\sqrt{-E}$, is the angular frequency mentioned above \cite{Arnold1989}. For the KWI model, however, or indeed even the binary mixture, the potential generated by the conformal transform is much more complicated  and neither reduces the dimensionality of the problem or leads to a conservation law, {\it{unless}} $E=0$. This means that we cannot use the analytical methods developed here to study confinement between two planar walls in these models. Finally, we stress that, within the density functional square-gradient theory, the meaning of the conformal mapping is of particular physical significance since the length of the line after the mapping is precisely the surface tension of the free interface. It is this that provides a direct link between microscopic density profiles and the macroscopic shape of a drop near the contact line.

 After the completion of our work, we realised that some of the methods we have developed have appeared in an entirely different area of physics. In particular, studies of domain walls (also referred to as kinks and anti-kinks and closely related to solitons) which appear in classical two-component Wess-Zumino field theories with multiple vacua. There is, in fact, a direct correspondence between mean-field, square-gradient, studies of wetting in ternary fluid mixtures and the physics of domain wall states -- these are persistent particle-like excitations in one space and time dimension -- in models of (bosonic) fields with three vacuua/ground states (which play the role of the bulk phases). In particular, the surface tension is the same as the kink energy, wetting transitions are equivalent to kink/anti-kink splitting, and the triple line where three phases meet is referred to as a Y junction \cite{Campos1998,Townsend1999}. Building on earlier work by Bogomolny \cite{Bog76} and Prasad and Sommerfeld \cite{Prasad75} on monopoles,  it has been noted that certain potentials in Wess-Zumino models lead to integrable domain wall configurations, called BPS states \cite{Chibisov1997,Chen2016,Abel2019,Alonso2021}. This happens when the potential can be written $\omega(x,y)=(\nabla v(x,y))^2/2$, in which case the second-order equations of motion become first-order equations (Bogomolny equations) and the kink energy is a topological invariant determined by the difference in the values of the {\it{superpotential}}
$v(x,y)$ at the two vacua connected by the domain wall. This  trick relies on simply completing the square in the variational functional (the classical action) to be minimised. Seen this way, the KWI model is equivalent to an example of such a two-component Wess-Zumino field theory with a superpotential identified as the harmonic conjugate function $v(x,y)$ given by (\ref{vharm}). The Bogomolny trick is very specific to local functionals and only works for the restricted class of potentials defined by a superpotential -- which is presumably why these methods have been largely ignored by the wetting and density functional community, which often is more focused on more microscopic non-local density functional theories, which include volume exclusion effects, and also long-ranged intermolecular forces. The physics occurring  beyond mean-field/classical is also very different in the statistical physics/field theory contexts, which for fluids is associated with incorporating the capillary-wave excitations of the unbinding interfaces. This returns us to issues mentioned at the beginning of the paper regarding fluctuation effects at 3D wetting transitions \cite{Squarcini2022} -- something which has not been considered in any equivalent detail in the study of domain walls and kinks.  Nevertheless, there may well be further fruitful connections between these two different approaches which will be helpful, for example, for calculating the correlation functions of the KWI model.

 \section*{Appendix C: The Sullivan local density functional model.}

 Here we reproduce some details of the analysis of Ding and Hauge who showed how a Sullivan local density functional can be recast as a square-gradient model with a mechanical analogy \cite{Ding1989}. Setting $k_BT=1$ for convenience, the grand potential functional is written
 \begin{equation}
 \Omega[n_1,n_2]=\int_0^\infty\!\!\! dt\; \Big( f_h(n_1,n_2)+\sum_{i=1,2}(V_i(t)-\mu_i)\;n_i(t)\Big)\,+\;\frac{1}{2}\int_0^\infty \!\!\int_0^\infty\!\!\! dt\, dt'\;n_i (t)\phi_{ij}(t-t)n_j(t')
 \end{equation}
 where, in the last term, a summation over repeated indices is understood. In this expression, $n_1(t)$ and $n_2(t)$ are the microscopic component density profiles with corresponding chemical potentials $\mu_1$, $\mu_2$ and $f_h(n_1,n_2)$ is the two-component hard-sphere contribution to the free-energy. External fields, $V_1(t)$ and $V_2(t)$, act on each component which, in the Sullivan model, decay exponentially as 
 \begin{equation}
 V_i(t)=-\epsilon_i\, e^{-t}
 \end{equation}
 with strengths $\epsilon_1,\epsilon_2$. Similarly, $\phi_{11}(t)$, $\phi_{12}(t)$ and $\phi_{22}(t)$ are the attractive contributions to the three intermolecular potentials which have the same short-range specified as
\begin{equation}
\phi_{ij}(t)=-\frac{\alpha_{ij}}{2}\,e ^{-|t|}
\end{equation}
The equilibrium density profiles are found from minimization of $\Omega$ and setting $\delta\Omega/\delta n_i=0$ gives the pair of integral equations
\begin{equation}
\mu_i=\mu_{hi}(t)+V_i(t)+\int dt \;\phi_{ij}(t-t')\;n_j (t'),\hspace{1cm} i=1,2
\end{equation}
 where the $\mu_{hi}=\partial f_h/\partial n_i$ are the two chemical potentials of the hard-core systems. The choice of exponential potentials in the Sullivan model is particularly convenient because differentiating twice with respect to the distance from the wall, $t$, reduces these integral equations to 
\begin{equation}
\frac{d^2 \mu_{hi}}{dt^2}=\mu_{hi}(t)-\mu_i-\alpha_{ij}\; n_j(t),\hspace{1cm} i=1,2
\end{equation}
which are already  similar to the Euler-Lagrange equations of a square-gradient theory. In fact, Ding and Hauge show how this can be recast as a 2D mechanical problem identifying  {\it{exactly}} the potential $\omega(x,y)$ which should appear in the equivalent square-gradient theory with isotropic gradient terms. This is achieved via a linear transformation of the hard-sphere chemical potentials
\begin{equation}
x=T_{11}\; (\mu_{h1}-\mu_1)+T_{12}\; (\mu_{h2}-\mu_2),\hspace{1cm}
y=T_{21}\; (\mu_{h1}-\mu_1)+T_{22}\; (\mu_{h2}-\mu_2)
\end{equation}
where $T_{ij}$ is the matrix, defined by $T_{ij}\alpha_{jk}T_{lk}=\delta_{lk}$, that diagonalizes the $2 \times 2$ matrix of the component interaction strengths. In this way, the profiles $x(t)$ and $y(t)$ of the transformed variables are solutions of the Euler-Lagrange equation
  \begin{equation}
  \ddot x=\omega_x(x,y), \hspace{1cm} \ddot y=\omega_y(x,y)
  \end{equation}
  where the square-gradient potential $\omega(x,y)$ is given by
  \begin{equation}
  \omega(x,y)=\frac{1}{2}(x^2+y^2)+p -p_h
  \end{equation}
  where $p$ is the pressure, and $p_h$ the hard-sphere pressure expressed in these new variables. We note that boundary conditions at the wall are uncoupled, with $\dot x(0)=x(0)-2(T_{11}\epsilon_1+T_{12}\epsilon_2)$ and $\dot y(0)=y(0)-2(T_{12}\epsilon_1+T_{22}\epsilon_2)$.
  
  As a specific example, Ding and Hauge suppose that the hard-sphere mixture is modeled by an ideal lattice gas, as used in the mean-field version of the Blume-Emery-Griffiths model \cite{Blume1971}, so that
  \begin{equation}
  \mu_{hi}=\ln\frac{n_i}{1-n_1-n_2},\hspace{1cm} p_h=-\ln(1-n_1-n_2)
  \end{equation}
 and further suppose that the binary mixture is symmetric with interaction strengths
 \begin{equation}
 \alpha_{11}=\alpha_{22}\equiv\alpha,\hspace{1cm}\alpha_{12}=k\alpha
 \end{equation}
 We shall also only be interested in bulk coexistence and associated critical points, and write $\mu_1=\mu_2\equiv\mu$. In this case, the bulk potential $\omega(x,y)$ is
 \begin{equation}
 \omega(x,y)=p+\frac{1}{2}(x^2+y^2) -\ln\Big(1+2\; e^{\mu+ax}\;\cosh by\Big)
 \end{equation}
 where $a=\sqrt{(1+k)\alpha/2}$ and $b=\sqrt{(1-k)\alpha/2}$. This potential describes gas-liquid and liquid-liquid phase coexistence bounded by critical points and critical end points. For example, at the critical point where the distinction between a low density gas and higher density liquid phase ends, occurring at $x=1$ and $y=0$, the potential has the expansion
 \begin{equation}
 \omega(x,y)=\omega_{40}\; (x-1)^4+\omega_{02}\;y^2+\cdots
 \end{equation}
where $\omega_{40}=a^4/192$ and $\omega_{02}=b^2/4-1/2$ are both positive, demonstrating the presence of anisotropy. 

The mapping onto a two-component square-gradient theory fails if the component interaction strengths satisfies $\alpha_{12}=\sqrt{\alpha_{11}\alpha_{22}}$, since the determinant of the matrix of coefficients vanishes. This case is actually much simpler to study since, as shown by Sullivan for free interfaces \cite{Sullivan1982}, the two hard-sphere chemical potentials decouple. Later, Telo da Gama and Evans extended this to adsorption near a wall \cite{TeloDaGama1983}, for which the wall-fields are chosen such that $\epsilon_2=\sqrt{\alpha_{22}/\alpha_{11}}\;\epsilon_1$. With these conditions the hard-sphere chemical potentials are related via $\mu_{h2}(t)-\mu_2=\sqrt{\alpha_{11}/\alpha_{22}}\;(\mu_{h1}-\mu_1)$, i.e.\ the analysis reduces to that of an effective one-component theory. Consequently, the surface phase diagram for the binary mixture near its consolute/critical point shows lines of second-order wetting transitions by $\alpha$ when $\alpha_{11}n_1^\alpha+\alpha_{12}n_2^\alpha=2\epsilon_1$, and similarly wetting $\beta$ when $\alpha_{12}n_1^\beta+\alpha_{22}n_2^\beta=2\epsilon_1$. Here, $n_1^\mu$ and $n_1^\mu$ are the values of the bulk component densities at coexistence for each phase. The two lines of wetting transitions converge to an ordinary surface phase transition at $T_c$, at $\epsilon_1=\alpha_{11}n_1^c+\alpha_{12}n_2^c$, determined by the critical values of the bulk components, where both component profiles are flat exactly as for the one-component Nakanishi-Fisher phase diagram, Fig.\ \ref{Fig25}. We note that the decoupling of the hard-sphere chemical potentials means the potential of the equivalent two-component square-gradient theory at criticality is $\omega(x,y)=\omega_{40}x^4+\omega_{04}y^4$, where we can identify that $x=\mu_{h1}-\mu_{h1}^c$ and $y=\mu_{h2}-\mu_{h2}^c$. This potential does not have an XY symmetry which is why critical point wetting must occur.

\bibliography{wetting}

\end{document}